\documentclass[graybox]{svmult}

\usepackage{type1cm}        % activate if the above 3 fonts are
\usepackage{makeidx}         % allows index generation
\usepackage{graphicx}        % standard LaTeX graphics tool
\usepackage{multicol}        % used for the two-column index
\usepackage[bottom]{footmisc}% places footnotes at page bottom

\usepackage{newtxtext}       % 
\usepackage{newtxmath}       % selects Times Roman as basic font

\usepackage{epsfig,graphics,empheq, graphicx,amsmath,amssymb,amsfonts}
\usepackage{epstopdf}

\usepackage{color}
\usepackage{amsmath}
\usepackage{cases}
\usepackage[position=bottom]{subfig}
\usepackage{paralist}
\usepackage{enumitem}
\usepackage{graphicx}
\usepackage[percent]{overpic}
\usepackage{tikz}

\newsavebox{\largestimage}

\setlist{nolistsep}

\newcommand{\bm}[1]{\mbox{\boldmath{$#1$}}}

\def\x{{\bm{x}}}

\def\0{{\bm 0}}

\def\cl {\nonumber \\}
\def\el {\nonumber}

\makeindex             % used for the subject index
\begin{document}

\title*{A 2D kinetic model for crowd dynamics with disease contagion}
\author{Daewa Kim and Annalisa Quaini}
\institute{Daewa Kim \at Department of Mathematics, West Virginia University, 94 Beechurst Ave, Morgantown, WV 26506 \email{daewa.kim@mail.wvu.edu}
\and Annalisa Quaini \at Department of Mathematics, University of Houston, 3551 Cullen Blvd, Houston TX 77204 \email{quaini@math.uh.edu}}
\maketitle

\abstract*{
We focus on the modeling and simulation of an infectious disease spreading in a medium size
population occupying a confined environment, such as an airport terminal, for short periods of time. 
Because of the size of the crowd and venue, we opt for a kinetic type model. 
The paper is divided into two parts. In the first part, we adopt the simplifying assumption 
that people's walking speed and direction are given. The resulting kinetic model
features a variable that denotes the level of exposure to people spreading the disease, 
a parameter describing the contagion interaction strength, and a kernel function
that is a decreasing function of the distance between a person and a spreading individual.
Such model is tested on problems involving a small crowd in a square walkable domain.
In the second part, ideas from the simplified model are used to incorporate 
disease spreading in a kinetic theory approach for crowd dynamics, i.e.~the walking speed and direction
result from interaction with other people and the venue.
}

\abstract{
We focus on the modeling and simulation of an infectious disease spreading in a medium size
population occupying a confined environment, such as an airport terminal, for short periods of time. 
Because of the size of the crowd and venue, we opt for a kinetic type model. 
The paper is divided into two parts. In the first part, we adopt the simplifying assumption 
that people's walking speed and direction are given. The resulting kinetic model
features a variable that denotes the level of exposure to people spreading the disease, 
a parameter describing the contagion interaction strength, and a kernel function
that is a decreasing function of the distance between a person and a spreading individual.
Such model is tested on problems involving a small crowd in a square walkable domain.
In the second part, ideas from the simplified model are used to incorporate 
disease spreading in a kinetic theory approach for crowd dynamics, i.e.~the walking speed and direction
result from interaction with other people and the venue.
}

\section{Introduction}

The goal of this paper is to extend to 2D a 1D model presented in \cite{kim_quaini2021} to study 
the early stage of an infectious disease spreading in an intermediate size
population occupying a confined environment, e.g.~a ER wait room,
for a short period of time (minutes or hours). In this context, classical epidemiological models
fail as they rely on averaged large population behaviors over a long time span (weeks or months).
We focus on a kinetic theory approach \cite{Bellomo2017_book, Bellomo2021} for a
disease that it spreads with close proximity of individuals, like, e.g. measles, 
influenza or COVID-19. In the first part of the paper, we present a model for the disease
spreading based on the simplifying assumption that people's walking speed and direction are given.
In the second part, ideas from the simplified model are used to incorporate 
disease spreading in a kinetic theory approach for crowd dynamics.

%The onset of SARS-CoV-2, which was the cause of the initial COVID-19 outbreak, has caused a pandemic around the world and society. It has caused health problems and complex economical problems and affected social life in general. A detailed description of the biology of viral dynamics and useful indications for the global spread of infectious diseases is described in \cite{Estrada2020} . The origin and mutation of viruses is referred to \cite{Andersen2020, MacLean2020, Sanjuan2016, Doremalen2020}. See \cite{Twarock2018} for RNA virus evolution. 

%Various aspects of the immune competition are treated in \cite{Bellouquid2006, Matricardi2020, Doremalen2020}, general issues of the immune response \cite{Musiani2018}, specifically, the activation of the immune response from the innate to the activated immunity \cite{Cecconi2020Summer, Cooper2010}. A topic somehow related to the search of vaccines is referred \cite{Callaway2021, Forni2021, Faraz2020}. Additional topics on the biology of the virus are reported in \cite{Cecconi2020April, Cyranoski2020, Kissler2020, Moore2020, Twarock2018_2}. Various aspects of contagion dynamics have been studied in \cite{Gao2021}. In addition, the impact on the economy of nations is treated in various research contributions, e.g. \cite{Avishai2020, Baldwin2020,  Dimarco2020, Dosi2020, Hardy2021}. These citations do not claim to be exhaustive. Indeed, the research activity in the field keeps growing motivated by the impact of the virus spread. 

The COVID-19 pandemic has motivated a large body of literature in mathematical epidemiology.
Several possible models have been investigated, including individual-based Markov models (see, e.g., \cite{Flandoli2021}), 
Susceptible, Exposed, Infectious or Recovered (SEIR) models (see, e.g., \cite{Hethcote2000}), and
networks of nodes where each mode dynamics has a SEIR structure (see, e.g., \cite{Boscheri2021, Gatto2020}). 
Some works (e.g., \cite{Mpeshe2021}) model the evolution of the 
COVID-19 epidemics in combination with other phenomena, such as fear. 
Other works focus a multiscale modeling approach. A notable example that features
an interdisciplinary framework including applied mathematics, immunology, economics and virology 
is \cite{Bellomo_preprint}. Similarly, modeling and simulation of disease spreading in pedestrian 
crowds has recently become a topic of increasing relevance. In \cite{Derjany2020}, contact evolution
is combined with a stochastic infection spread model to simulate disease spreading in queuing 
pedestrians. Agent-based numerical simulations of pedestrian dynamics are used in \cite{Harweg2021} to 
assess the behavior of pedestrians in public places in the context of contact transmission of infectious diseases
and gather insights about exposure times and the overall effectiveness of distancing measures.
Finally, we mention the combination of a microscopic force-based model with a contact tracking method 
in \cite{Rathinakumar_quaini2020} to simulate the initial spreading of a highly infectious airborne disease 
in a confined environment.

%Crowd dynamics also develop to include social awareness \cite{Aylaj2020, Bellomo2019_new, Salam2021}. Modeling the contagion parameter is one of the key problem in crowds \cite{Echeverria2021}. 

The model that we propose for disease spreading in a walking crowd takes
inspiration from the work on emotional contagion (i.e., spreading of fear or panic) 
in Ref.~\cite{Bertozzi2014ContagionSI,Bertozzi2015}. In order to focus on the spreading mechanism, we first introduce 
a simplifying assumption: the walking speed and direction are given. The key features
of the resulting simplified model are:
\begin{itemize}
\item[-] a variable that denotes the level of exposure to people spreading the disease, with the underlying
idea that the more a person is exposed the more likely they are to get infected;
\item[-] a model parameter that describes the contagion interaction strength;
\item[-] a kernel function that is a decreasing function of the distance between
a person and a spreading individual.
\end{itemize}
We show preliminary results for a problem involving a small crowd in a square walkable
domain with no obstacles or walls.

In the second part of the paper, the simplifying assumption is removed: the walking speed and direction
result from a kinetic model of crowd dynamics, which incorporates a term to account for disease spreading. 
Kinetic models are derived from observing the system at the mesoscale (see, e.g.,~\cite{Agnelli2015,Bellomo2011383,Bellomo2017_book,Bellomo2013_new,Bellomo2015_new,Bellomo2016_new,Bellomo2019_new}),
i.e.~the intermediate scale between the macroscopic one  (see, e.g., \cite{Cristiani2014_book,Hughes2003,5773492})
and the microscopic one (see, e.g., \cite{ASANO2010842,Chraibi2019,DAI20132202,Helbing1995}).
The framework for mesoscale models comes from the kinetic theory of gases, with the main 
difference that interactions of ``active'' particles are irreversible, 
non-conservative and, in some cases, non-local and nonlinearly additive~\cite{Aristov_2019}.
In general, one derives the kinetic model from interactions at the microscopic scale and the 
same modeling principles can lead to macroscopic (or hydrodynamic) models. See, e.g., Ref.~\cite{Bellomo2020}.
We remark that the approach to model crowd dynamics with disease contagion presented here
is different from the one in Ref.~\cite{kim_quaini2020}, where two models (one for pedestrian dynamics
and one for disease spreading) are coupled. 

The particular kinetic model for crowd dynamics that we extend to account for disease spreading was first presented 
in \cite{kim_quaini}. 
Based on earlier works \cite{Agnelli2015,Bellomo2015_new}, the model's main features are
\begin{itemize}
\item[-] discrete walking directions;
\item[-] interactions modeled using tools of stochastic games;
\item[-] heuristic, deterministic modeling of the walking speed corroborated by
experimental evidence~\cite{Schadschneider2011545}.
\end{itemize} 
In \cite{kim_quaini}, the model has been shown to compare favorably against experimental
data related to egression from a facility of a medium-sized group of people
(40 to 138 pedestrians) \cite{Kemloh}. In addition, we demonstrated that 
realistic scenarios, such as passengers moving through one terminal of Hobby Airport in Houston (USA),
can be reproduced \cite{kim_quaini2021}.

%A multiscale model of virus pandemic accounting for the interaction of different spatial scales (from the
%small scale of the virus itself and cells, to the large scale of individuals and further up to
%the collective behavior of populations) is presented in Ref.~\cite{Bellomo_preprint}.
 
The paper is organized as follows. We introduce our simplified contagion model in Sec.~\ref{sec:contagion}
and its numerical discretization in Sec.~\ref{sec:discr}. Numerical results are shown in 
Sec.~\ref{sec:num_res}. Sec.~\ref{sec:complex} presents the kinetic theory approach for crowd dynamics
extended to incorporate disease spreading. Conclusions are drawn in Sec.~\ref{sec:concl}.

%%%%%%%%%%%%%%%%%%%%%%%%%%%%%%%%%%%%%%
\section{A simplified two-dimensional kinetic model}\label{sec:contagion}
% Always give a unique label
% and use \ref{<label>} for cross-references
% and \cite{<label>} for bibliographic references
% use \sectionmark{}
% to alter or adjust the section heading in the running head
We start from an agent-based model. At the microscopic level
we consider a group of $N$ people divided between healthy or not spreading the disease yet
($N_h$) and actively spreading ($N_s$), with $N = N_h + N_s$.
If person $n$ belongs to the group of healthy (or not spreading) people, we denote with $q_n \in [0, 1)$
their level of exposure to people spreading the disease, with the underlying
idea that the more a person is exposed the more likely they are to get infected.
If person $n$ is actively spreading the disease, then $q_n = 1$ and this value stays constant throughout the entire simulation time. 

Let $[t_0, t_f]$ be a time interval of interest. 
Let  $\x_n(t)=(x_n(t), y_n(t))^T$ and $v_n(t)$ denote the position and speed of person $n$
for $t \in [t_0, t_f]$, with initial position $\x_n(t_0)$ and initial speed $v_n(t_0)$ being given.
The microscopic model reads for $n= 1,2,3, \dots, N$: 
\begin{align}
&\frac{d\x_n}{dt}=v_n(\cos\theta, \sin\theta)^T, ~ \frac{dq_n}{dt}=\gamma \max\{(q_n^*-q_n), 0\},~ q_n^* =\frac{\sum_{m=1}^{N} \kappa_{n,m}q_m}{\sum_{m=1}^{N_s}\kappa_{n,m}}, \label{eq:diseasef}
\end{align}
where the walking speed $v_n$ and walking direction $\theta$ are assumed to be given.
In model \eqref{eq:diseasef}, $q_n^{\ast}$
corresponds to a weighted average ``level of sickness'' surrounding person $n$, with 
$\kappa_{n,m}$ serving as the weight in the average. 
We define $\kappa_{n,m}$ as follow
\begin{equation}\label{kappa}
\kappa_{n,m} = 
\left\{
\begin{split}
& \kappa(|\x_n-\x_m|) = \dfrac{R}{(|\x_n-\x_m|^2+R^2) \pi} \quad \text{if }m\text{ is spreading}   \\
&0 \hspace{6.2cm} \text{otherwise}
\end{split}
\right.
\end{equation}
Notice that if person $m$ is spreading the disease, the interaction kernel 
is a decreasing function of mutual distance between two people and is parametrized by an interaction distance $R$.
The value of $R$ is set so that the value of $\kappa_{n,m}$ is ``small'' at about 6 ft or 2 m. More details are  
given in Sec.~\ref{sec:num_res}. The meaning of
parameter $\gamma$ in \eqref{eq:diseasef} is contagion strength: 
for $\gamma = 0$ there is no contagion, while for
$\gamma \neq 0$ the contagion occurs and it is faster for larger the values of $\gamma$. 
Note that obviously the level of exposure can only increase over time. In addition, the second equation in \eqref{eq:diseasef}
ensures that the people spreading the disease will constantly have $q_n = 1$ in time.

In order to derive a kinetic model from the agent-based model \eqref{eq:diseasef}, we introduce the empirical distribution:
\begin{equation}
h^N=\frac{1}{N}\sum_{n=1}^N\delta(\x-\x_n(t))\delta(q-q_n(t)), \cl
\end{equation}
where $\delta$ is the Dirac delta measure.
We assume that all people remain in a fixed compact domain $\Omega$ for the entire time interval under consideration:
$(\x_n(t), q_n(t)) \in \Omega \subset \mathbb{R}^3$, for $n= 1,2,3, \dots, N$ and $t \in [t_0, t_f]$.
Let $\mathcal{P}(\mathbb{R}^3$) be the space of probability measures on $\mathbb{R}^3$.
The sequence $\{h^N\}$ is relatively compact in the weak$^{\ast}$ sense (see, e.g., \cite{Billingsley1999}). 
Therefore, there exists a subsequence $\{h^{N_k}\}_k$ such that $h^{N_k}$ converges to $h$ with weak$^{\ast}$-convergence in $\mathcal{P}(\mathbb{R}^3$) and pointwise convergence in time as $k \rightarrow \infty$.  

Let $\psi \in C_0^1(\mathbb{R}^3)$ be a test function. We have
\begin{align}\label{eq:testfunction}
\frac{d}{dt} \langle h^N, \psi \rangle_{\x,q} 
&=\frac{d}{dt} \biggl< \frac{1}{N}\sum_{n=1}^{N}\delta(\x-\x_n(t))\delta(q-q_n(t)), \psi \biggr>_{\x,q} \cl
&=\frac{d}{dt} \frac{1}{N}\sum_{n=1}^{N}\psi(\x_n(t), q_n(t)) \cl
&=\frac{1}{N}\sum_{n=1}^{N} \left( \psi_x v_n \cos \theta +  \psi_y v_n \sin \theta + \psi_q\gamma\max\{(q_n^*-q_n), 0\} \right) \cl
&= \langle h^N, \psi_x v \cos \theta \rangle_{\x,q} + \langle h^N, \psi_y v \sin \theta \rangle_{\x,q} \cl
& \quad \quad + \frac{\gamma}{N}\sum_{n=1}^{N}\psi_q \max\left\{\left( \frac{\sum_{m=1}^{N} \kappa_{n,m}q_n}{\sum_{m=1}^{N}\kappa_{n,m}}- q_n \right), 0 \right\},
\end{align} 
where $\langle \cdot \rangle_{\x,q}$ means integration against both $\x$ and $q$. 

Let us define
\begin{equation}
\rho(\x)=\frac{1}{N}\sum_{n=1}^{N}\delta(\x-\x_n) \cl
\end{equation}
and
\begin{equation}
m(\x)=\biggl< q, \frac{1}{N}\sum_{m=1}^{N}\delta(\x-\x_m)\delta(q-q_m) \biggr>_{\x,q}
=\frac{1}{N}\sum_{m=1}^{N}\delta(\x-\x_m)q_m, \cl
\end{equation}
We have
\begin{align}
\frac{1}{N}\sum_{m=1}^{N}\kappa(|\x_n-\x_m|)&= \biggl< \kappa(|\x_n-\tilde{\x}|), \frac{1}{N}\sum_{m=1}^{N}\delta(\tilde{\x}-\x_m) \biggr>_{\x}
=\kappa \ast\rho(\x_n), \cl
\frac{1}{N}\sum_{m=1}^{N}\kappa(|\x_n-\x_m|)q_m&= \biggl< \kappa(|\x_n-\tilde{\x}|), \frac{1}{N}\sum_{m=1}^{N}\delta(\tilde{\x}-\x_m)q_m \biggr>_{\x}
=\kappa \ast m(\x_n), \el
\end{align}
where $\langle \cdot \rangle_\x$ means integration only in $\x$.
Then, we can rewrite eq.~($\ref{eq:testfunction}$) as
\begin{align}
\frac{d}{dt} \langle h^{N}, \psi \rangle_{\x,q} &= \langle h^N, \psi_x v \cos \theta \rangle_{\x,q} + \langle h^N, \psi_y v \sin \theta \rangle_{\x,q} \cl
& \quad \quad + \gamma \biggl< h^{N}, \psi_{q} \max \left\{ \frac{\kappa \ast m_{s}}{\kappa \ast \rho_{s}}-q, 0 \right\} \biggr>_{\x,q}. \label{eq3_rewritten}
\end{align}
Via integration by parts, eq.~\eqref{eq3_rewritten} leads to
\begin{equation}\label{eq:Nkineticsystem}
h_t^N+\nabla \cdot (v(\cos\theta, \sin\theta)^T h^N)+\gamma(\max\{(q^\ast-q),0\}h^N)_q = 0, 
\end{equation}
where $q^*$ is the local \emph{average} sickness level weighted by \eqref{kappa}:
\begin{equation} \label{q_act}
q^{\ast}(t,\x)= \frac{\iint\kappa(|\x-\overline{\x}|) h(t,\overline{\x},q)qdqd\overline{\x}}{\iint \kappa(|\x-\overline{\x}|)h(t,\overline{\x},q)dqd\overline{\x}}.
\end{equation}

Letting $k \rightarrow \infty$, we obtain the limiting kinetic model from the subsequence $h^{N_k}$: 
\begin{equation} \label{eq:2kineticeq}
h_t+ \nabla \cdot (v(\cos\theta, \sin\theta)^T h)+\gamma(\max \{(q^\ast-q), 0\} h)_q = 0, 
\end{equation}
where $h(t,\x,q)$ is the probability of finding at time $t$ and position $\x$ 
a person with level of exposure $q$ if $q \in [0, 1)$
or a person spreading the disease if $q=1$.  

It is convenient to switch to non-dimensional variables. Let $D$ be the  
largest distance a pedestrian can cover in domain $\Omega$ and $V_M$ the largest speed a person can reach.
Then, we can define characteristic time $T=D/V_M$. Moreover, let  $\rho_M$ be the maximum people density per square meter. 
The non-dimensional quantities are: position $\hat{\x}=\x/D$, walking speed $\hat{v}=v/V_M$, time $\hat{t}=t/T$, 
and distribution function $\hat{h}=h/ \rho_M$.
For ease of notation though, we will drop the hat 
with the understanding that all quantities are non-dimensional unless otherwise specified.

\section{Discretization in space and time}\label{sec:discr}

In this section, we present the numerical discretization of model (\ref{eq:2kineticeq}).

Let us start from the space discretization. 
Divide the spatial domain into a number of cells 
$[x_{i-\frac{1}{2}}, x_{i+\frac{1}{2}}] \times [y_{j-\frac{1}{2}}, y_{j+\frac{1}{2}}]$ 
of length $\Delta x$ and $\Delta y$, respectively. 
The discrete mesh points $x_{i}$ and $y_j$ are given by
\begin{align}%\label{eq:x_i_y_j} 
x_i =i \Delta x, \quad x_{i+1/2}=x_{i}+ \frac{\Delta x}{2}=\Big(i+\frac{1}{2}\Big)\Delta x, \cl 
y_j =j \Delta y, \quad y_{j+1/2}=y_{j}+ \frac{\Delta y}{2}=\Big(j+\frac{1}{2}\Big)\Delta y, \el
\end{align}
for $i= 0, 1, \dots, N_x$ and $j = 0, 1, \dots, N_y$,
The contagion level domain is partitioned into subdomains $[q_{k-\frac{1}{2}}, q_{k+\frac{1}{2}}]$ of length of $\Delta q$ with $k \in 1,2,\dots, N_q$.
\begin{align}
q_k =k \Delta q, \quad q_{k+1/2}=q_{k}+ \frac{\Delta q}{2}=\Big(k+\frac{1}{2}\Big)\Delta q. \el 
\end{align}
Here, $N_x$, $N_y$ and $N_q$ are the total number of points in $x-$, $y-$ and $q-$ directions, respectively. 
%For simplicity, we assume each cell is centered at $x_{i}$, $y_{j}$ or $q_k$ with a uniform length $\Delta x$, $\Delta y$ and $\Delta q$. 

Let us denote $h_{i, j, k}=h(t, x_{i}, y_{j}, q_k)$. The average level of sickness
is computed using a midpoint rule for the integrals in \eqref{q_act}:
\begin{align}
q^{\ast}(t,x_i, y_j) \approx q^{\ast}_{i,j} &= \frac{\sum_k \sum_{\overline{j}} \sum_{\overline{i}} \kappa_{\overline{i},\overline{j}} h_{\overline{i}, \overline{j},k} q_k \Delta q \Delta x \Delta y}{\sum_k \sum_{\overline{j}} \sum_{\overline{i}} \kappa_{\overline{i},\overline{j}} h_{\overline{i}, \overline{j},k} \Delta q \Delta x \Delta y}, \label{q_sum}
\end{align}
where if at $(x_{\overline{i}}, y_{\overline{j}})$ there is nonzero probability of finding a person that is spreading the disease: 
\begin{equation*}
\kappa_{\overline{i},\overline{j}}  = \frac{R}{((x_i - x_{\overline{i}})^2+(y_j - y_{\overline{j}})^2+R^2)\pi},
\end{equation*}
otherwise $\kappa_{\overline{i},\overline{j}} = 0$.

%\anna{In 1D
%\begin{align}
%q^{\ast}(t,x_i) \approx q^{\ast}_{i} &= \frac{\sum_k \sum_{\overline{i}} \kappa_{\overline{i}} h_{\overline{i},k} q_k \Delta q \Delta x }{\sum_k \sum_{\overline{i}} \kappa_{\overline{i}} h_{\overline{i},k} \Delta q \Delta x}, \el
%\end{align}
%where if at $x_{\overline{i}}$ there is nonzero probability of finding a person that is spreading the disease (i.e., at $x_{\overline{i}}$,
% NOT $x_i$, there is a person with $q = 1$): 
%\begin{equation}\label{kappa}
%\kappa_{\overline{i}}  = \frac{R}{((x_i - x_{\overline{i}})^2+R^2)\pi},
%\end{equation}
%otherwise $\kappa_{\overline{i},\overline{j}} = 0$. Since at the numerator in $q^{\ast}$ we are effectively only 
%counting the people with $q_k = 1$, the numerator and denominator should coincide. 
%}

We start with a first-order semi-discrete upwind scheme for problem (\ref{eq:2kineticeq}), which reads:
\begin{align} \label{eq:first_o}
\partial_t h_{i, j, k}&+\frac{{\eta}_{i, j, k}-{\eta}_{i-1, j, k}}{\Delta x}
+\frac{{\phi}_{i, j, k}-{\phi}_{i, j-1, k}}{\Delta y} +\gamma \frac{\xi_{i, j, k+\frac{1}{2}}-\xi_{i, j, k-\frac{1}{2}}}{\Delta q}= 0,
\end{align}
where
\begin{align}
\eta_{i, j, k}  &=v \cos \theta \,h_{i,j,k}, \quad \phi_{i, j, k} =v \sin \theta\,h_{i,j,k}, \label{eq:eta_phi} \\
\xi_{i, j, k+\frac{1}{2}} & = \max \left\{  q_{i, j}^\ast - q_{k+\frac{1}{2}}, 0 \right\}h_{i, j, k}. \label{eq:xi}
\end{align}
Next, let us discretize in time.
Let $\Delta t \in \mathbb{R}$, $t^n = t_0 + n \Delta t$, with $n = 0, ..., N_T$ and $t_f = t_0 + N_T \Delta t$,
where $t_f$ is the end of the time interval under consideration. 
Moreover, we denote by $y^n$ the approximation of a generic quantity $y$ at the time $t^n$.
For the time discretization of problem \eqref{eq:first_o}, we use the forward Euler scheme:
\begin{align} \label{eq:full_discretization1}
h_{i, j, k}^{l+1} = h_{i, j, k}^l - \Delta t
\left( \dfrac{{\eta}^l_{i, j, k}-{\eta}^l_{i-1, j, k}}{\Delta x}+\dfrac{{\phi}^l_{i, j, k}-{\phi}^l_{i, j-1, k}}{\Delta y} + 
\gamma \dfrac{\xi^l_{i, j, k+\frac{1}{2}}-\xi^l_{i, j, k-\frac{1}{2}}}{\Delta q} \right).
\end{align}

To construct a scheme that is of second order, we add a flux limiter. Let $\varphi$ be a slope limiter function. 
for example, one could choose the Van Leer function:
\begin{equation}\label{van_leer}
\varphi(\theta)=\frac{|\theta|+\theta}{1+|\theta|}.
\end{equation}
The space discretized  eq.~\eqref{eq:2kineticeq} now reads:
%\anna{The rest is from the paper with Will that we'll use to write the second order scheme}
\begin{align} \label{eq:disease_modified}
\partial_t h_{i, j, k}&+\frac{\overline{\eta}_{i, j, k}-\overline{\eta}_{i-1, j, k}}{\Delta x}
+\frac{\overline{\phi}_{i, j, k}-\overline{\phi}_{i, j-1, k}}{\Delta y} \cl
&+\gamma \frac{\xi_{i, j, k+\frac{1}{2}}-\xi_{i, j, k-\frac{1}{2}}}{\Delta q}
+\gamma \frac{C_{i, j, k+\frac{1}{2}}-C_{i, j, k-\frac{1}{2}}}{\Delta q}= 0,
\end{align}
where
\begin{align}
\overline{\eta}_{i, j, k} & =\eta_{i, j, k} + \frac{\eta_{i+1, j, k} - \eta_{i, j, k}}{2} \varphi\left( \frac{\eta_{i, j, k} - \eta_{i-1, j, k}}{\eta_{i+1, j, k} - \eta_{i, j, k}} \right), \label{eq:eta_bar} \\
\overline{\phi}_{i, j, k} &= \phi_{i, j, k} + \frac{\phi_{i, j+1, k} - \phi_{i, j, k}}{2} \varphi\left( \frac{\phi_{i, j, k} - \phi_{i, j-1, k}}{\phi_{i, j+1, k} - \phi_{i, j, k}} \right), \label{eq:phi_bar} \\
q_{k+\frac{1}{2}} &= \frac{q_k + q_{k+1}}{2}, ~ s_{i, j, k+\frac{1}{2}}= q^{\ast}_{i, j}-q_{k+\frac{1}{2}} \cl
W_{i, j, k-\frac{1}{2}}&=h_{i, j,k}-h_{i, j,k-\frac{1}{2}} \cl
C_{i, j, k+\frac{1}{2}} & =\dfrac{1}{2} \left| s_{i, j,k+\frac{1}{2}} \right| \left(1-\dfrac{\Delta t}{\Delta q} \left| s_{i, j, k+\frac{1}{2}} \right| \right)
W_{i, j,k-\frac{1}{2}}\varphi \left( \dfrac{W_{i,j,\textbf{k}-\frac{1}{2}}}{W_{i, j, k-\frac{1}{2}}} \right). 
\label{eq:eta_xi_C}
\end{align}
In eq.~\eqref{eq:eta_xi_C}, the subscript \textbf{k} is k-1 if $s_{i, j, k-\frac{1}{2}} > 0$ and k+1 if $s_{i, j, k-\frac{1}{2}} < 0$. 
Recall that $\eta_{i, j, k}$ in \eqref{eq:eta_bar} and $\phi_{i, j, k}$ in \eqref{eq:phi_bar} are defined in \eqref{eq:eta_phi}, 
while $\xi_{i, j, k+\frac{1}{2}}$ is defined in \eqref{eq:xi}. Thanks to the flux limiter, 
scheme \eqref{eq:disease_modified} is of second-order scheme in velocity.

For the time discretization of problem \eqref{eq:disease_modified}, we adopt again  the forward Euler scheme 
\begin{align} \label{eq:full_discretization2}
h_{i, j, k}^{l+1} = h_{i, j, k}^l - \Delta t
\Big( & \dfrac{\overline{\eta}^l_{i, j, k}-\overline{\eta}^l_{i-1, j, k}}{\Delta x}
+\dfrac{\overline{\phi}^l_{i, j, k}-\overline{\phi}^l_{i, j-1, k}}{\Delta y} \cl
+& \gamma \dfrac{\xi^l_{i, j, k+\frac{1}{2}}-\xi^l_{i, j, k-\frac{1}{2}}}{\Delta q}
+\gamma \dfrac{C^l_{i, j, k+\frac{1}{2}}-C^l_{i, j, k-\frac{1}{2}}}{\Delta q} \Big)
\end{align}

The time step $\Delta t$ for scheme \eqref{eq:full_discretization1} or \eqref{eq:full_discretization2}  is set as 
\begin{align}\label{eq:CFL}
\Delta t = \frac{1}{2}\min\Bigg\{ \frac{\Delta x}{\max_{k}q_{k}},  \frac{\Delta y}{\max_{k}q_{k}},
\frac{\Delta q}{2\gamma \max_{k}q_{k}} \Bigg\}
\end{align} 
to satisfy the Courant-Friedrichs-Lewy (CFL) condition.

\section{Numerical results}\label{sec:num_res}

We assess the approach presented in Sec.~\ref{sec:discr} through a series of tests. 
For all the problems, the computational domain in the $xy$-plane is
$[0, 10] \times [0, 10]$ m$^2$ before non-dimensionalization, while $q \in [0, 1]$. 
The dimensionless quantities are obtained by using the following reference quantities: $D=10\sqrt{2}$ m,
%\anna{(Daewa, I see a problem with this: in Fig.~\ref{fig:v0} and following the square is $[0, 1] \times [0, 1]$, so 
%$D = 10$. In addition, the section in Fig.~\ref{fig:v0} is over space interval $[0,1]$, while the section in 
%Fig.~\ref{fig:v1_IC} (and \ref{fig:ICbis} and \ref{fig:v1_ICbis}) in over $[0,\sqrt{2}]$. 
%Right now the comparison of the sections in Fig.~\ref{fig:v0}
%and Fig.~\ref{fig:v1_IC} is misleading)} 
$V_M= 1$ m/s, $T = 10\sqrt{2}$ s, $\rho_M = 7$ people/m$^2$.
In all the tests, we take the initial density to be constant in space
and equal to $\rho_M$. So, the group occupying the walkable domain consists of 700 people.

\begin{figure}[htb]
\centering
\begin{overpic}[width=0.55\textwidth, grid=false]{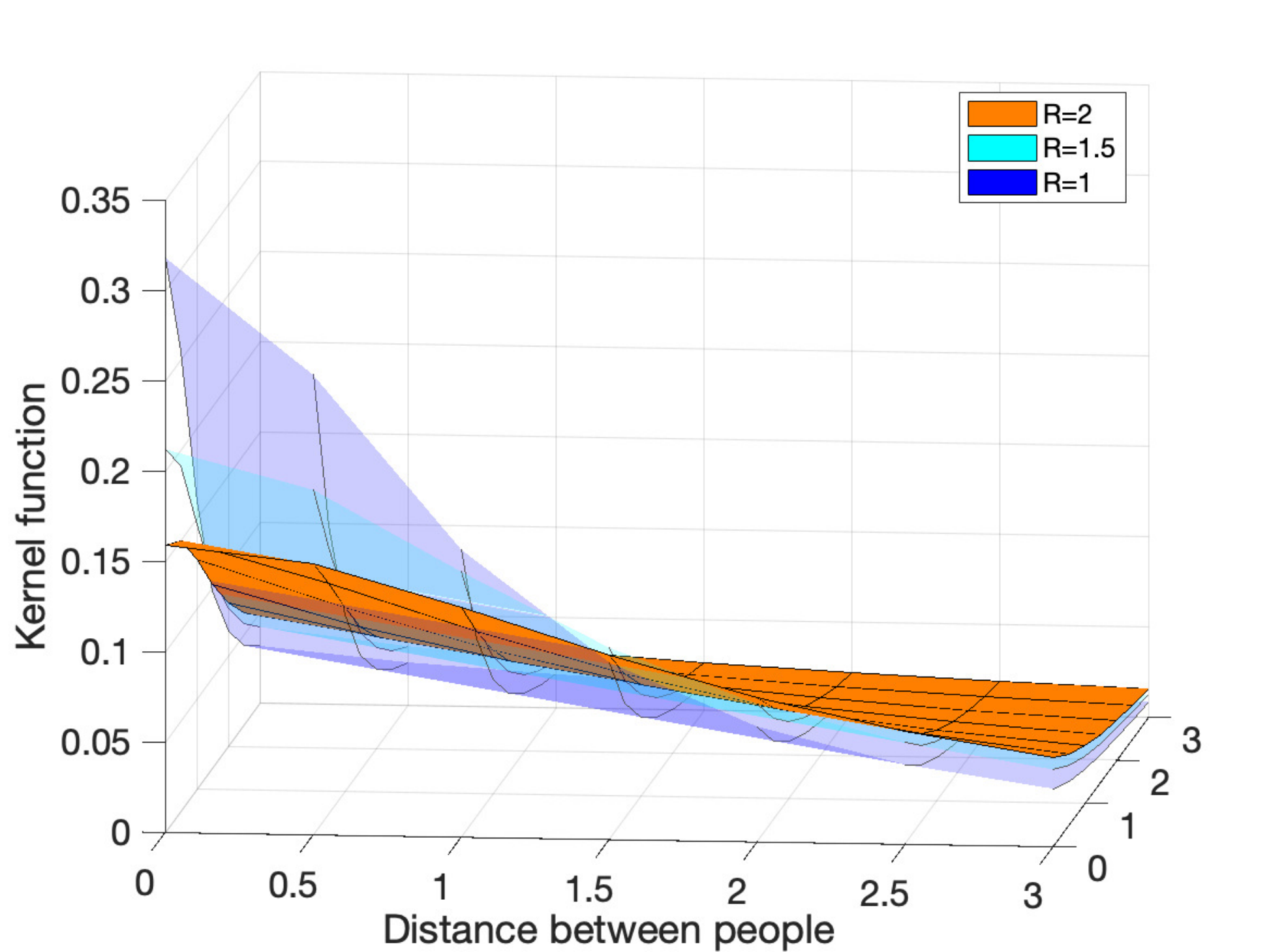}
\end{overpic}
\caption{Kernel function vs the distance between people for interaction radius $R = 1, 1.5, 2$.} \label{fig:kappa}
\end{figure}

We set the interaction distance $R = 1$ m since this choice makes the value of the kernel function relatively small at a distance of 2 m (or about 6 ft). 
See Fig.~\ref{fig:kappa}.

To discretize the walkable domain, we take $\Delta x = \Delta y = 0.1$ m, while we set $\Delta q = 0.01$. 
We will consider two values for the contagion strength: $\gamma = 50$ and $\gamma = 10$. 
The time step is set to $\Delta t=5 \cdot 10^{-5}$ s for $\gamma = 50$ and to 
$\Delta t=25 \cdot 10^{-5}$ s for $\gamma = 10$ according to \eqref{eq:CFL}.
First, we keep the group of people still (i.e., $v = 0$) to make sure that the 
level of exposure evolves as expected.
Then, in a second set of tests, we let people walk and observe how the motion affects the spreading
of the disease. We run each simulation for $t \in (0, 10]$ s.

\subsection{Tests with $\boldsymbol{v = 0}$}\label{sec:v0}

\begin{figure}
\centering
\begin{overpic}[width=0.42\textwidth,grid=false,tics=10]{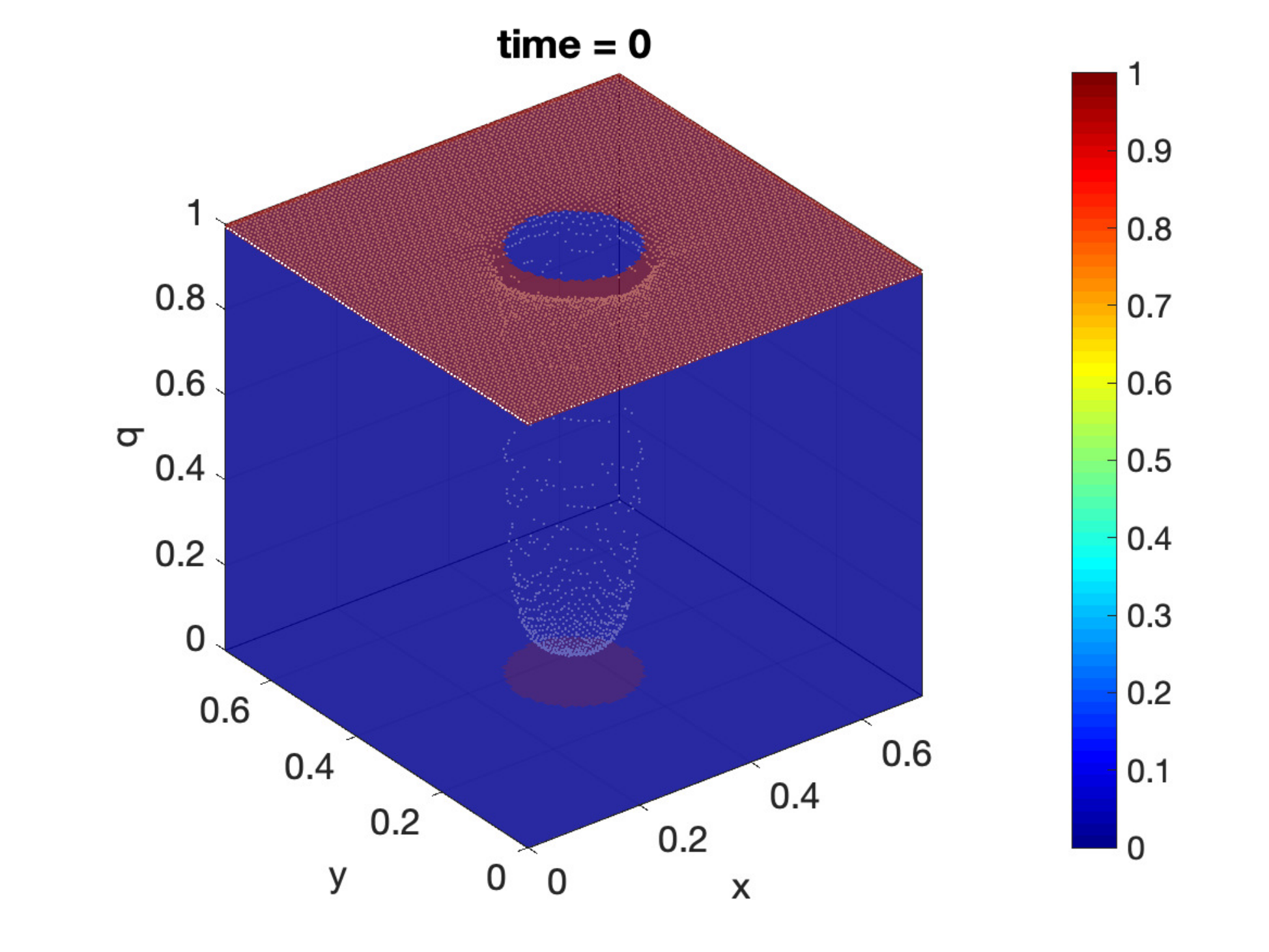}
\end{overpic} 
\begin{overpic}[width=0.42\textwidth,grid=false,tics=10]{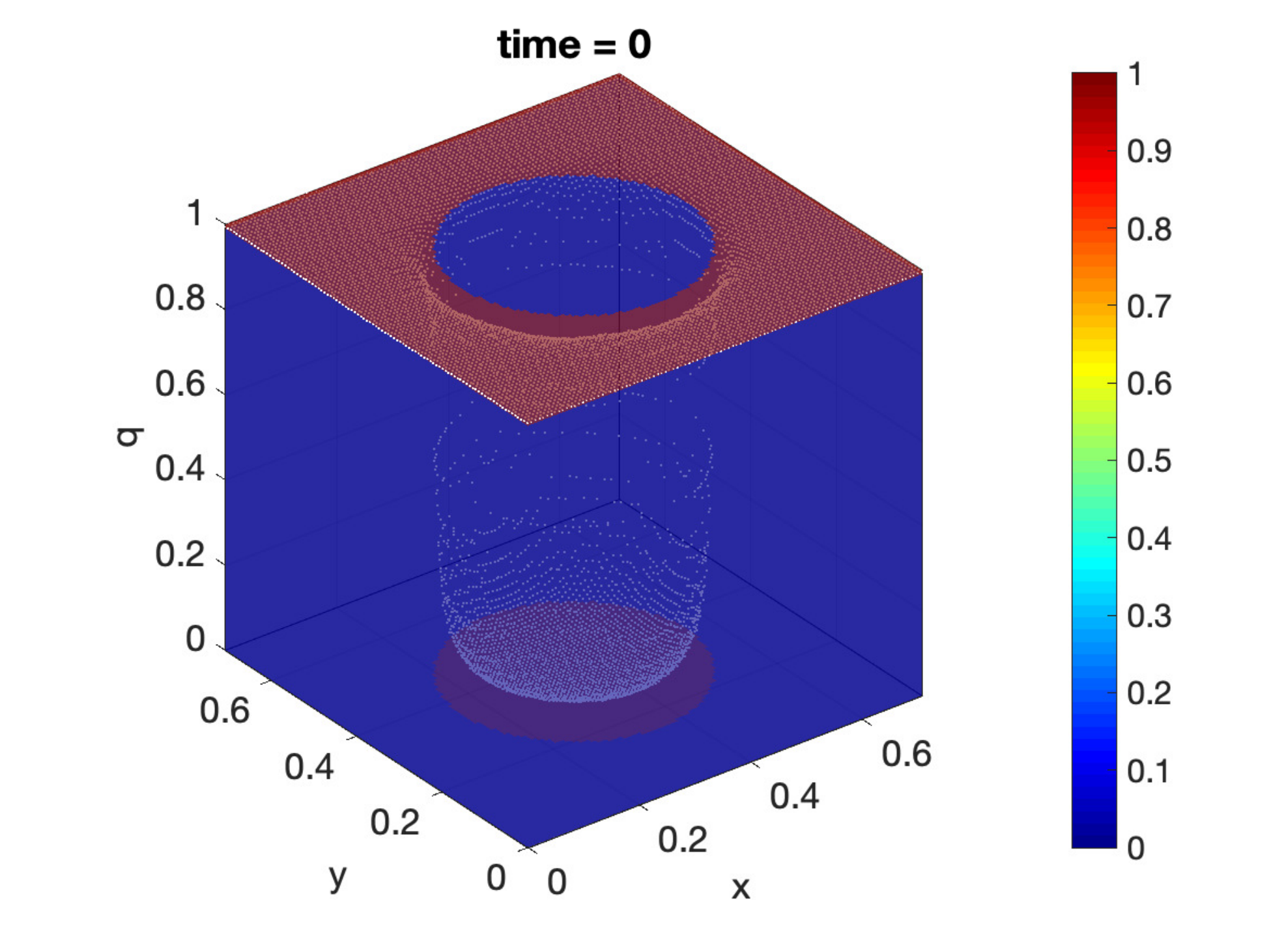}
\end{overpic} 
\caption{Initial conditions IC1 (left) and IC2 (right).}\label{fig:IC}
\end{figure}

We adapt two initial conditions from \cite{kim_quaini2021}:
\begin{itemize}
\item[-] IC1: people that are certainly spreading (i.e., $q = 1$) are located outside the circle of radius 1 m
centered at $(5,5)$ m, while within that circle we place people that have certainly not been exposed (i.e, $q = 0$). 
%\anna{Daewa, is that correct?}
See Fig.~\ref{fig:IC} (left).
\item[-] IC2: people that are certainly spreading (i.e., $q = 1$) are located outside the circle of radius 2 m
centered at $(5,5)$ m, while the rest of the people located within that circle have certainly not been exposed (i.e, $q = 0$). 
%\anna{Daewa, is that correct?}
See Fig.~\ref{fig:IC} (right).
\end{itemize}
The difference between the two boundary conditions is that 
all the healthy people in IC1 are exposed to spreading people, while some 
healthy people in IC2  are not exposed to spreading people. Thus, we expect 
that the level of exposure for the people that are centrally located rises faster 
for IC1 than for IC2.

\begin{figure}
\centering
\begin{overpic}[width=0.31\textwidth,grid=false,tics=10]{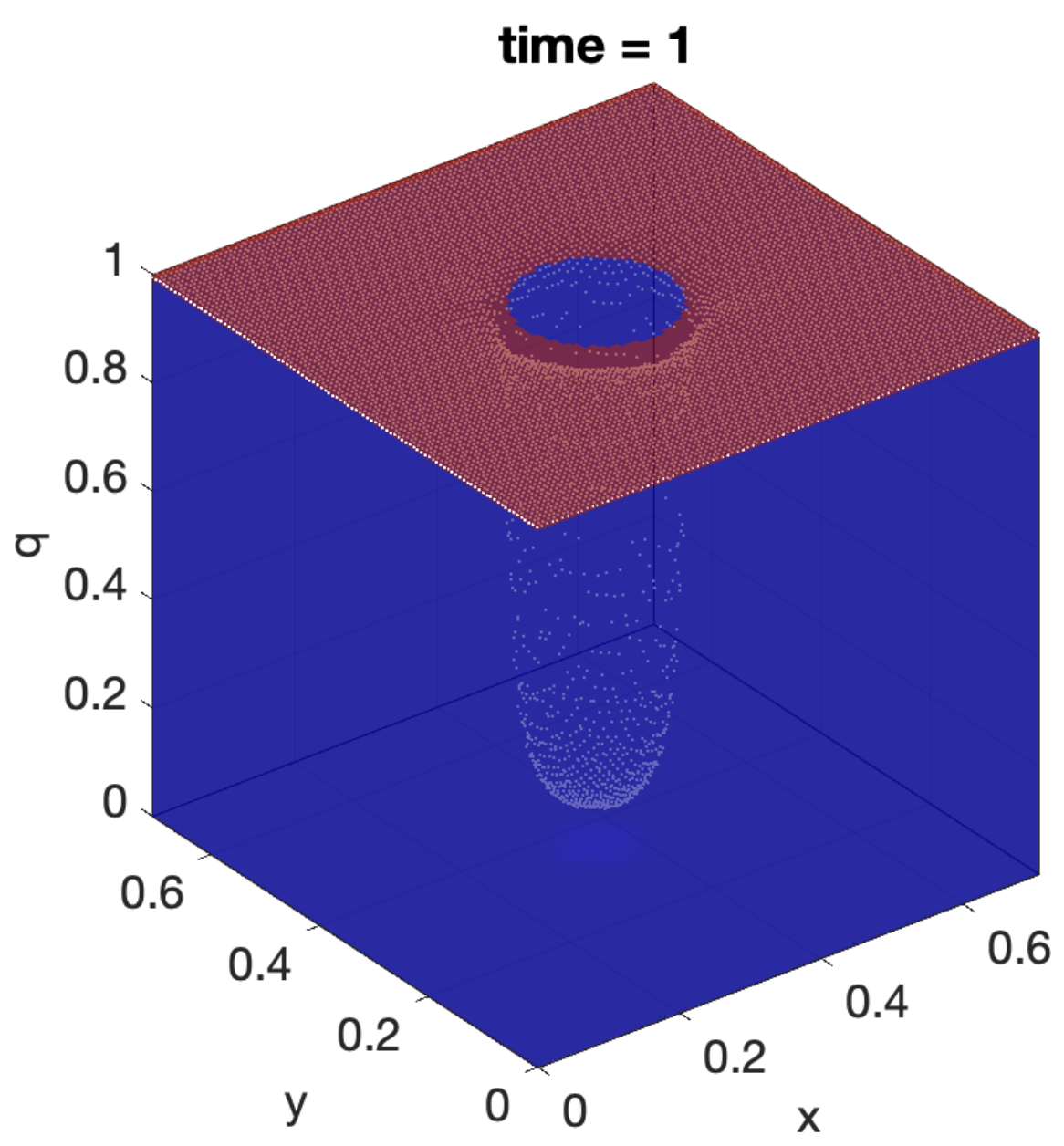}
\put(-18,45){IC1}.
\end{overpic} 
\begin{overpic}[width=0.31\textwidth,grid=false,tics=10]{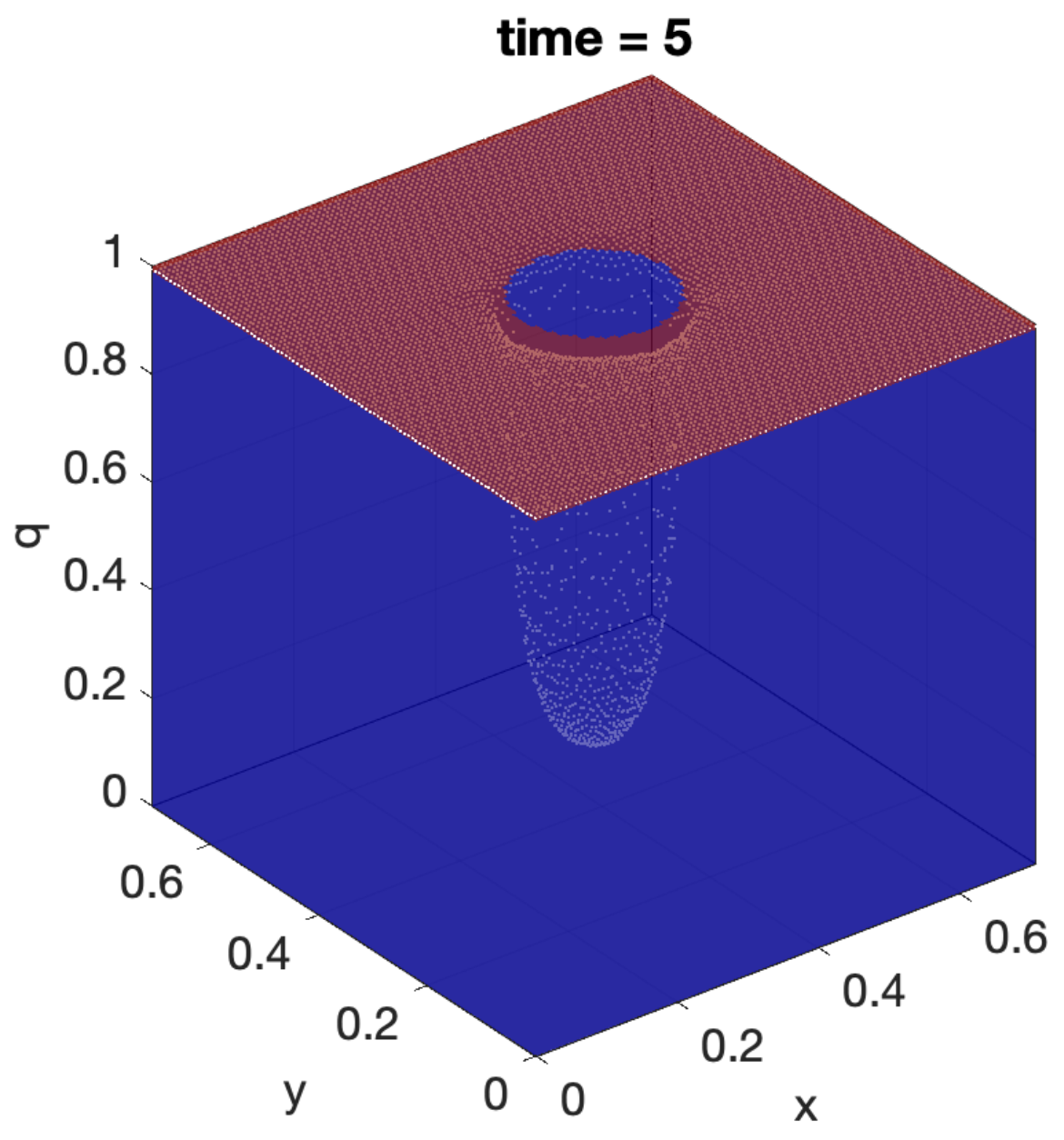}
\end{overpic}
\begin{overpic}[width=0.31\textwidth,grid=false,tics=10]{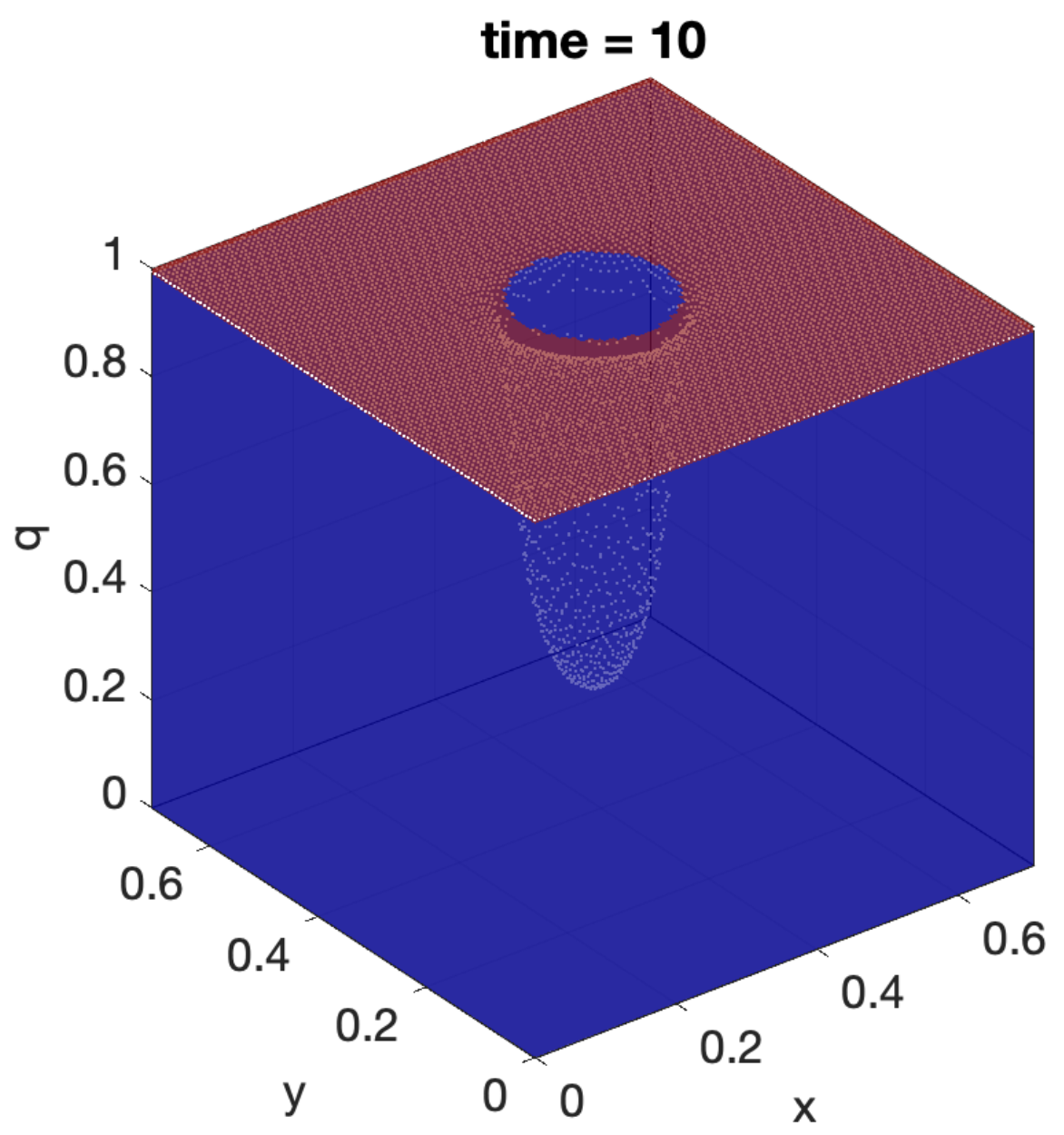}
\end{overpic} 
\begin{overpic}[width=0.31\textwidth,grid=false,tics=10]{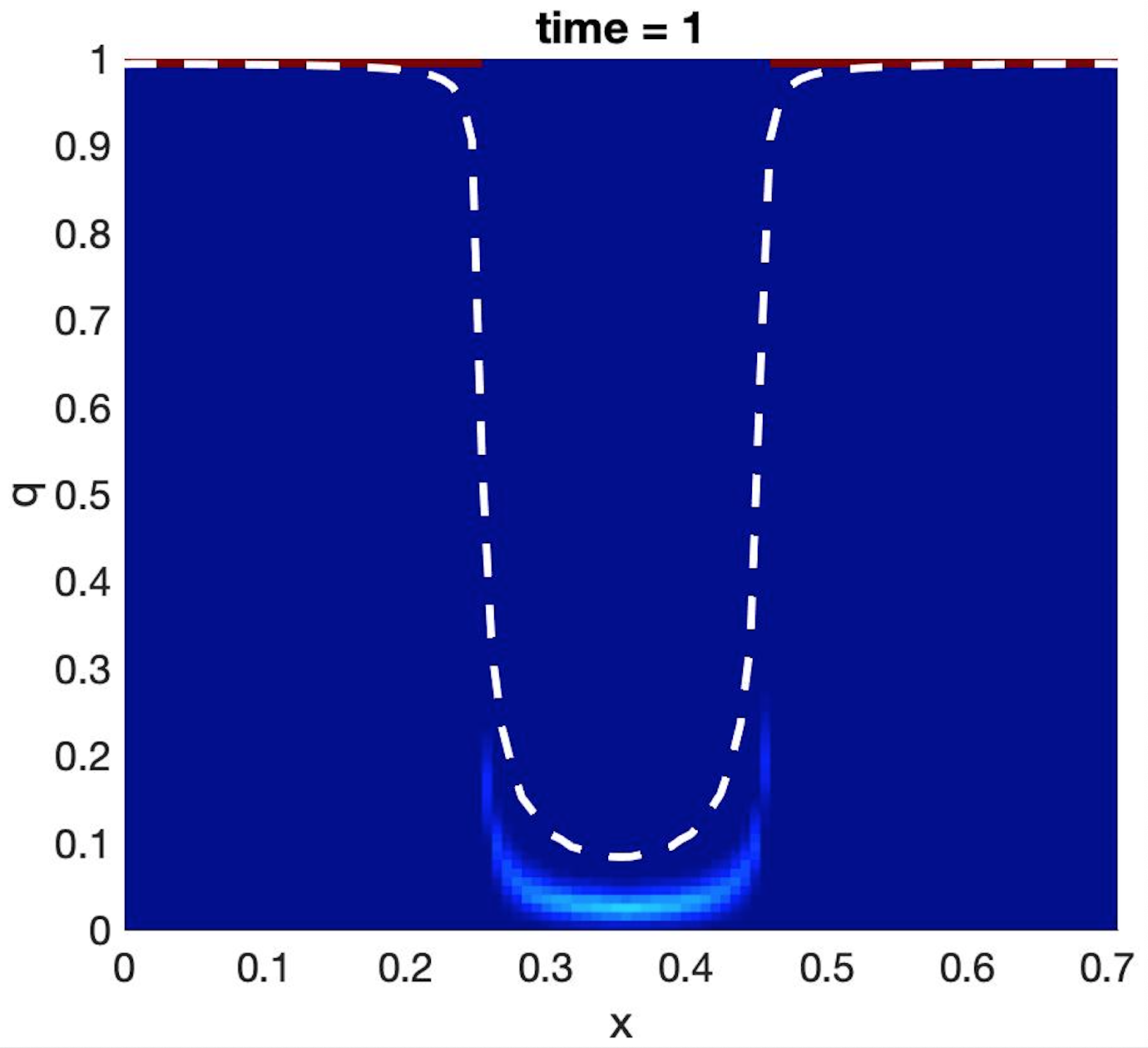}
\put(-18,45){IC1}
\end{overpic} 
\begin{overpic}[width=0.31\textwidth,grid=false,tics=10]{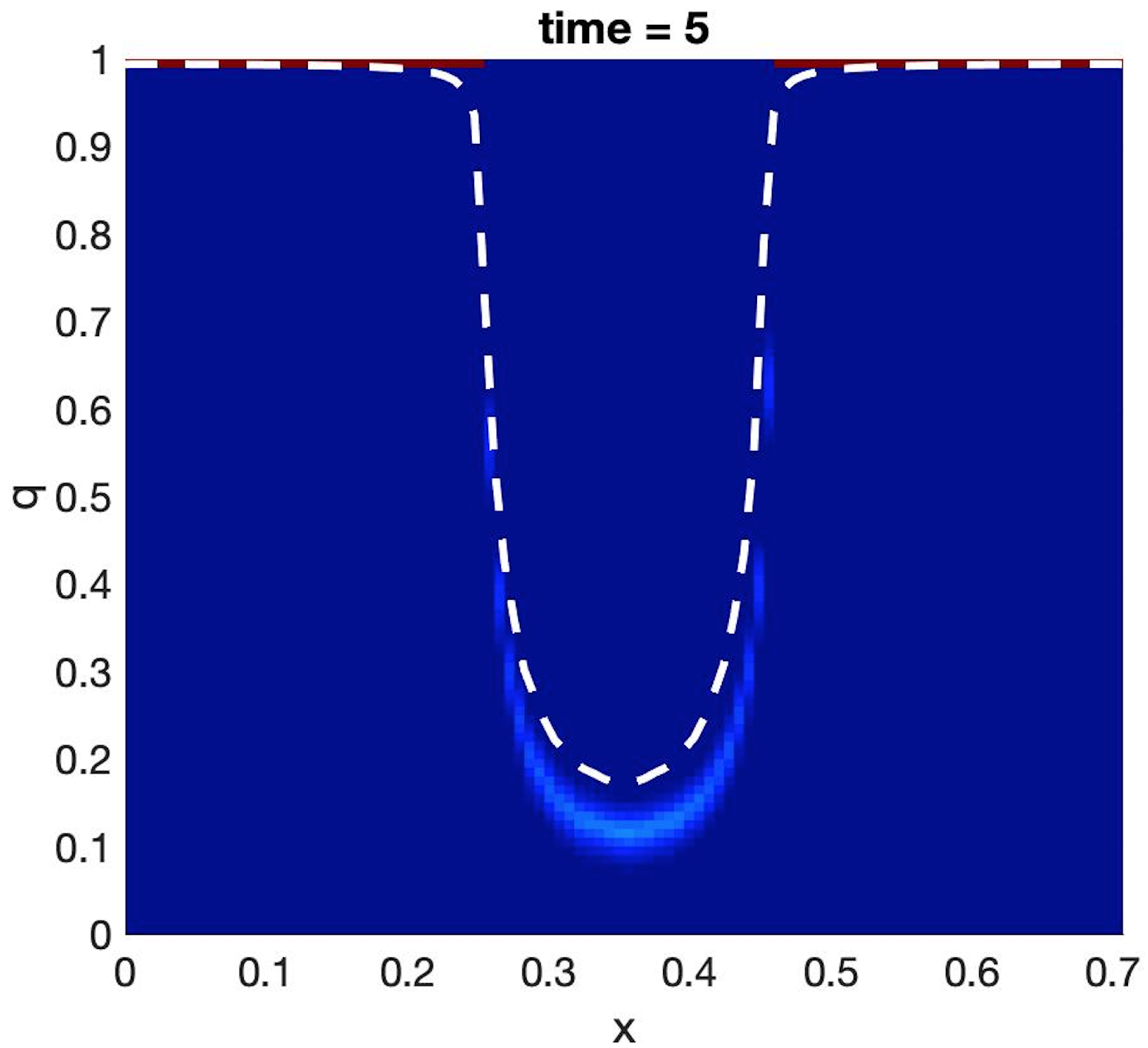}
\end{overpic}
\begin{overpic}[width=0.31\textwidth,grid=false,tics=10]{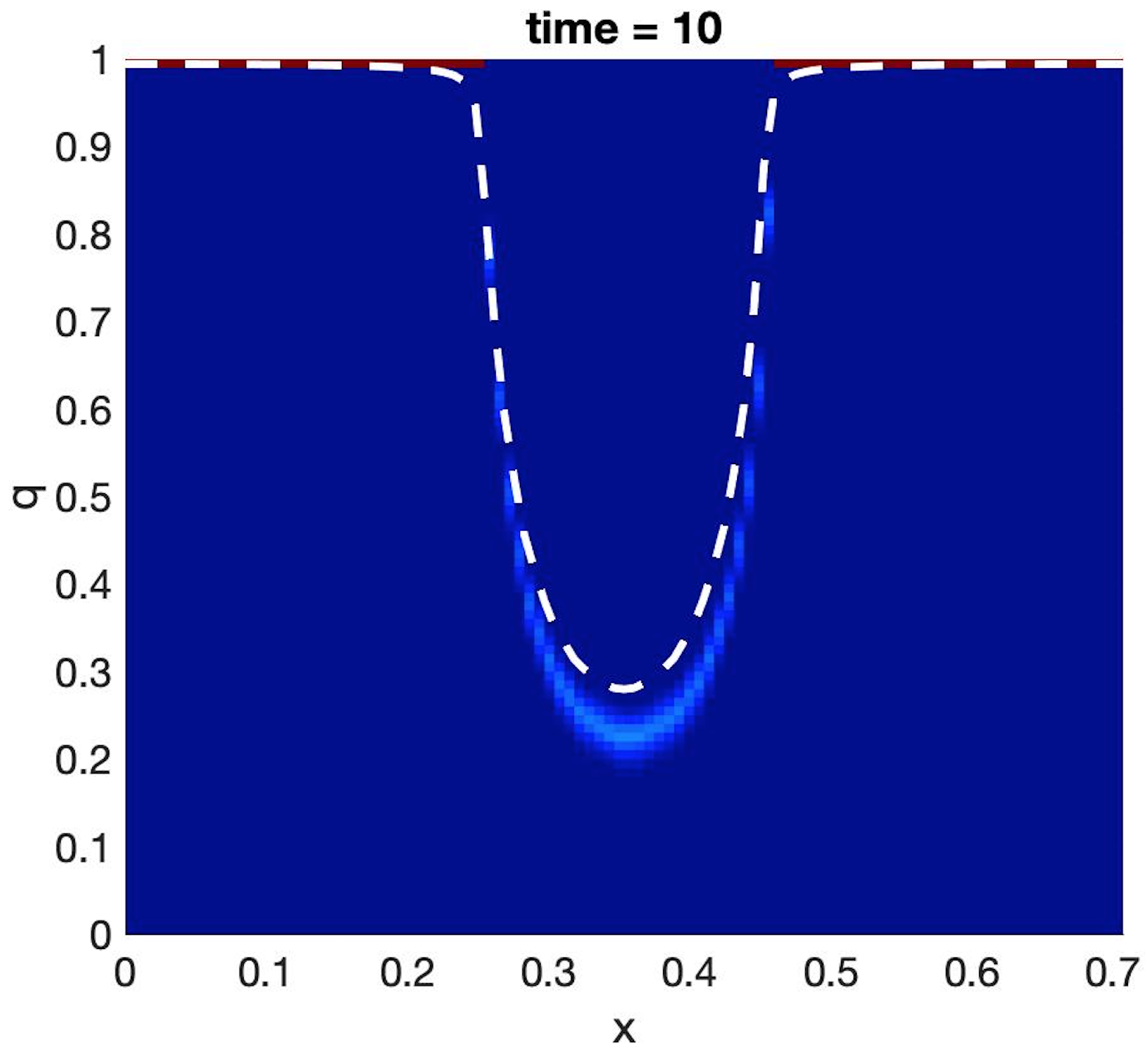}
\end{overpic}
\begin{overpic}[width=0.31\textwidth,grid=false,tics=10]{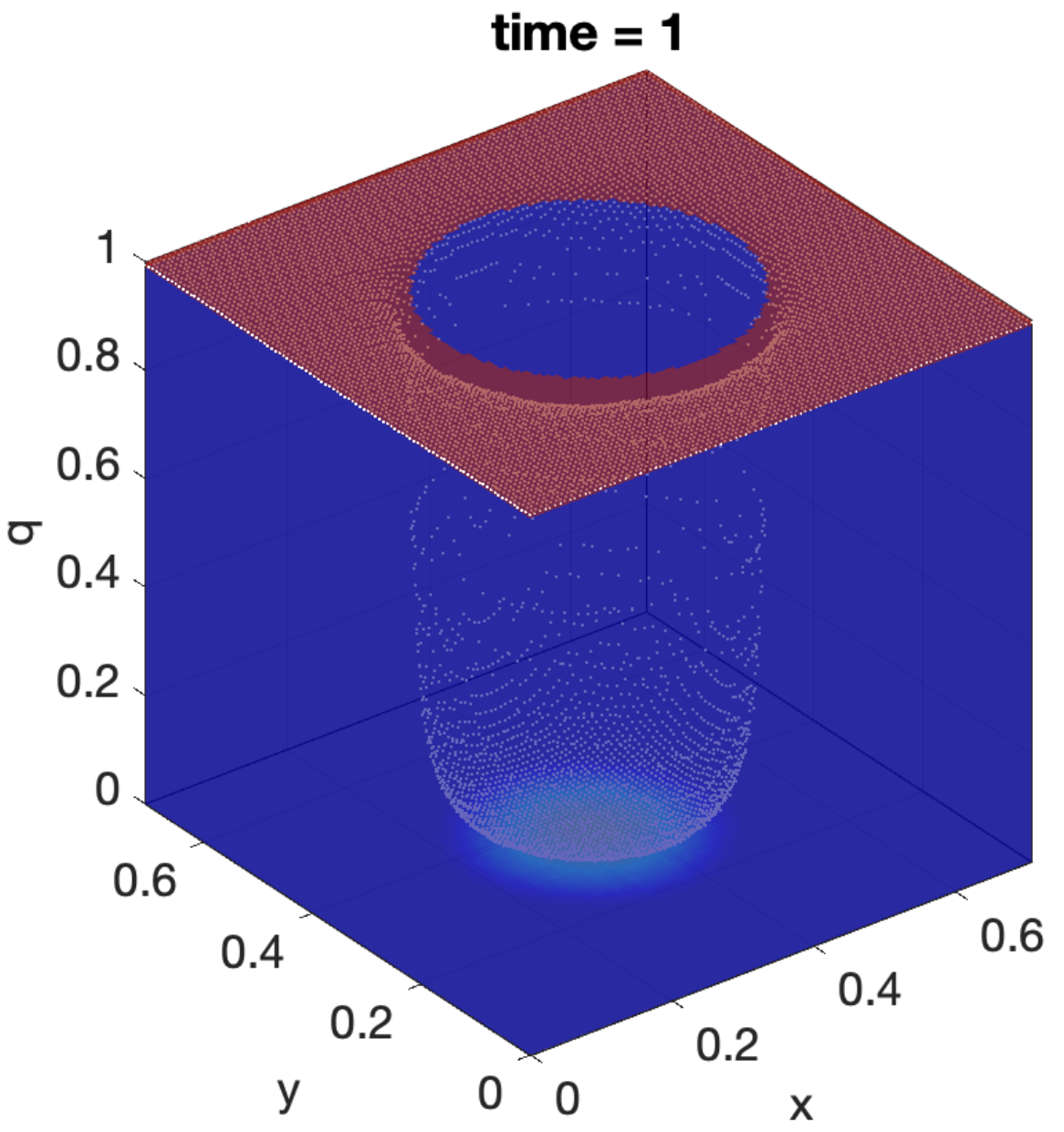}
\put(-18,45){IC2}
\end{overpic} 
\begin{overpic}[width=0.31\textwidth,grid=false,tics=10]{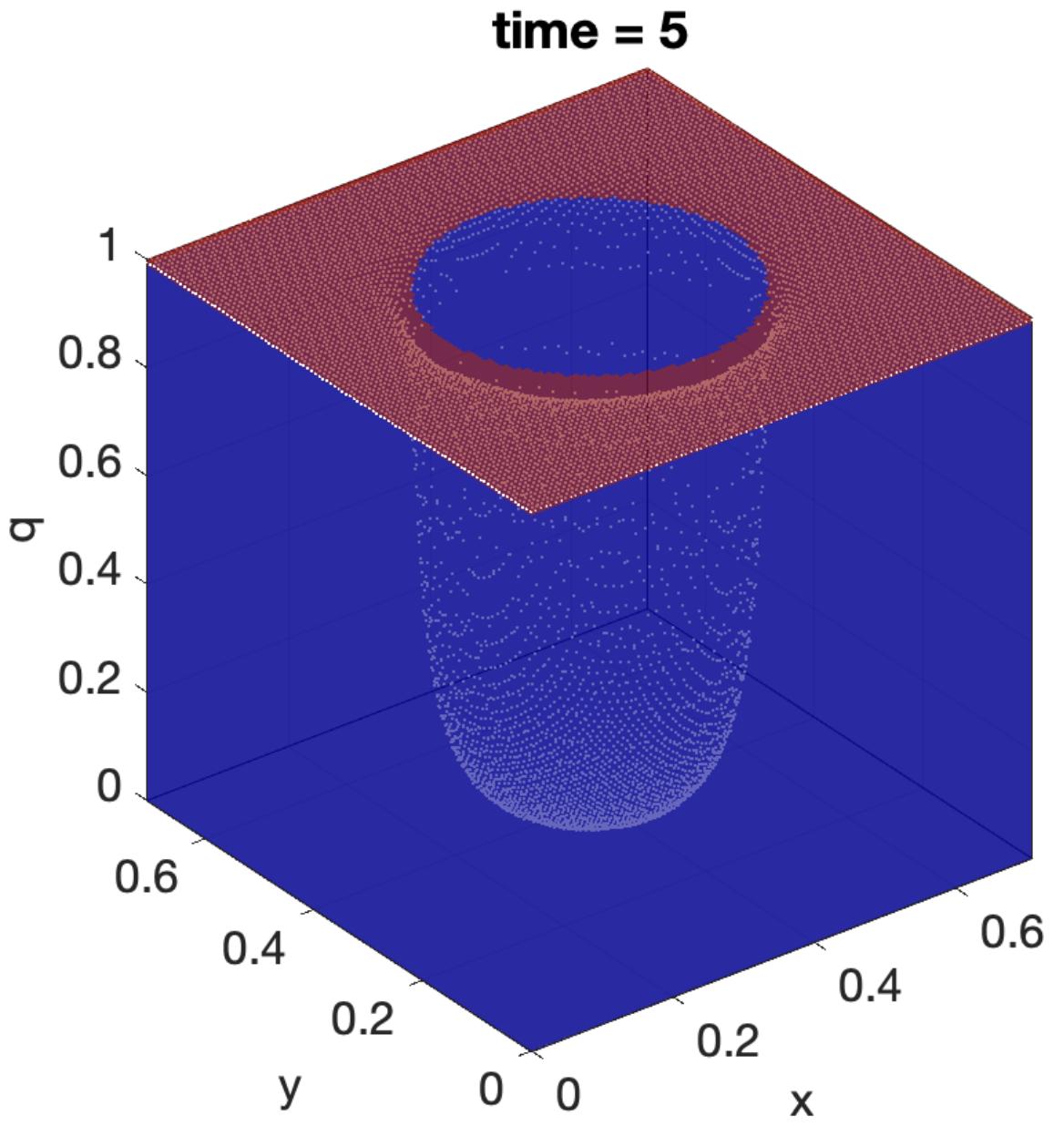}
\end{overpic}
\begin{overpic}[width=0.31\textwidth,grid=false,tics=10]{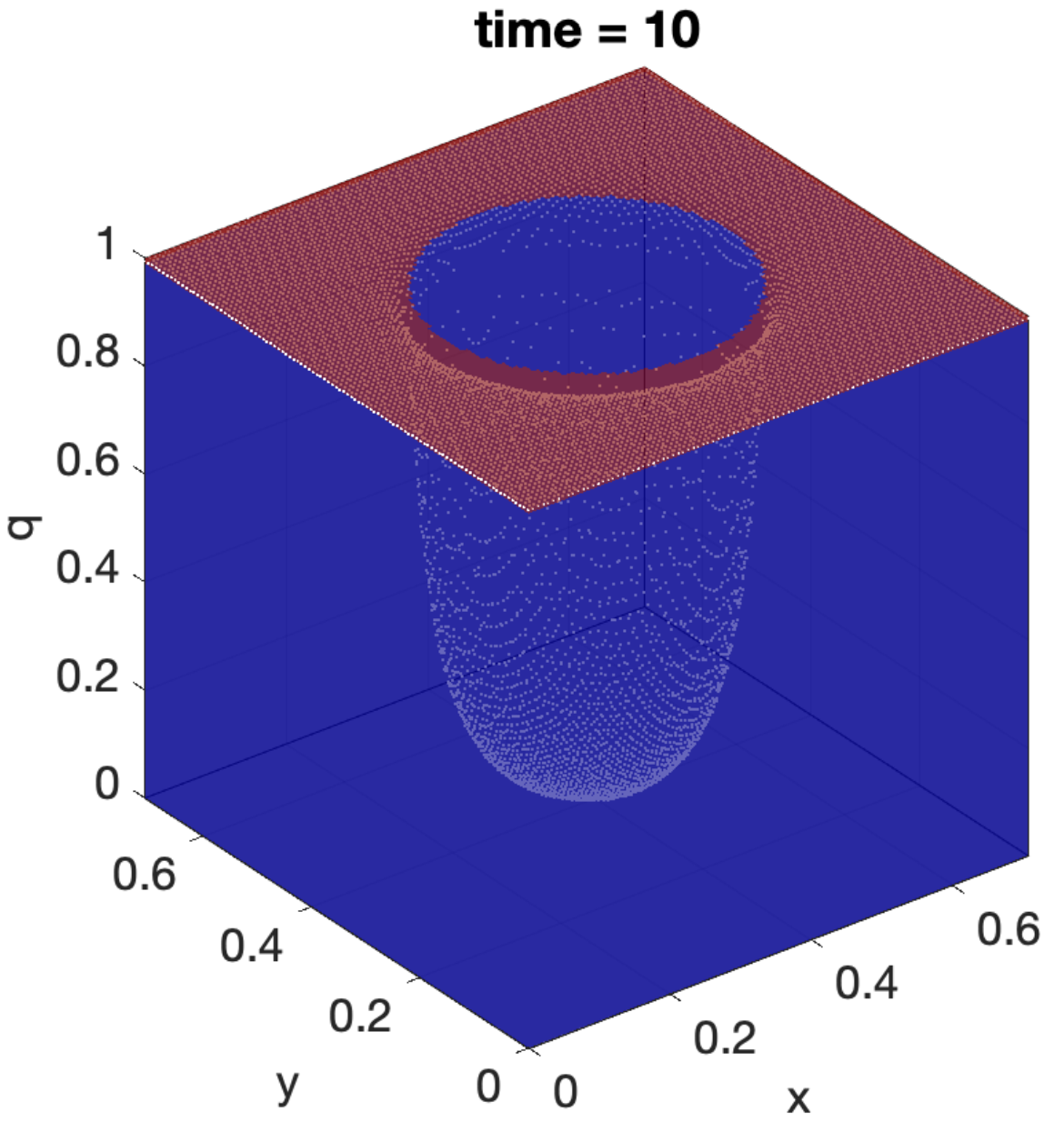}
\end{overpic} 
\begin{overpic}[width=0.31\textwidth,grid=false,tics=10]{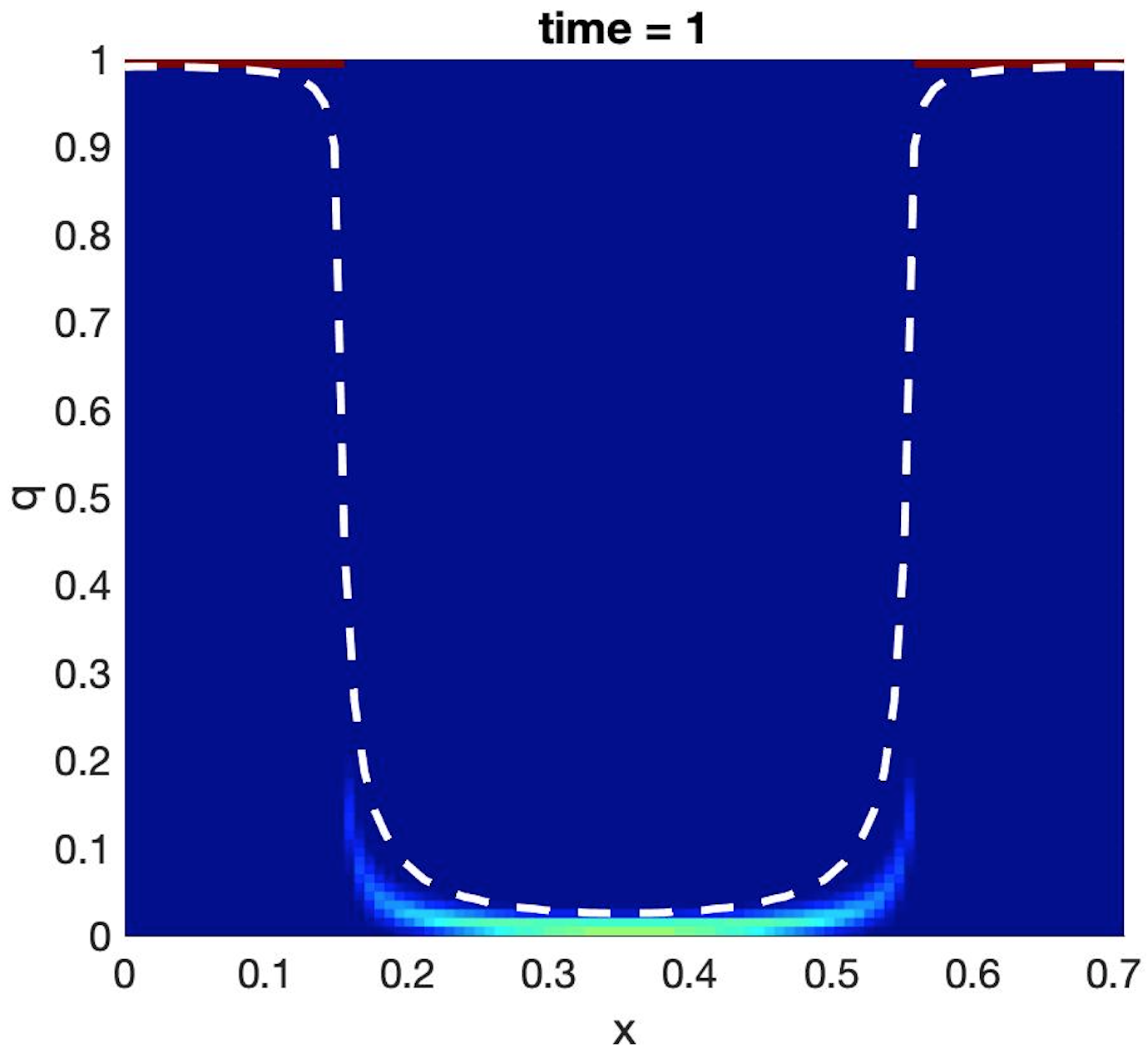}
\put(-18,45){IC2}
\end{overpic} 
\begin{overpic}[width=0.31\textwidth,grid=false,tics=10]{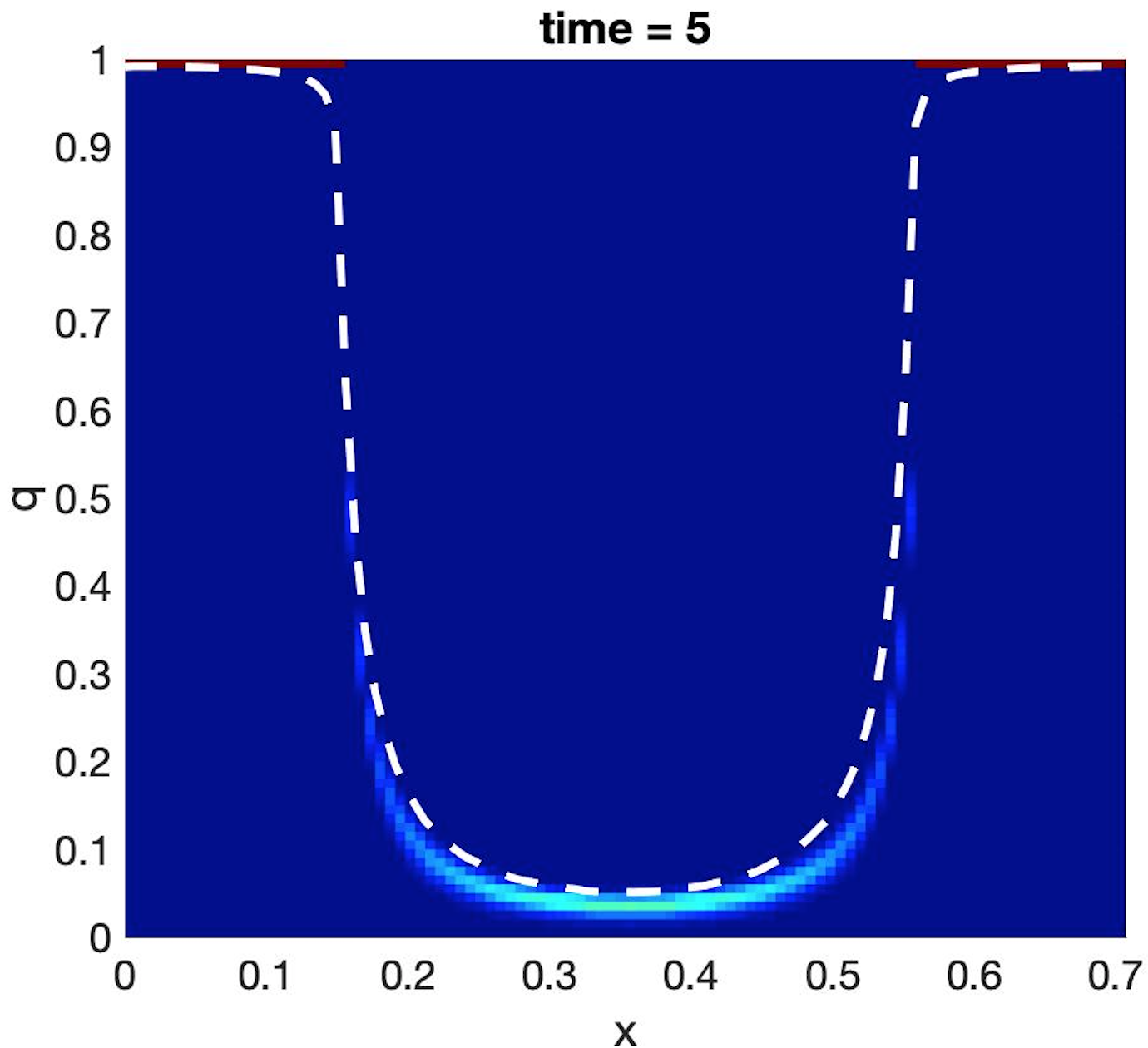}
\end{overpic}
\begin{overpic}[width=0.31\textwidth,grid=false,tics=10]{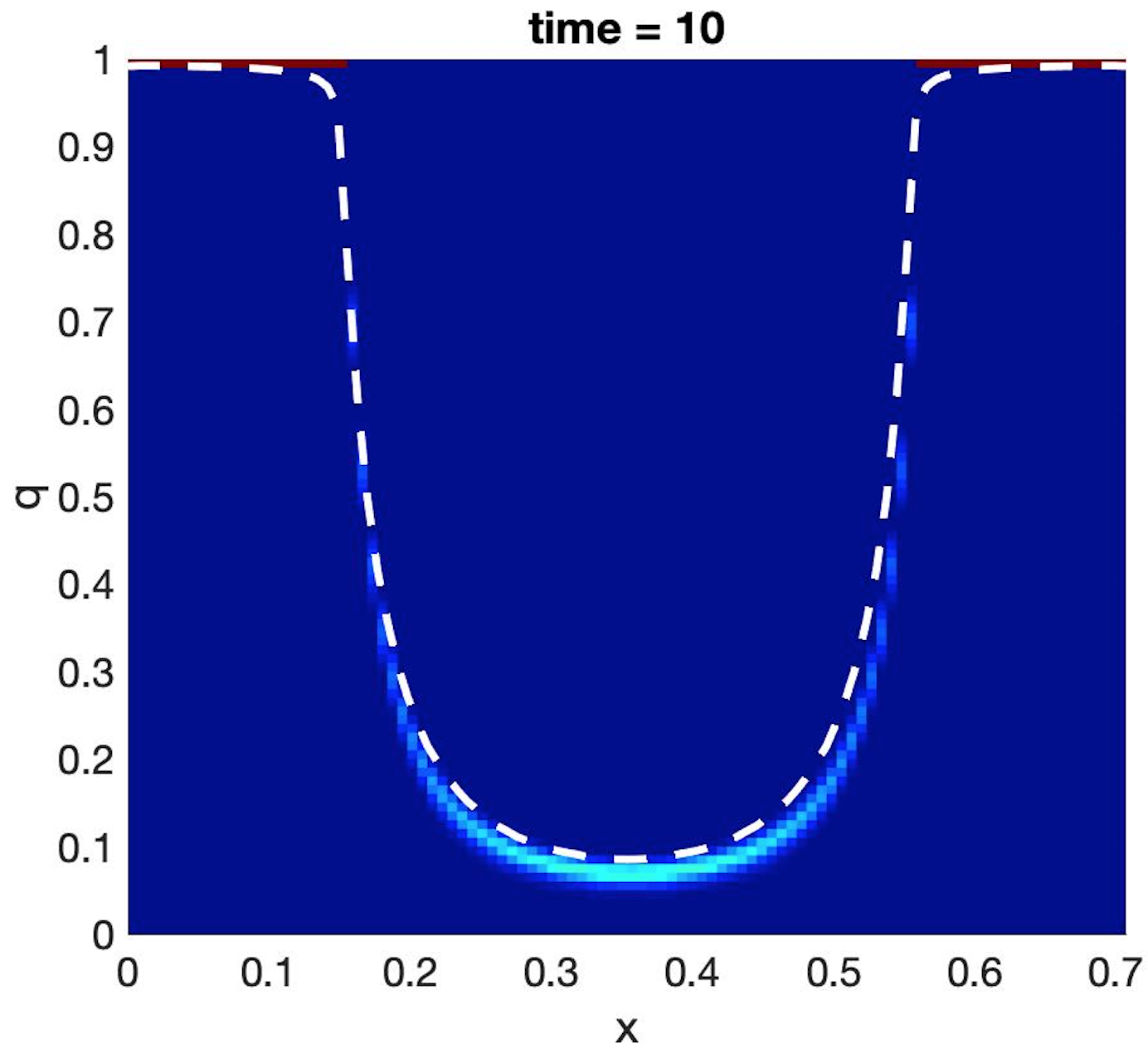}
\end{overpic}
\caption{Tests with $v = 0$: evolution of the distribution density $h$ for initial condition IC1 (first row)
and corresponding results on section $y = 0.5$ (second row), 
evolution of $h$ for initial condition IC2 (third row)
and corresponding results on section $y = 0.5$ (fourth row).
Note that in all the subfigures time is dimensional while space is non-dimensional.
In both cases, we set $\gamma = 50$.
The white dashed line in the images on row two and four represents $q^*$.
The legend for all the images is the same as in Fig.~\ref{fig:IC}.}
\label{fig:v0}
\end{figure}

Fig.~\ref{fig:v0} shows the evolution of the distribution density $h$ for initial conditions IC1 and IC2
with $\gamma  = 50$,
together with a view of the results on a section of the 3D domain. 
We observe that the level of exposure of the central group of healthy people in IC1 increases quickly. 
In particular, we see that the the shape from $q^*$ quickly moves away from being close to a sharp discontinuity 
and gets closer to a paraboloid, since the level of exposure increases faster for the people closer to
the circle separating healthy people from spreading people. As expected, the rise in the level of exposure
is much slower for the simulation with initial condition IC2. This is evident when comparing the second and 
fourth row in Fig.~\ref{fig:v0}. Moreover, notice that the healthy people close to the center of the circle in 
IC2 get very little exposure to spreading people.

\begin{figure}
\centering
\begin{overpic}[width=0.31\textwidth,grid=false,tics=10]{./test1_contagion_time_30000}
\put(-23,45){$\gamma = 50$}
\end{overpic} 
\begin{overpic}[width=0.31\textwidth,grid=false,tics=10]{./test1_contagion_time_150000}
\end{overpic}
\begin{overpic}[width=0.31\textwidth,grid=false,tics=10]{./test1_contagion_time_300000}
\end{overpic} 
\begin{overpic}[width=0.31\textwidth,grid=false,tics=10]{./test1_slice_time_30000}
\put(-23,45){$\gamma = 50$}
\end{overpic} 
\begin{overpic}[width=0.31\textwidth,grid=false,tics=10]{./test1_slice_time_150000}
\end{overpic}
\begin{overpic}[width=0.31\textwidth,grid=false,tics=10]{./test1_slice_time_300000}
\end{overpic}
\begin{overpic}[width=0.31\textwidth,grid=false,tics=10]{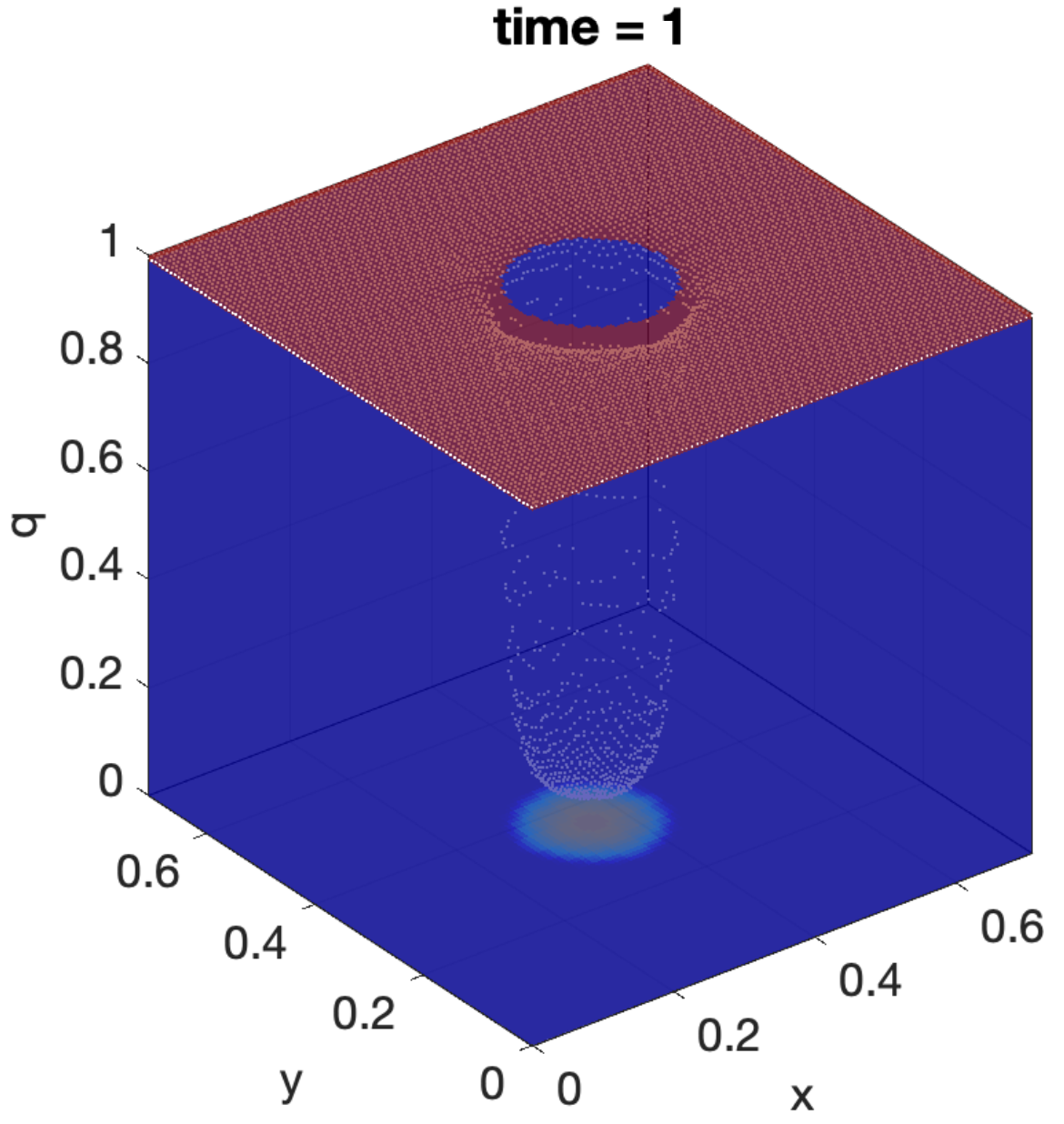}
\put(-23,45){$\gamma = 10$}
\end{overpic} 
\begin{overpic}[width=0.31\textwidth,grid=false,tics=10]{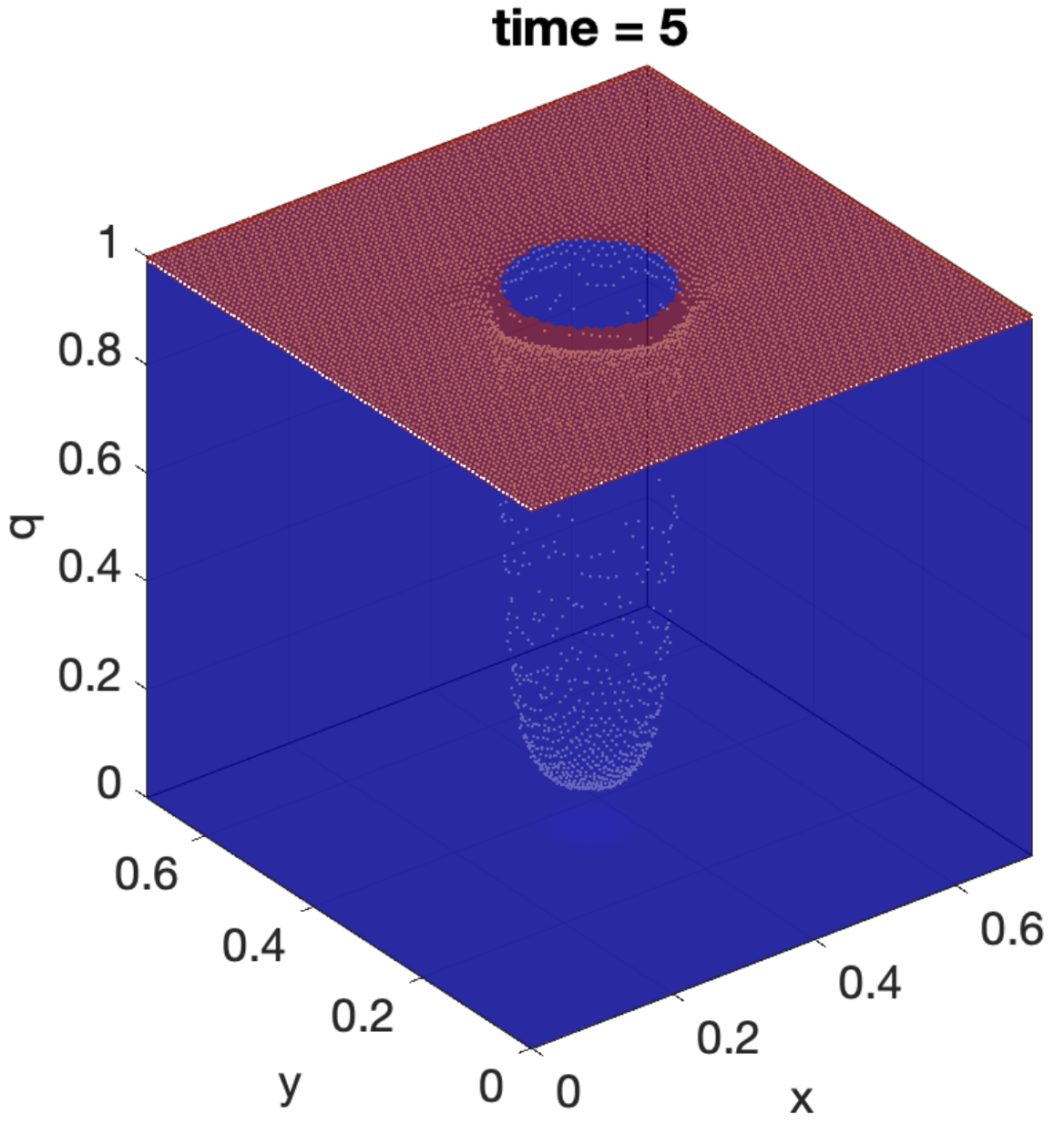}
\end{overpic}
\begin{overpic}[width=0.31\textwidth,grid=false,tics=10]{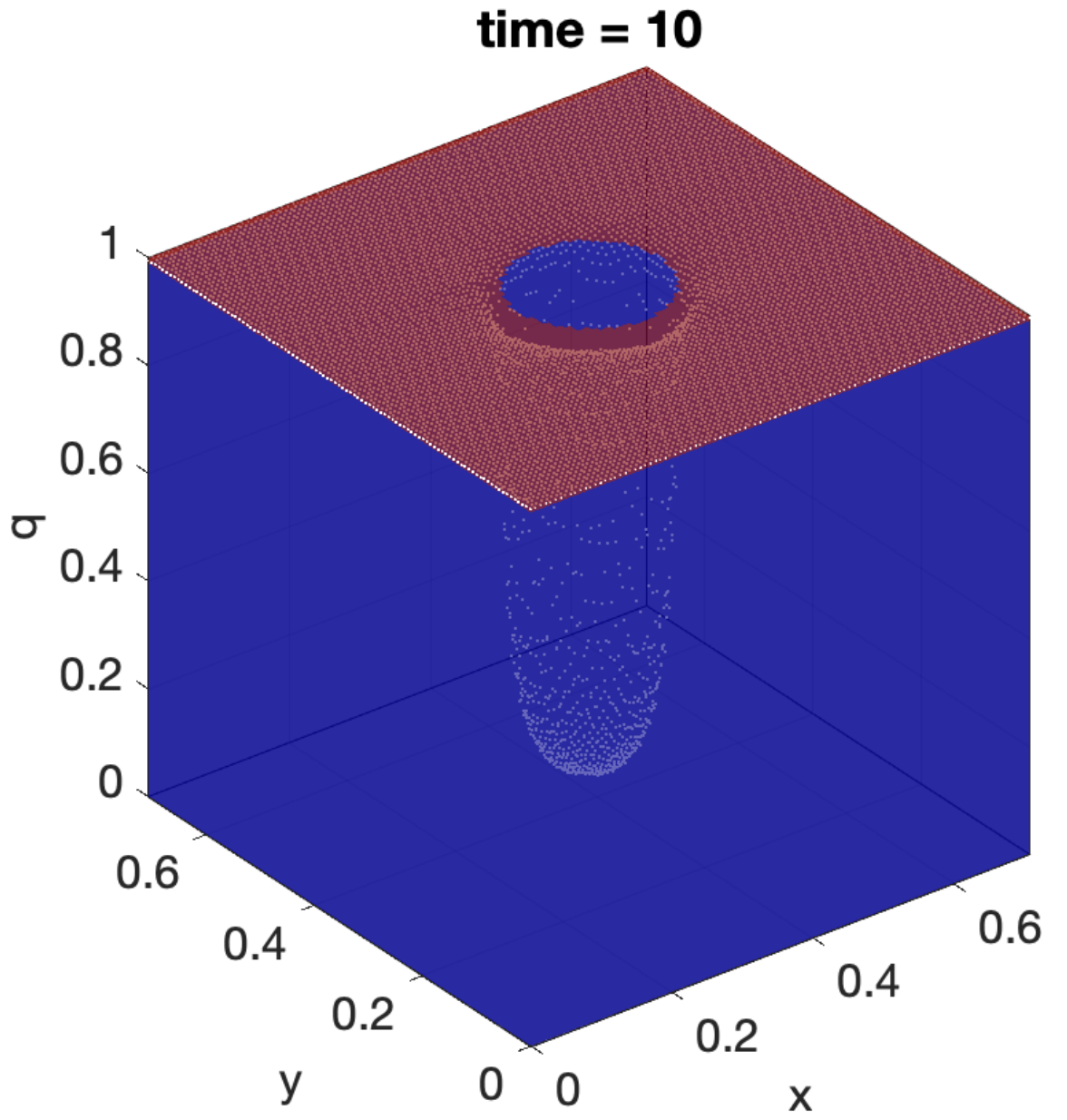}
\end{overpic} 
\begin{overpic}[width=0.31\textwidth,grid=false,tics=10]{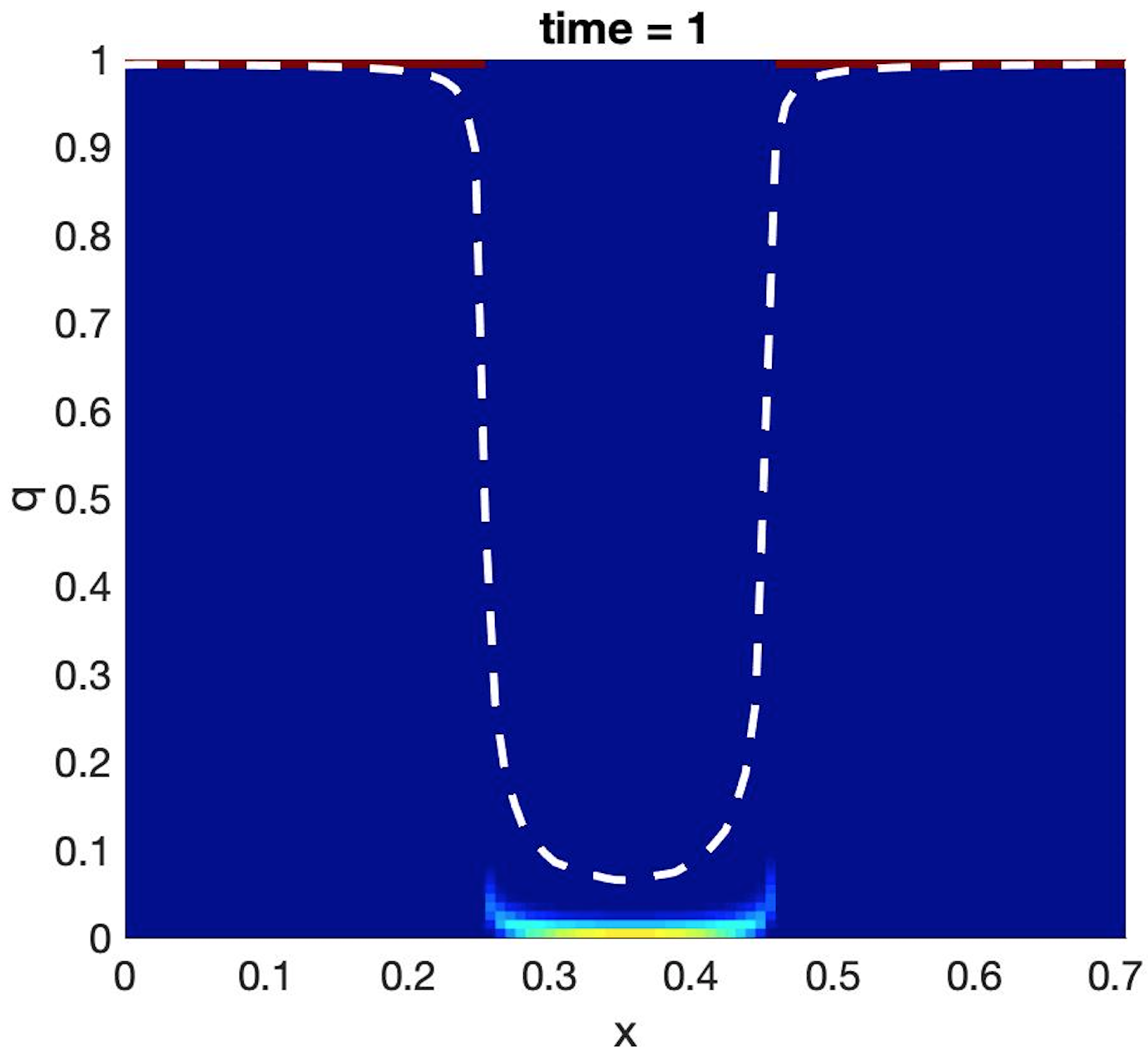}
\put(-23,45){$\gamma = 10$}
\end{overpic} 
\begin{overpic}[width=0.31\textwidth,grid=false,tics=10]{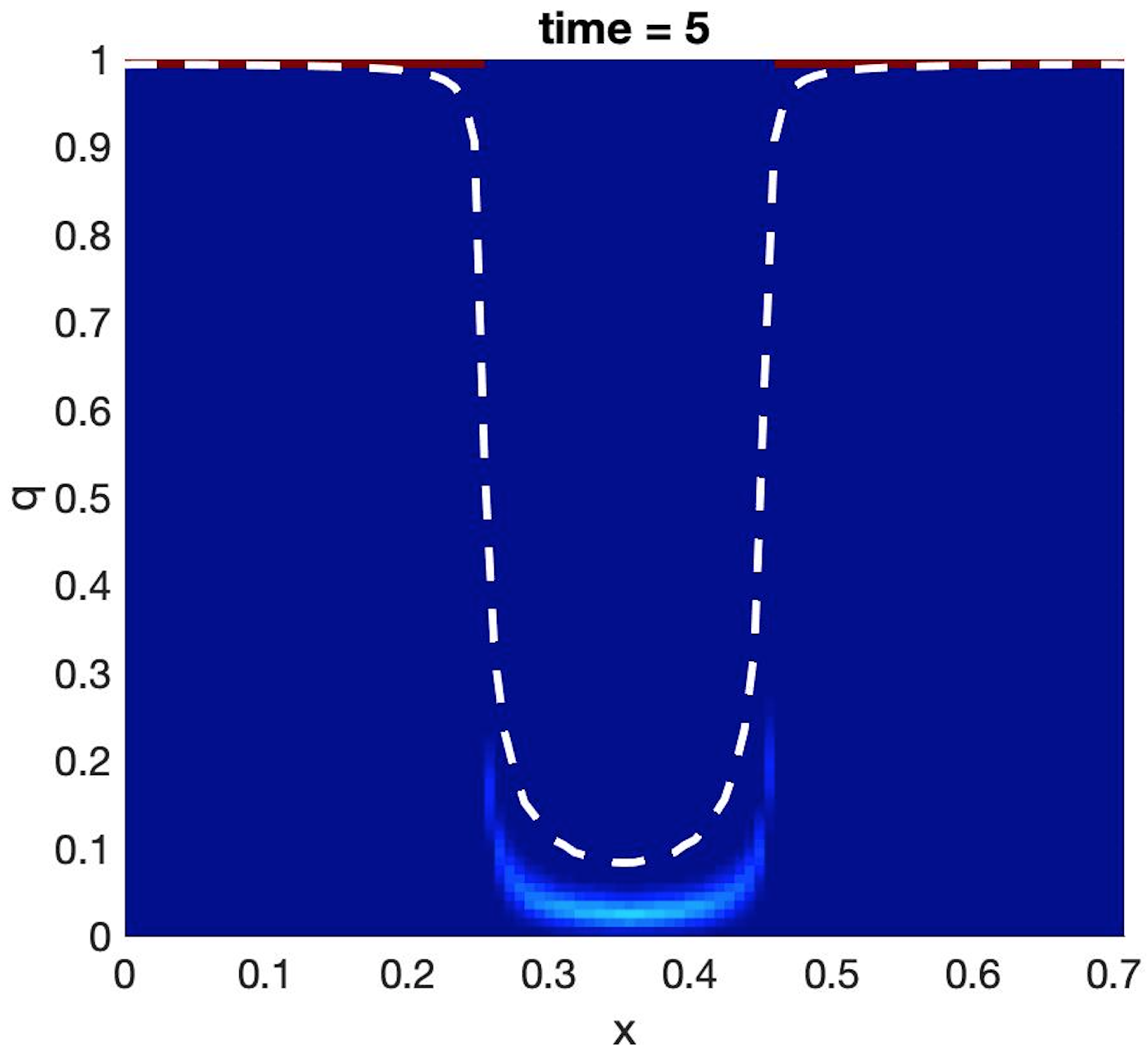}
\end{overpic}
\begin{overpic}[width=0.31\textwidth,grid=false,tics=10]{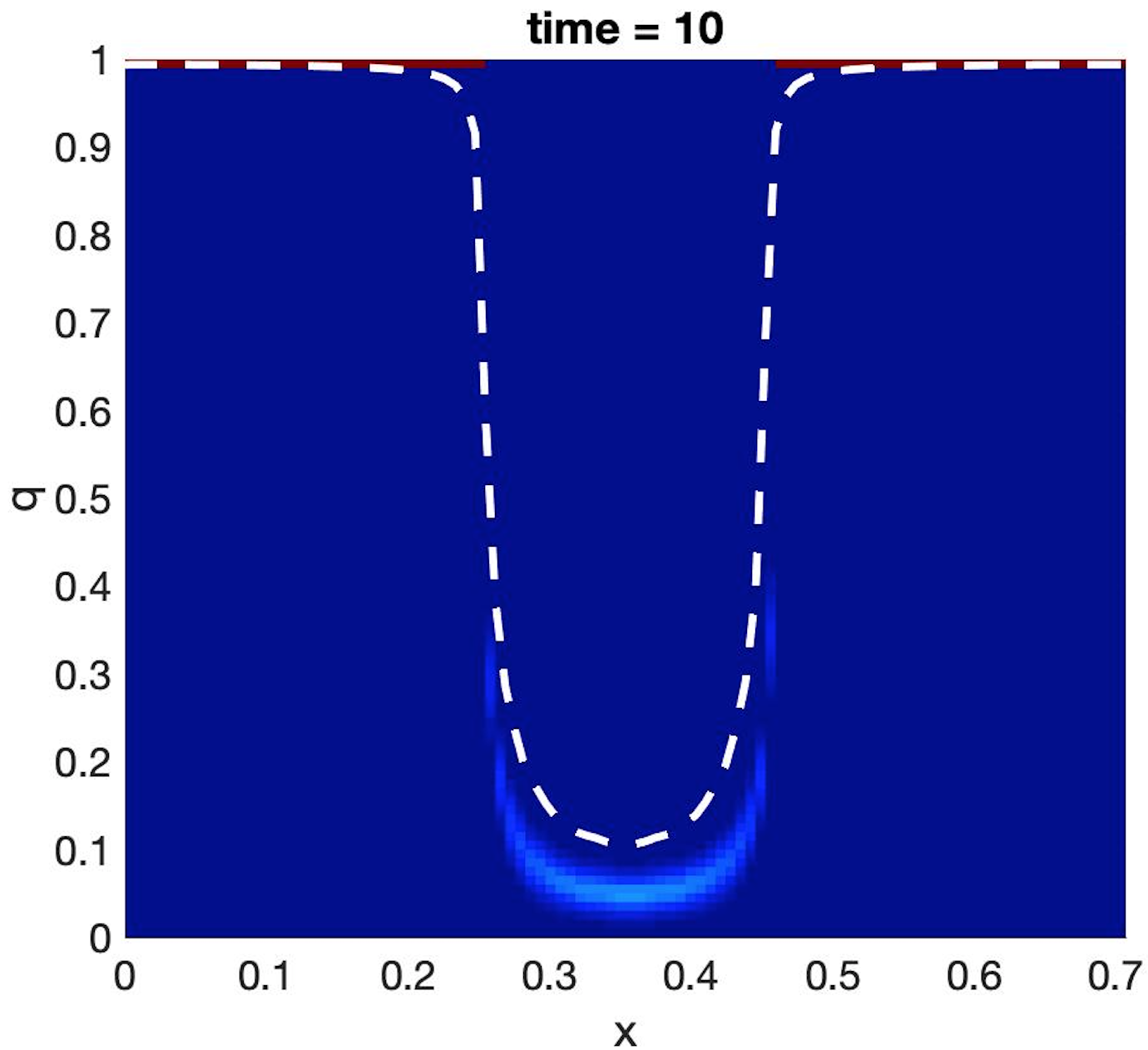}
\end{overpic}
\caption{
Tests with $v = 0$: evolution of the distribution density $h$ for $\gamma = 50$ (first row)
and corresponding results on section $y = 0.5$ (second row), 
evolution of $h$ for $\gamma = 10$ (third row)
and corresponding results on section $y = 0.5$ (fourth row). In both cases, the
initial condition is IC1.
Note that in all the subfigures time is dimensional while space is non-dimensional.
The white dashed line in the images on row two and four represents $q^*$.
The legend for all the images is the same as in Fig.~\ref{fig:IC}.}
\label{fig:v0_gamma}
\end{figure}

Next, we consider initial condition IC1 and compare the evolution of the distribution density $h$
for $\gamma = 50$ and $\gamma = 10$. See Fig.~\ref{fig:v0_gamma}. We see that
the level of exposure of the central group of healthy people increases faster for larger values
of $\gamma$, as expected. Parameter $\gamma$ plays a key role in the 
evolution of the disease spreading according to model \eqref{eq:2kineticeq}. If one was interested
in modeling a realistic scenario, such as the spreading of COVID-19 in a ER waiting room, 
$\gamma$ would have to be carefully tuned.

This first set of tests was meant to verify our implementation of method described in 
Sec.~\ref{sec:discr} and to check that the disease spreading term in eq.~\eqref{eq:2kineticeq} 
(i.e., the third term on the left-hand side) produced the expected outcomes.
Next, we will set people in motion. 

\subsection{Tests with prescribed walking velocity} 
We assign to all people walking speed $v = 1$ m/s and walking direction $\theta = \pi/4$, as if
they were headed to the upper right corner of the domain in the $xy$-plane. 
Once the people in the spreading phase of the disease have left the domain, we assume they
cannot spread to the people in the walkable domain anymore.

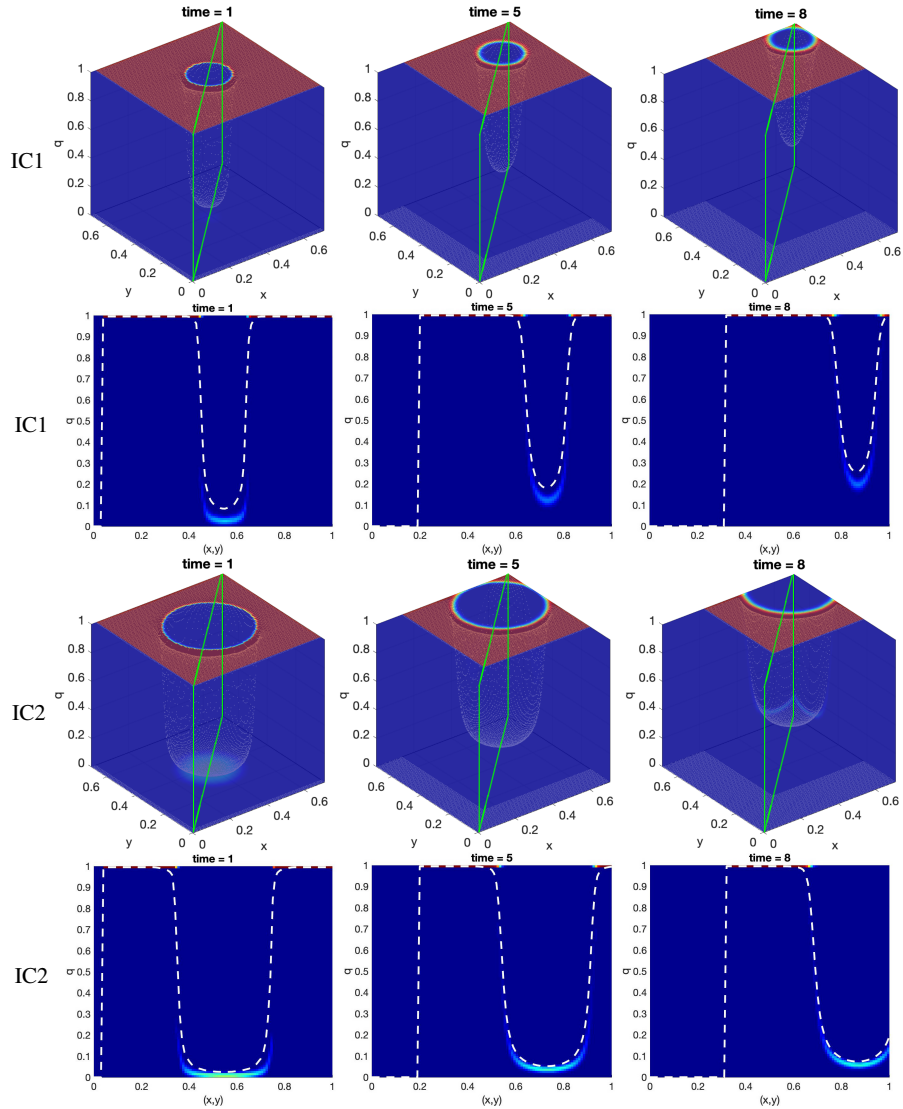
\begin{figure}
\centering
\begin{overpic}[width=0.32\textwidth,grid=false,tics=10]{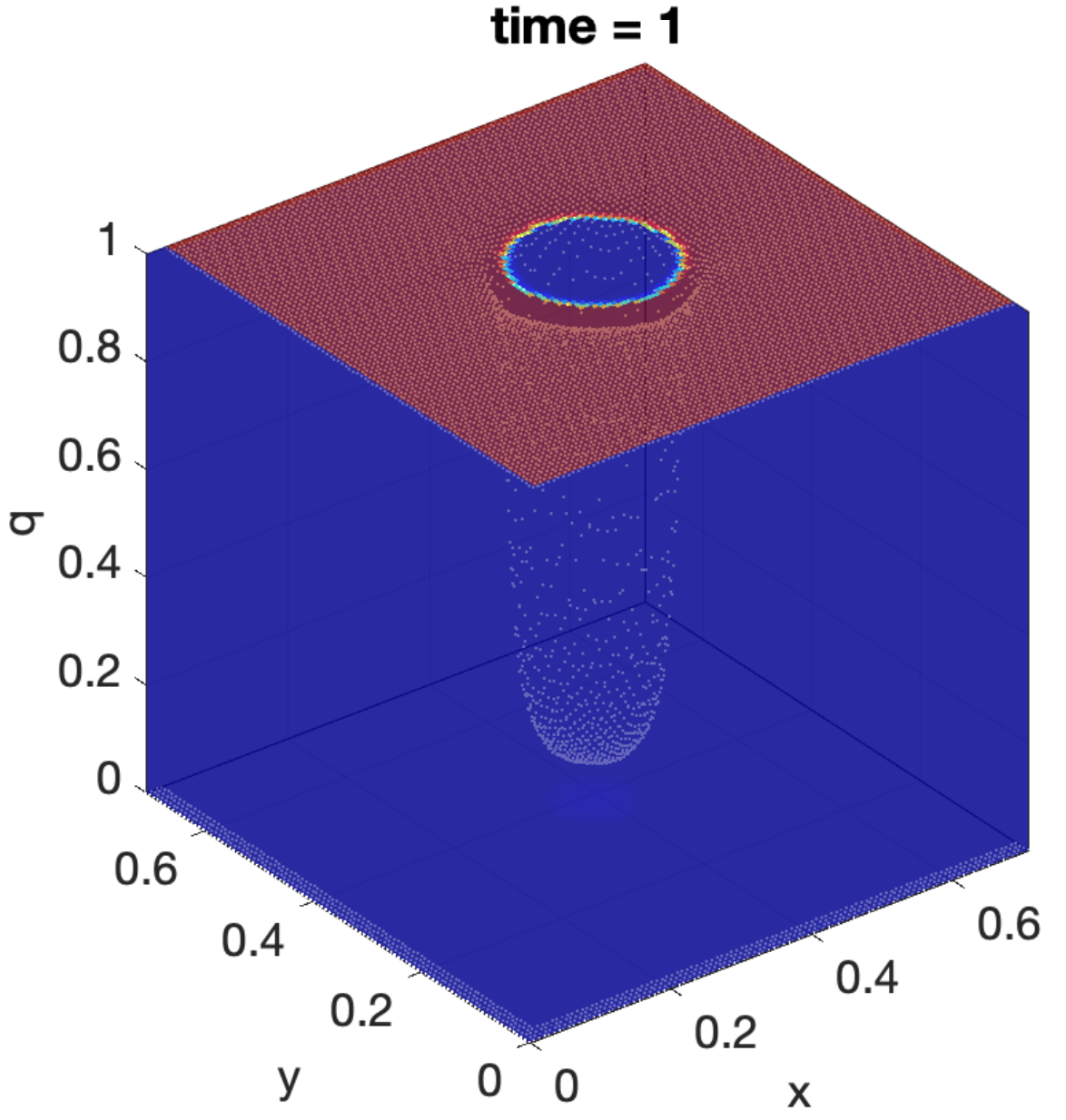}
\put(57.4,46){\color{green}\line(0,1){48.5}}
\put(57.4,46){\color{green}\line(-1,-4){9.8}}
\put(47.6,6.8){\color{green}\line(0,1){49.7}}
\put(57.1,94.2){\color{green}\line(-1,-4){9.5}}
\put(-14,45){IC1}
\end{overpic} 
\begin{overpic}[width=0.32\textwidth,grid=false,tics=10]{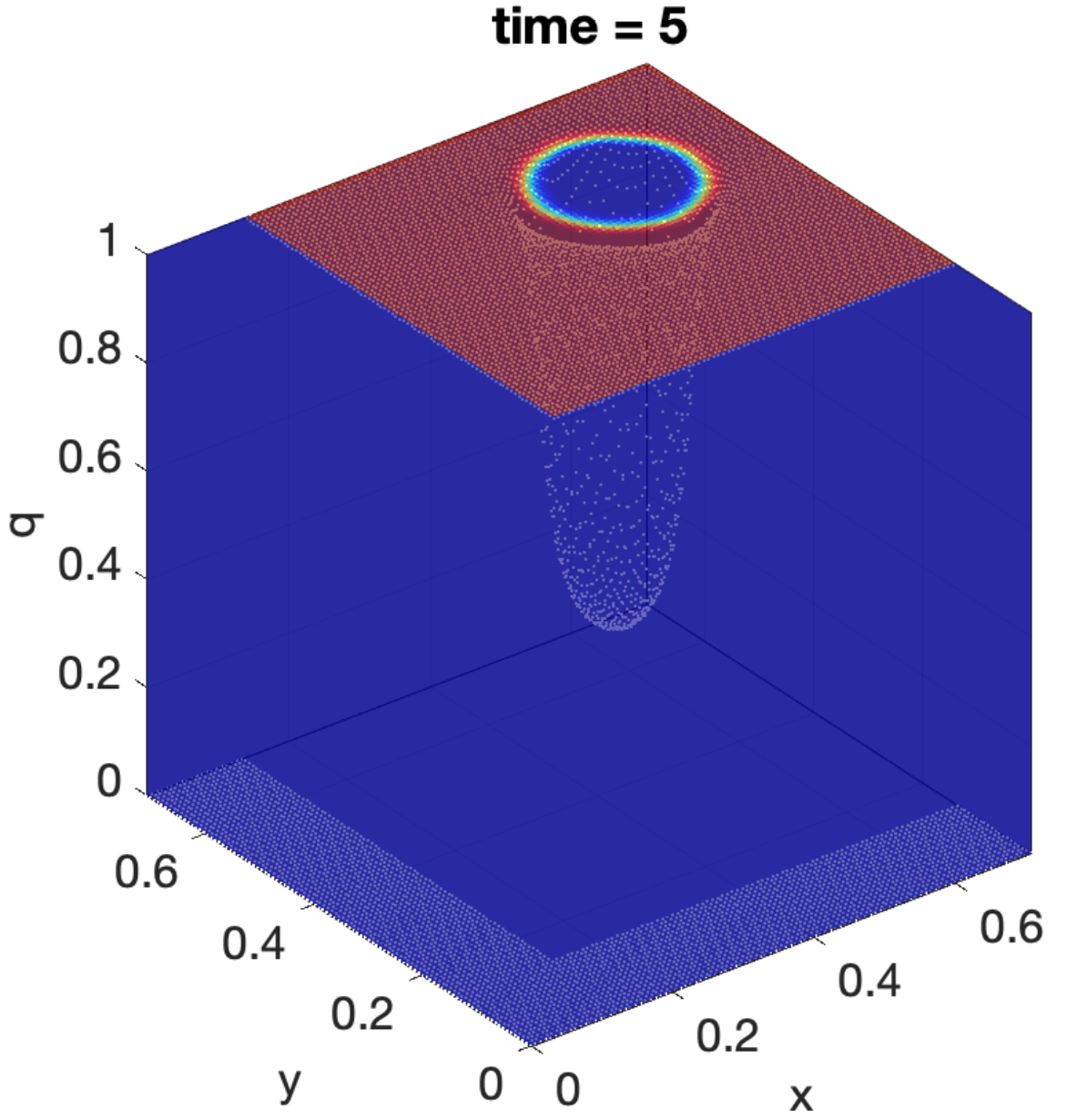}
\put(57.4,46){\color{green}\line(0,1){48.5}}
\put(57.4,46){\color{green}\line(-1,-4){9.8}}
\put(47.6,6.8){\color{green}\line(0,1){49.7}}
\put(57.1,94.2){\color{green}\line(-1,-4){9.5}}
\end{overpic}
\begin{overpic}[width=0.32\textwidth,grid=false,tics=10]{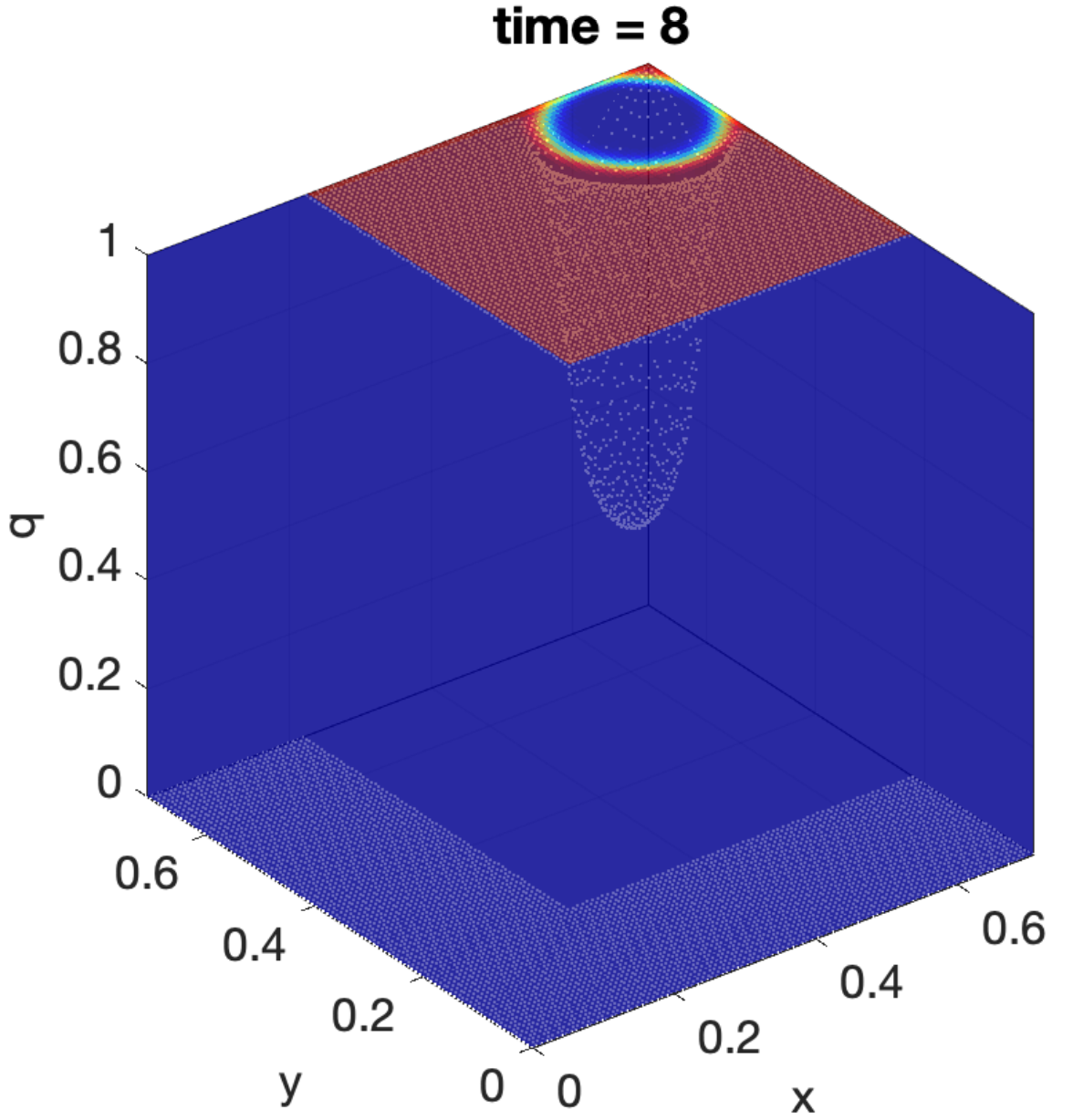}
\put(57.4,46){\color{green}\line(0,1){48.5}}
\put(57.4,46){\color{green}\line(-1,-4){9.8}}
\put(47.6,6.8){\color{green}\line(0,1){49.7}}
\put(57.1,94.2){\color{green}\line(-1,-4){9.5}}
\end{overpic}
\begin{overpic}[width=0.31\textwidth,grid=false,tics=10]{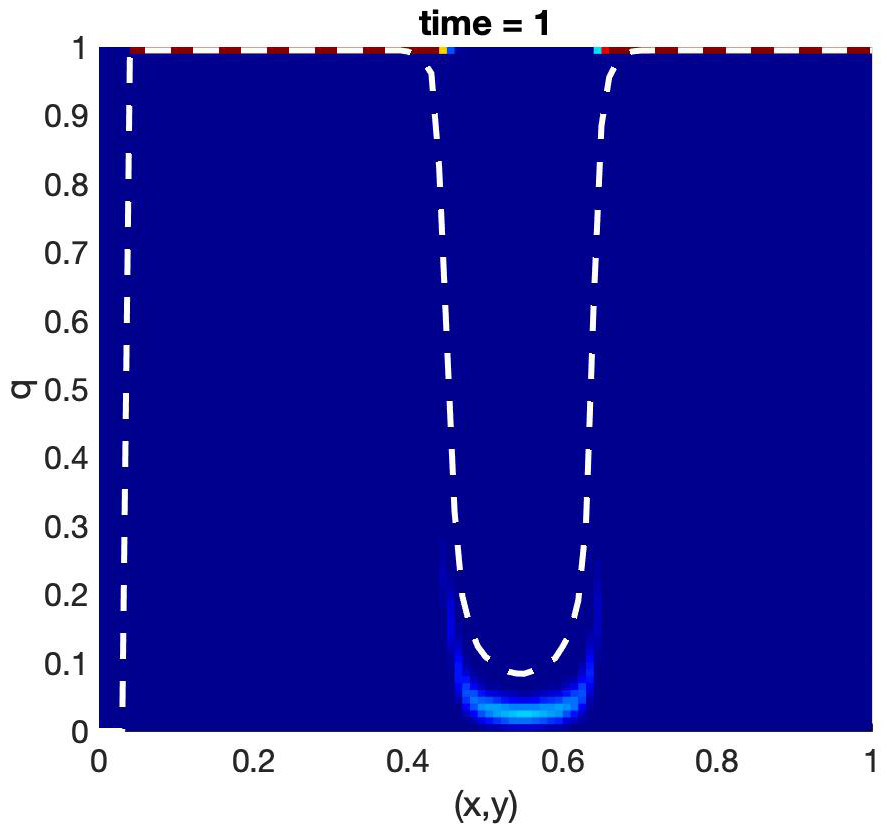}
\put(-18,45){IC1}
\end{overpic} 
\begin{overpic}[width=0.31\textwidth,grid=false,tics=10]{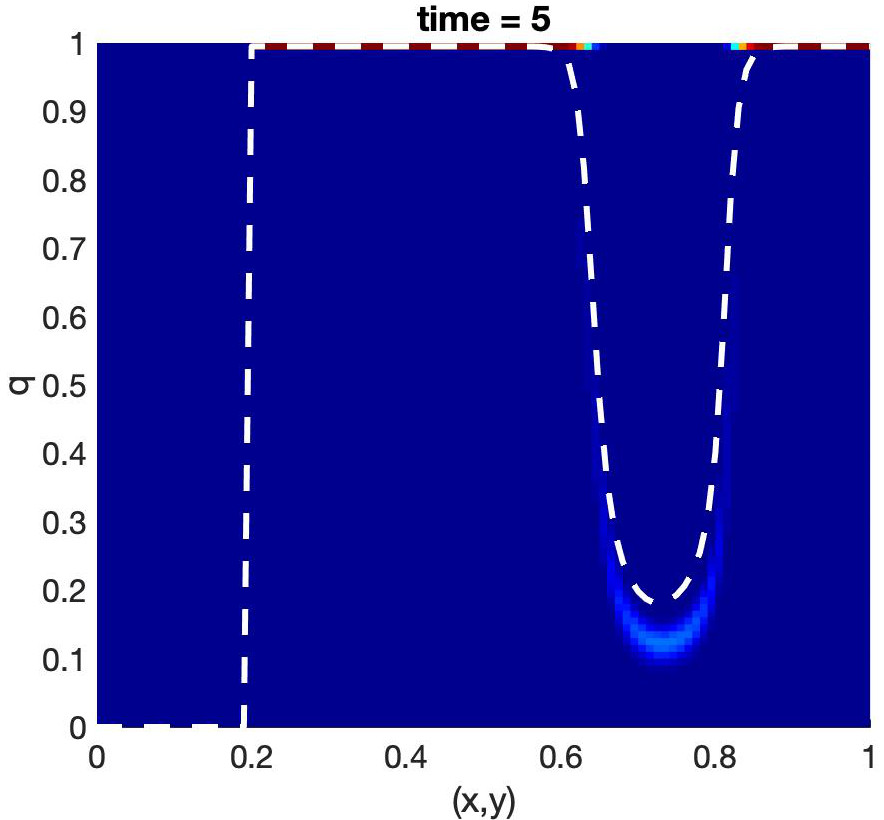}
\end{overpic}
\begin{overpic}[width=0.31\textwidth,grid=false,tics=10]{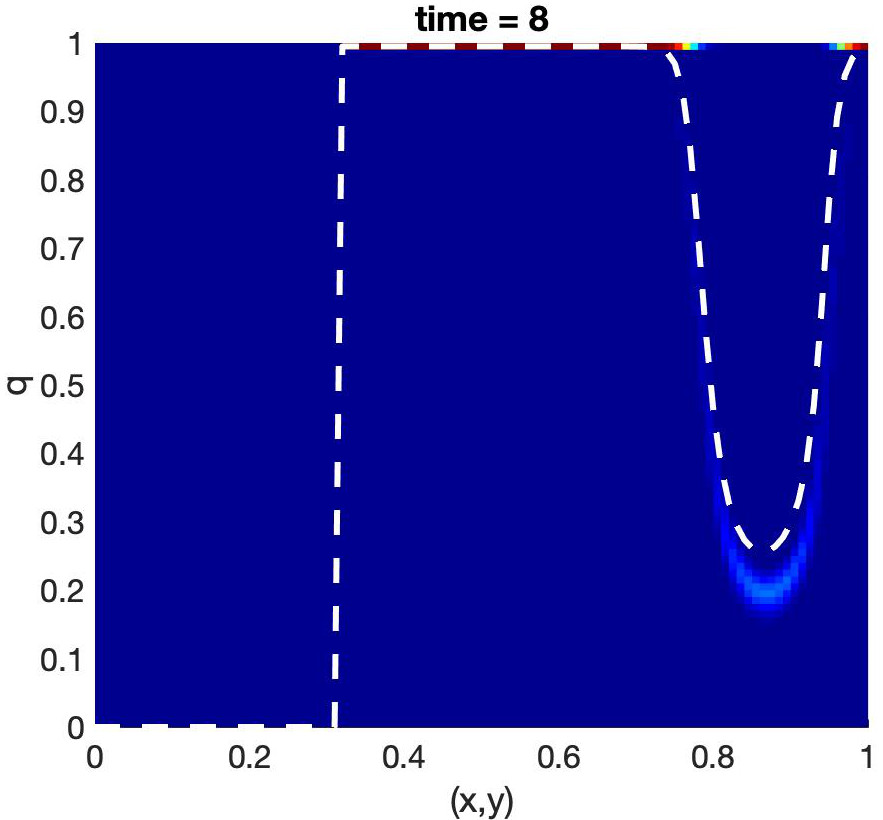}
\end{overpic} 
\begin{overpic}[width=0.32\textwidth,grid=false,tics=10]{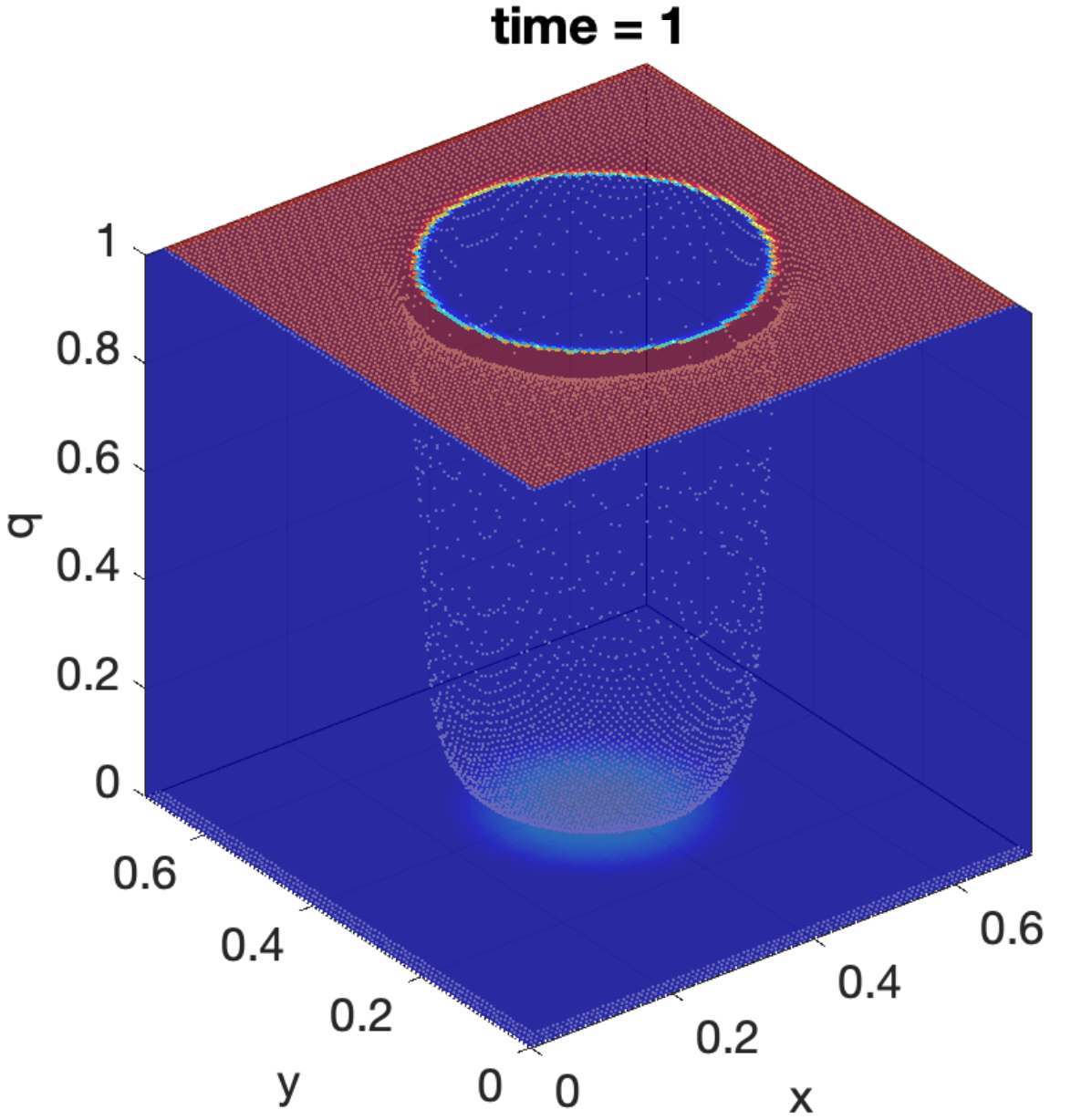}
\put(57.4,46){\color{green}\line(0,1){48.5}}
\put(57.4,46){\color{green}\line(-1,-4){9.8}}
\put(47.6,6.8){\color{green}\line(0,1){49.7}}
\put(57.1,94.2){\color{green}\line(-1,-4){9.5}}
\put(-14,45){IC2}
\end{overpic} 
\begin{overpic}[width=0.32\textwidth,grid=false,tics=10]{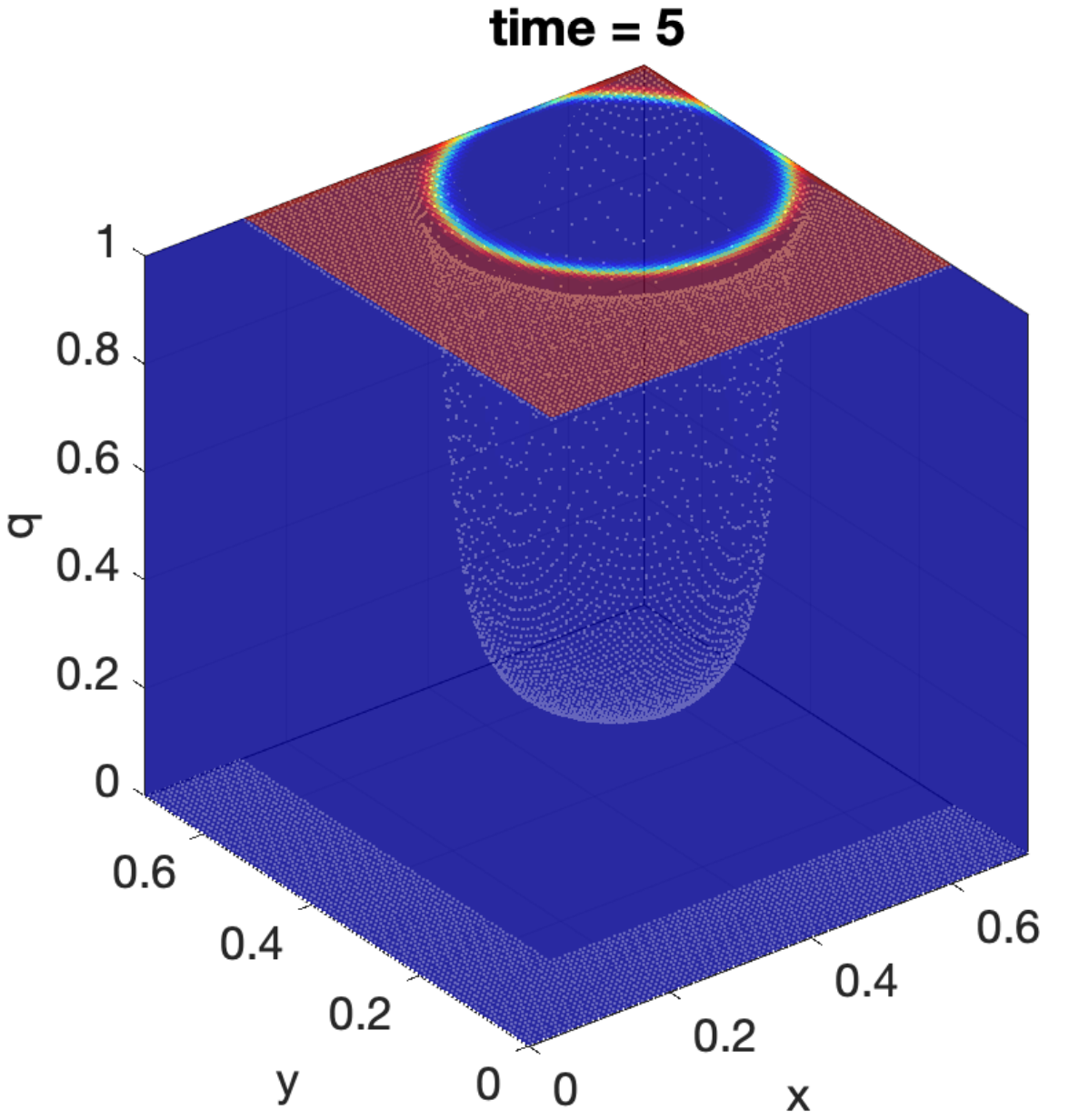}
\put(57.4,46){\color{green}\line(0,1){48.5}}
\put(57.4,46){\color{green}\line(-1,-4){9.8}}
\put(47.6,6.8){\color{green}\line(0,1){49.7}}
\put(57.1,94.2){\color{green}\line(-1,-4){9.5}}
\end{overpic}
\begin{overpic}[width=0.32\textwidth,grid=false,tics=10]{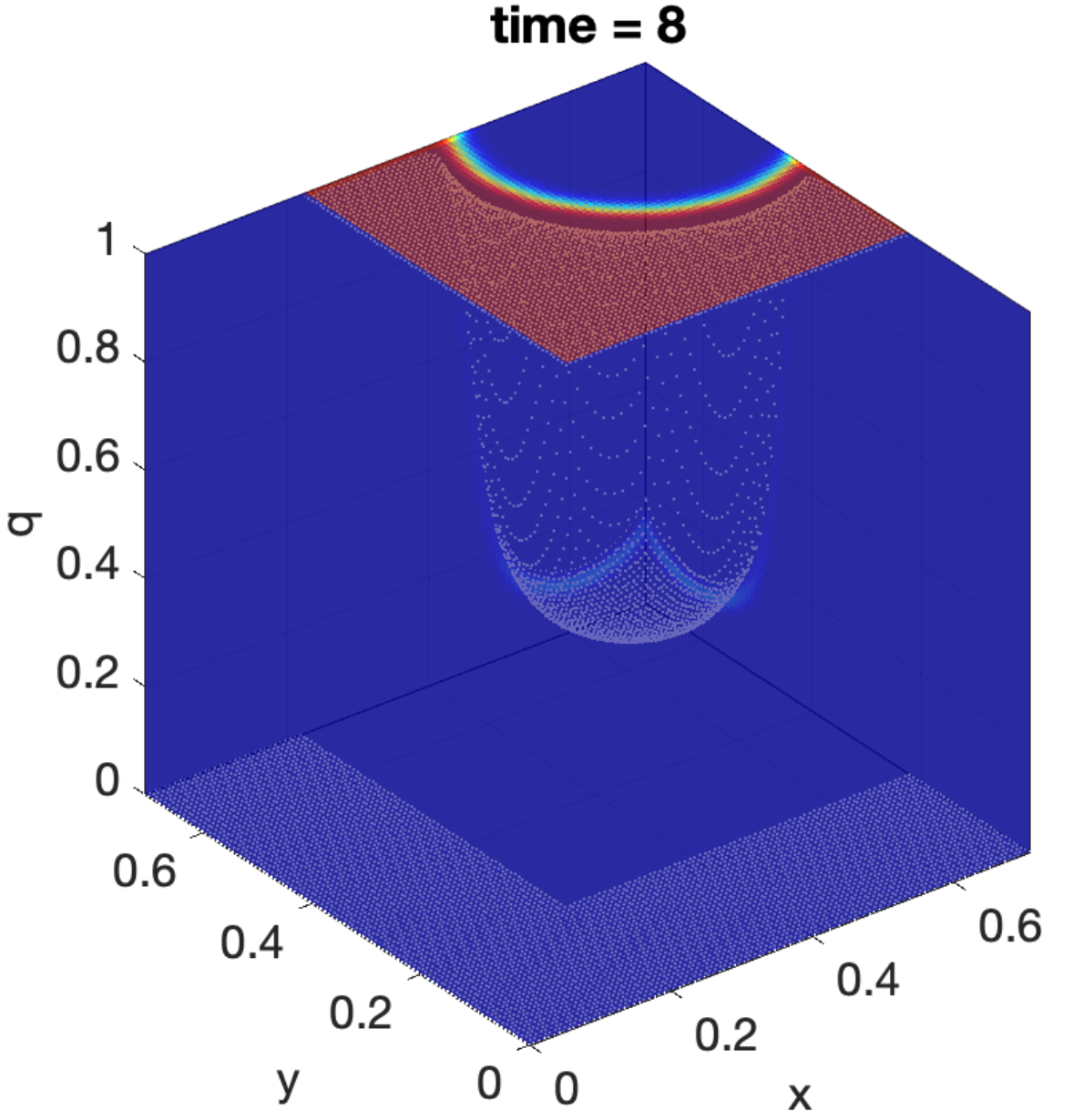}
\put(57.4,46){\color{green}\line(0,1){48.5}}
\put(57.4,46){\color{green}\line(-1,-4){9.8}}
\put(47.6,6.8){\color{green}\line(0,1){49.7}}
\put(57.1,94.2){\color{green}\line(-1,-4){9.5}}
\end{overpic}
\begin{overpic}[width=0.31\textwidth,grid=false,tics=10]{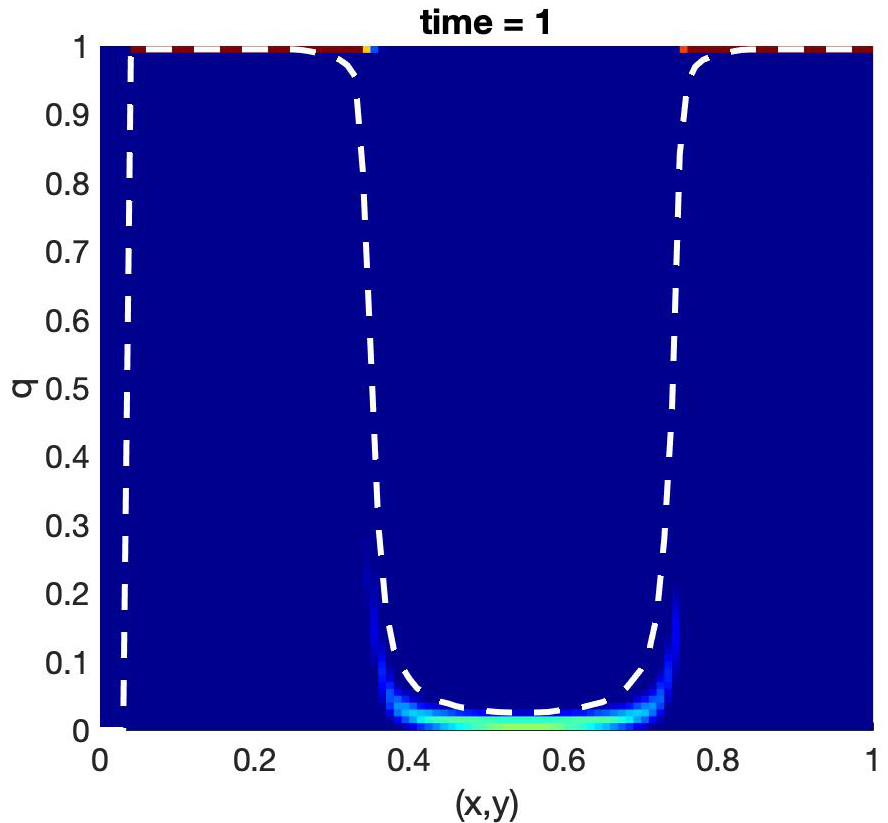}
\put(-18,45){IC2}
\end{overpic} 
\begin{overpic}[width=0.31\textwidth,grid=false,tics=10]{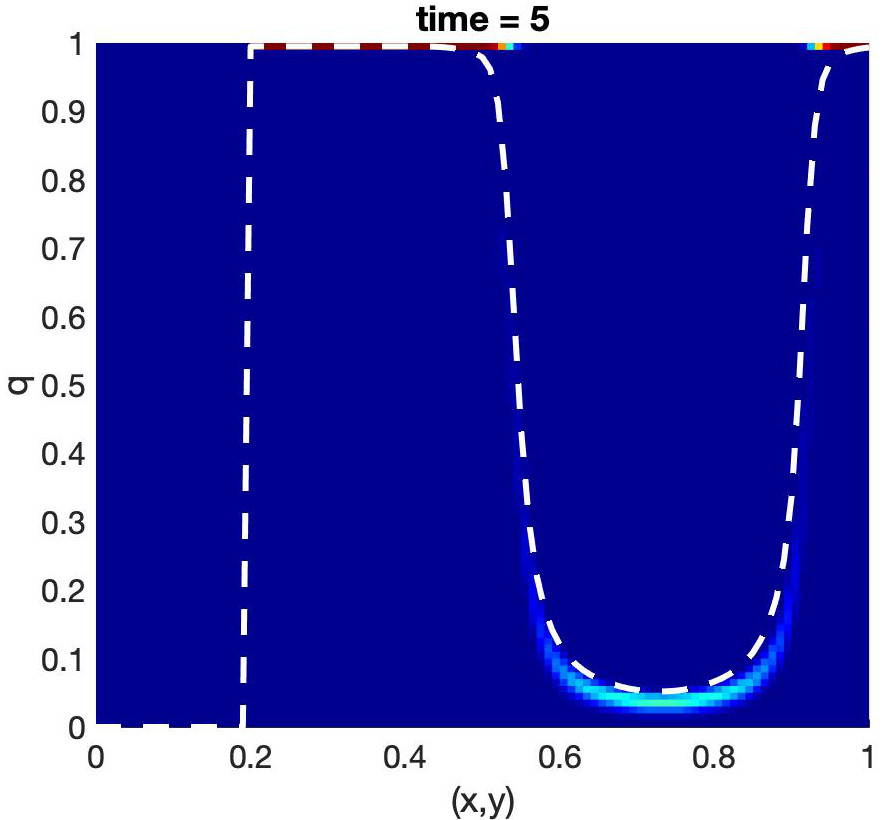}
\end{overpic}
\begin{overpic}[width=0.31\textwidth,grid=false,tics=10]{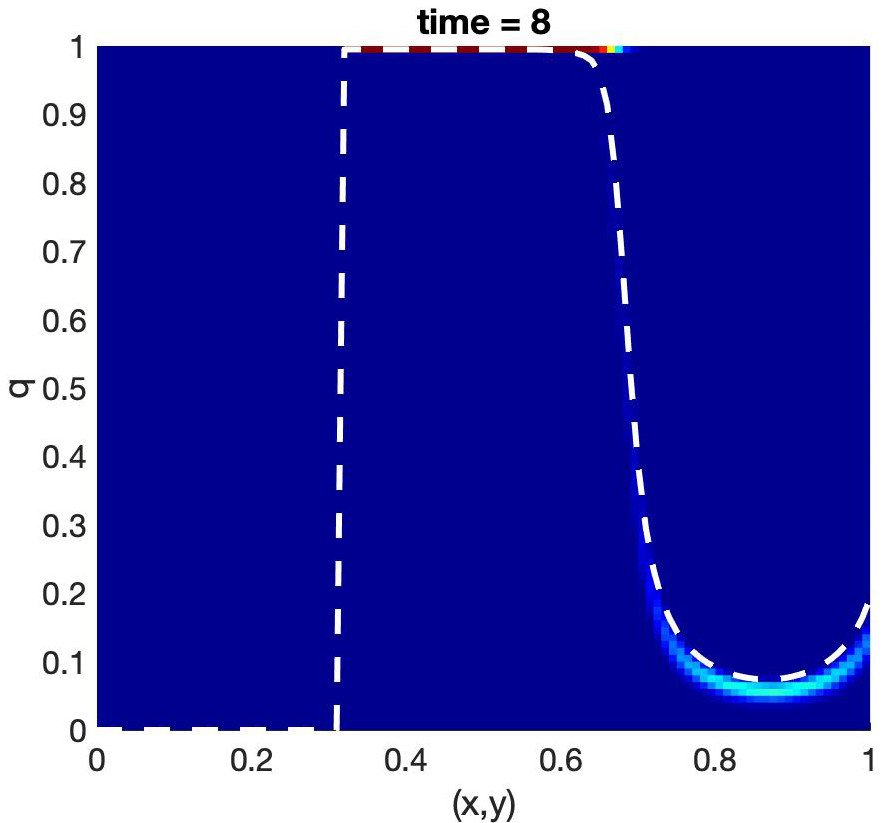}
\end{overpic} 
\caption{
Tests with prescribed walking velocity: 
the first row shows the evolution of the distribution density $h$ for initial condition IC1 and
highlighted in green is the section whose results are reported in the second row. 
The third row shows the evolution of $h$ for initial condition IC2 and
highlighted in green is the section whose results are reported in the fourth row. 
Note that in all the subfigures time is dimensional while space is non-dimensional.
For both cases, $\gamma = 50$.
The white dashed line in the images on row two and four represents $q^*$.
The legend for all the images is the same as in Fig.~\ref{fig:IC}.} \label{fig:v1_IC}
\end{figure}

We consider again IC1 and IC2 with $\gamma = 50$.
Fig.~\ref{fig:v1_IC} shows the evolution of the distribution density $h$ for both initial conditions. 
As one would expect, the motion contributes to lowering the exposure level in the both cases, 
since some of the spreading people leave the domain as soon as the motion starts. This difference
is obvious when one compares the second (resp., fourth) row in Fig.~\ref{fig:v0} with the second (resp., fourth)
row in Fig.~\ref{fig:v1_IC}. 

\begin{figure}
\centering
\begin{overpic}[width=0.42\textwidth,grid=false,tics=10]{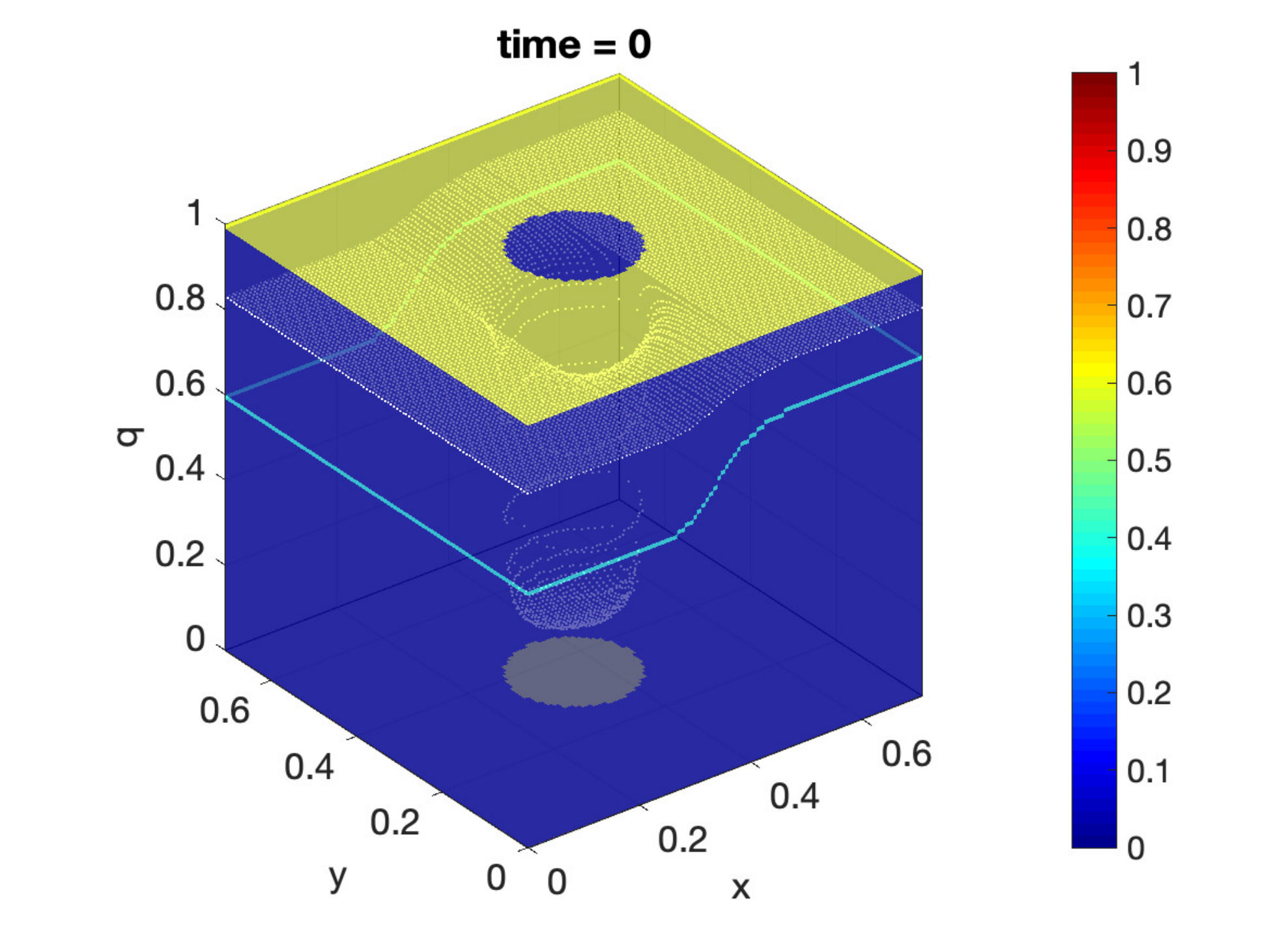}
\put(48.6,35.6){\color{green}\line(0,1){33.7}}
\put(48.5,69.2){\color{green}\line(-1,-4){6.9}}
\put(41.6,8.3){\color{green}\line(0,1){33.5}}
\put(48.6,35.6){\color{green}\line(-1,-4){6.8}}
\end{overpic} 
\begin{overpic}[width=0.42\textwidth,grid=false,tics=10]{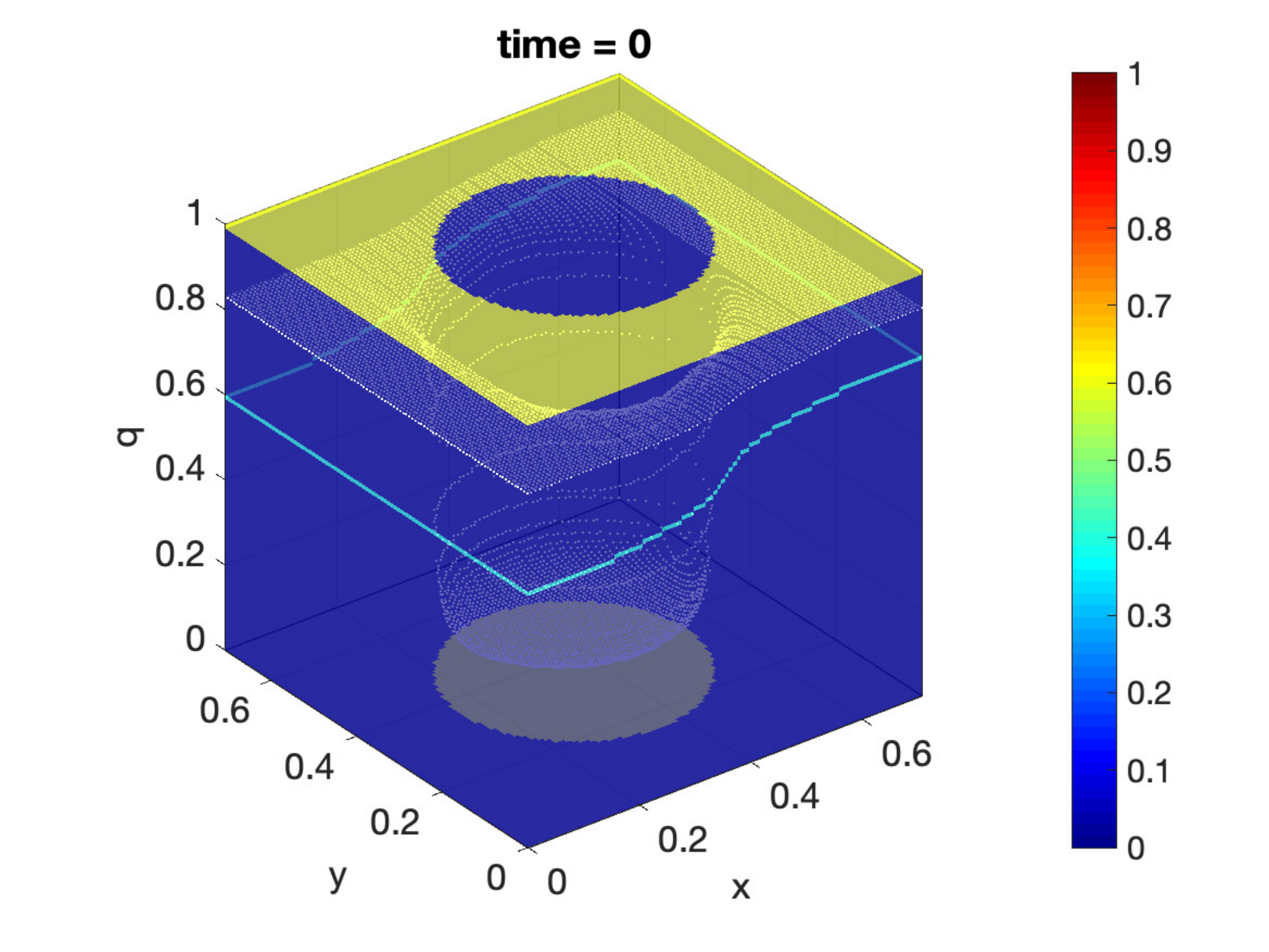}
\put(48.6,35.6){\color{green}\line(0,1){33.7}}
\put(48.5,69.2){\color{green}\line(-1,-4){6.9}}
\put(41.6,8.3){\color{green}\line(0,1){33.5}}
\put(48.6,35.6){\color{green}\line(-1,-4){6.8}}
\end{overpic} 
\\
\begin{overpic}[width=0.42\textwidth,grid=false,tics=10]{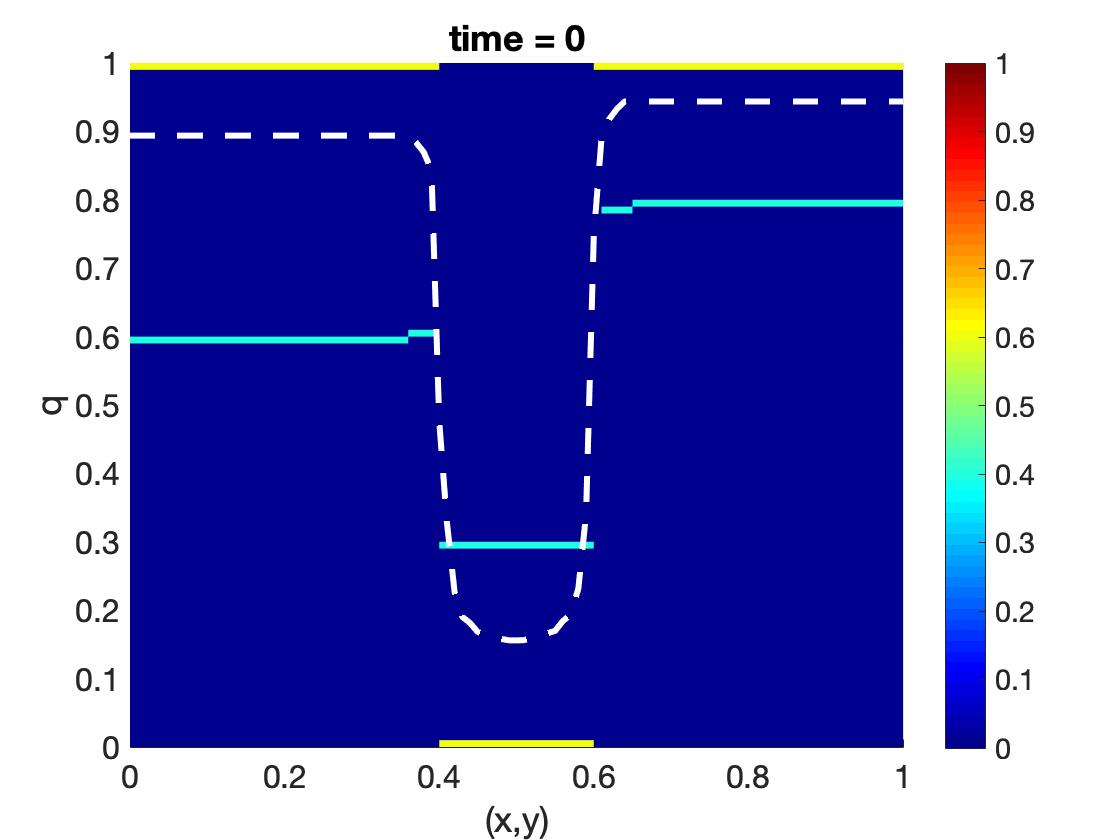}
\end{overpic} 
\begin{overpic}[width=0.42\textwidth,grid=false,tics=10]{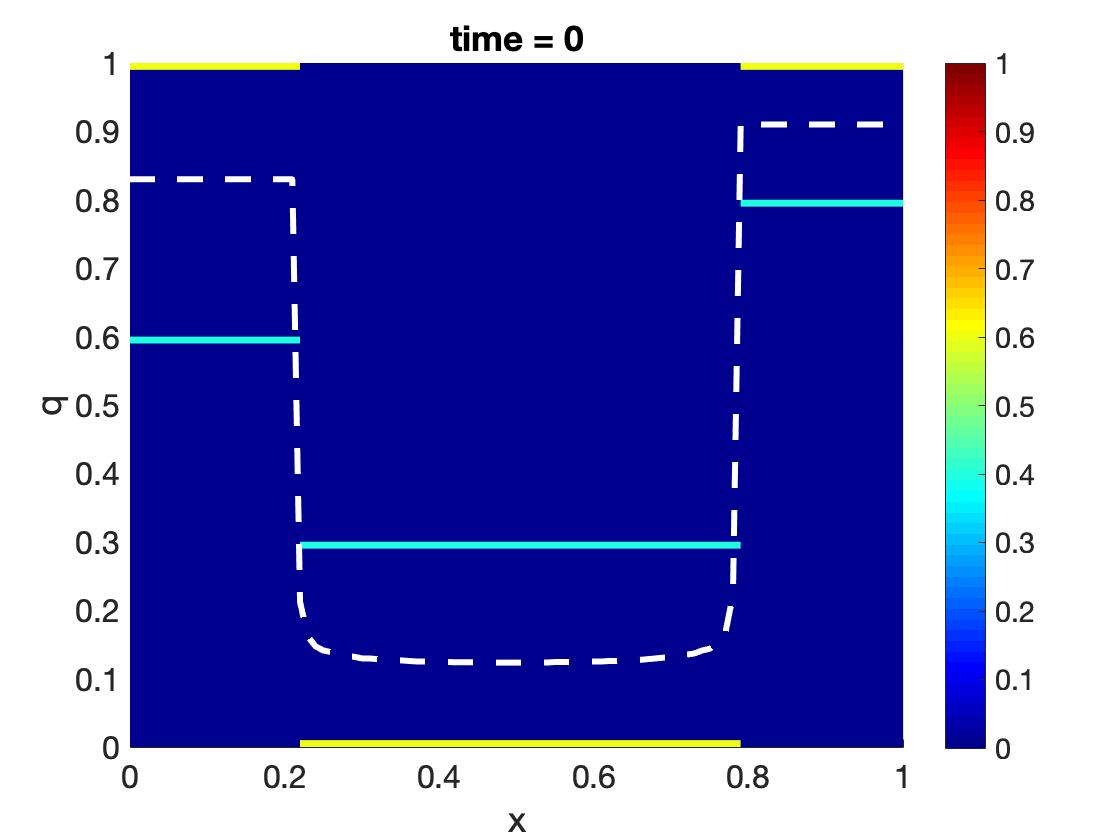}
\end{overpic} 
\caption{Tests with prescribed walking velocity: initial conditions IC1-bis (top left) and IC2-bis (top right).
The initial data on the section highlighted in green in the top panels is reported in the corresponding bottom panel. 
}\label{fig:ICbis}
\end{figure}

\begin{figure}
\centering
\begin{overpic}[width=0.32\textwidth,grid=false,tics=10]{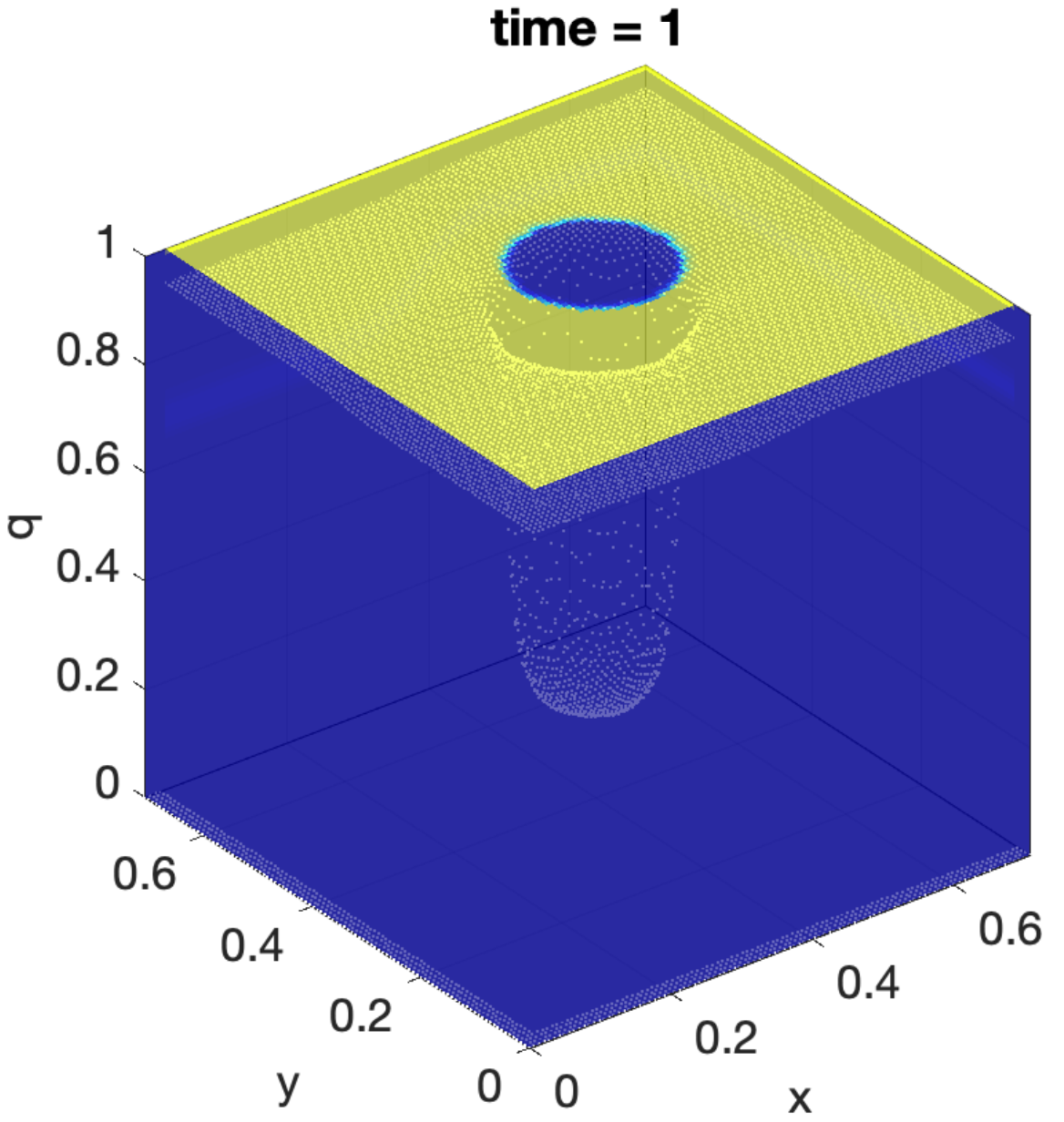}
\put(57.4,46){\color{green}\line(0,1){48.5}}
\put(57.4,46){\color{green}\line(-1,-4){9.8}}
\put(47.6,6.8){\color{green}\line(0,1){49.7}}
\put(57.1,94.2){\color{green}\line(-1,-4){9.5}}
\put(-18,45){IC1-bis}
\end{overpic} 
\begin{overpic}[width=0.32\textwidth,grid=false,tics=10]{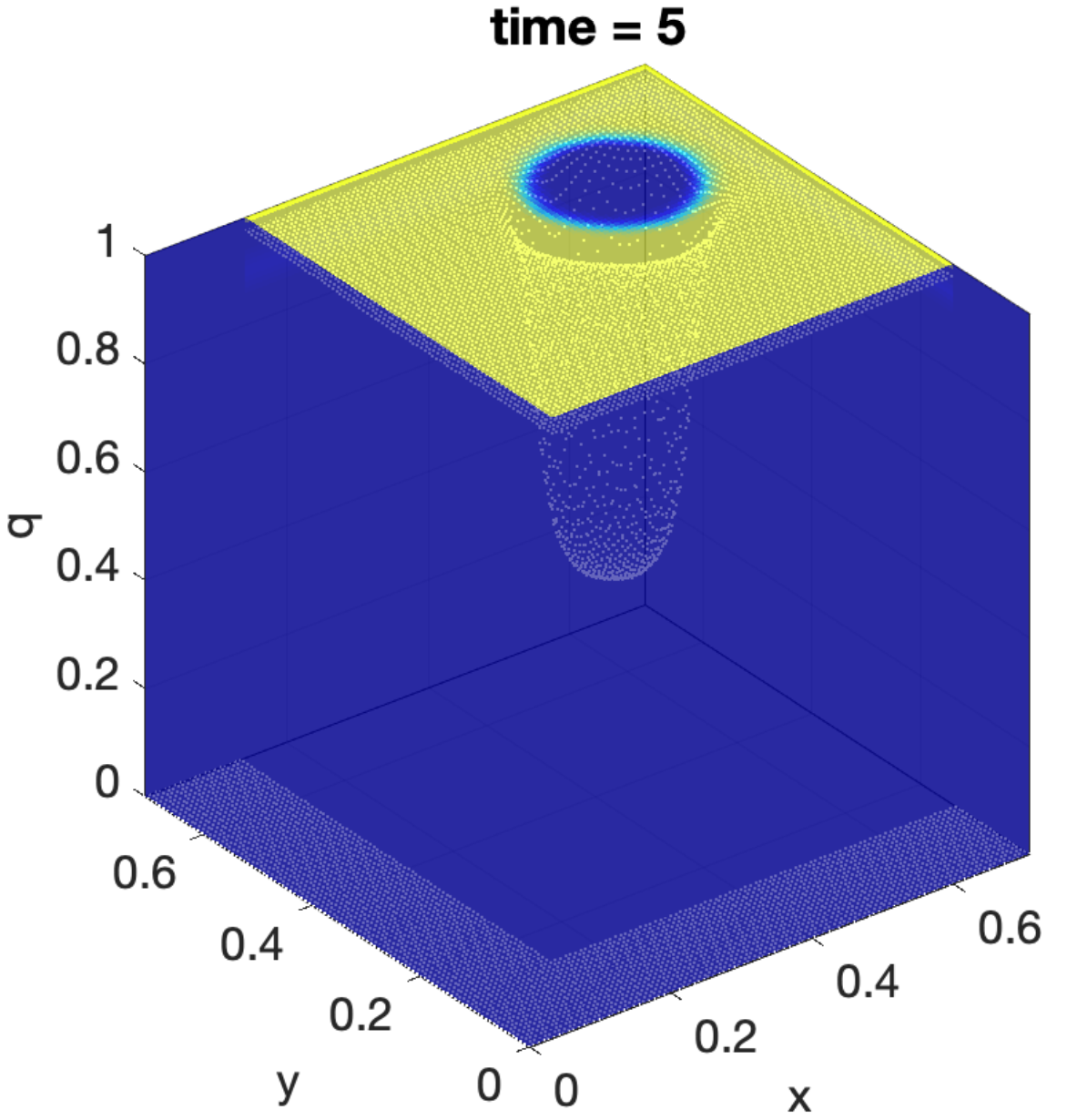}
\put(57.4,46){\color{green}\line(0,1){48.5}}
\put(57.4,46){\color{green}\line(-1,-4){9.8}}
\put(47.6,6.8){\color{green}\line(0,1){49.7}}
\put(57.1,94.2){\color{green}\line(-1,-4){9.5}}
\end{overpic}
\begin{overpic}[width=0.32\textwidth,grid=false,tics=10]{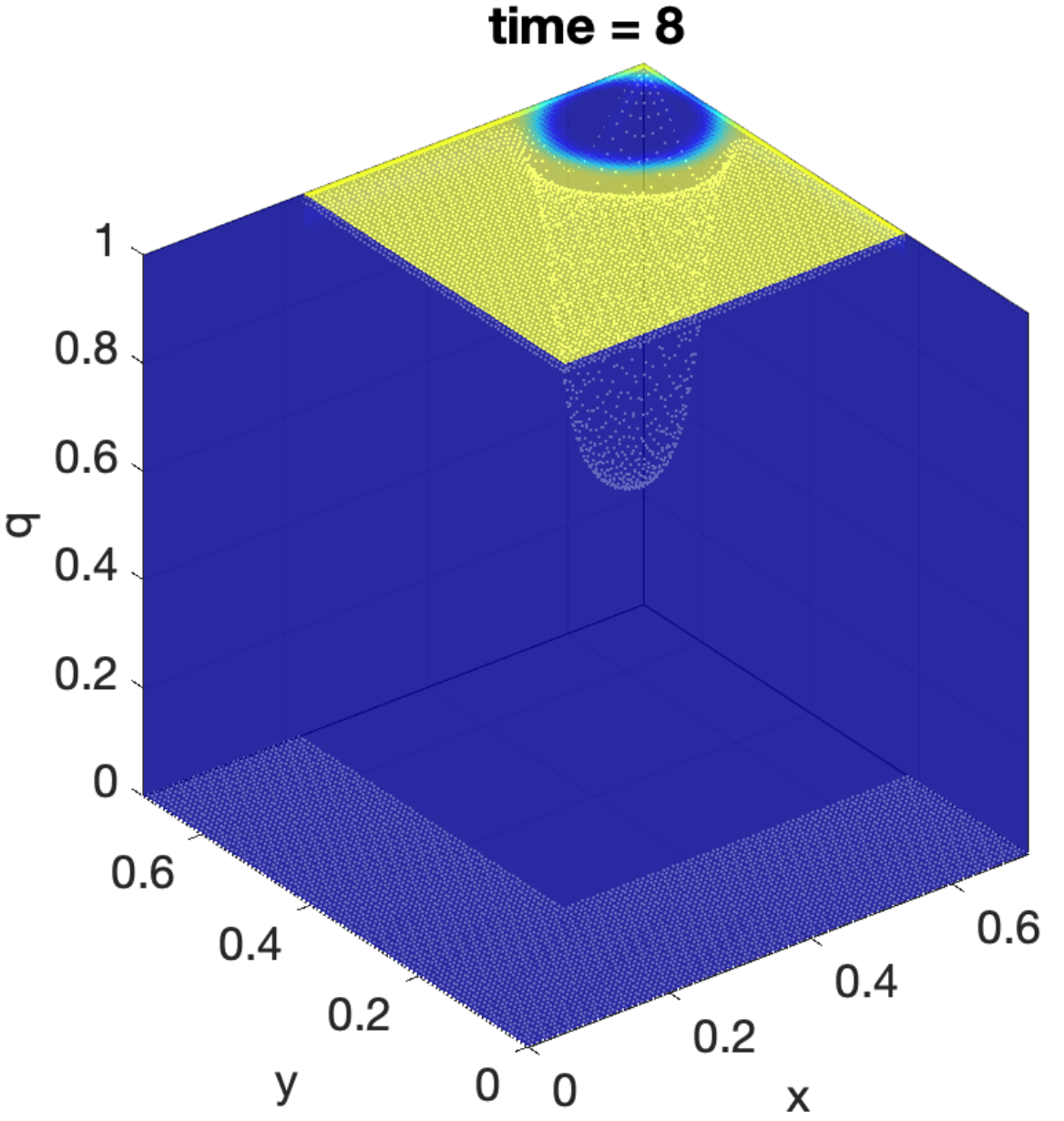}
\put(57.4,46){\color{green}\line(0,1){48.5}}
\put(57.4,46){\color{green}\line(-1,-4){9.8}}
\put(47.6,6.8){\color{green}\line(0,1){49.7}}
\put(57.1,94.2){\color{green}\line(-1,-4){9.5}}
\end{overpic}
\begin{overpic}[width=0.31\textwidth,grid=false,tics=10]{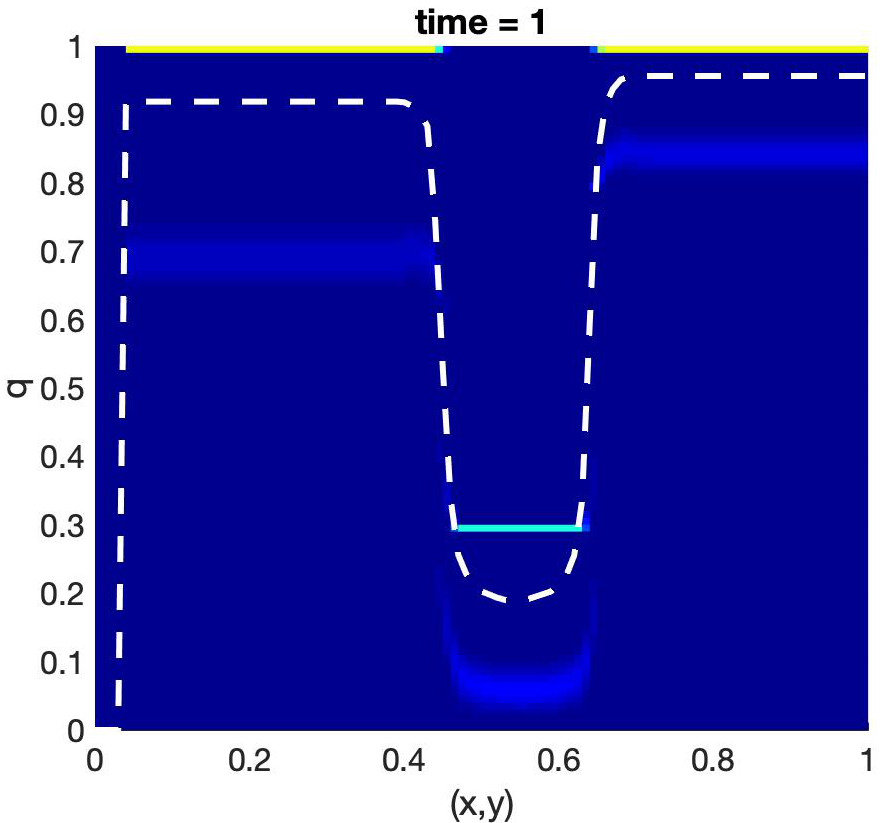}
\put(-22,43){IC1-bis}
\end{overpic} 
\begin{overpic}[width=0.31\textwidth,grid=false,tics=10]{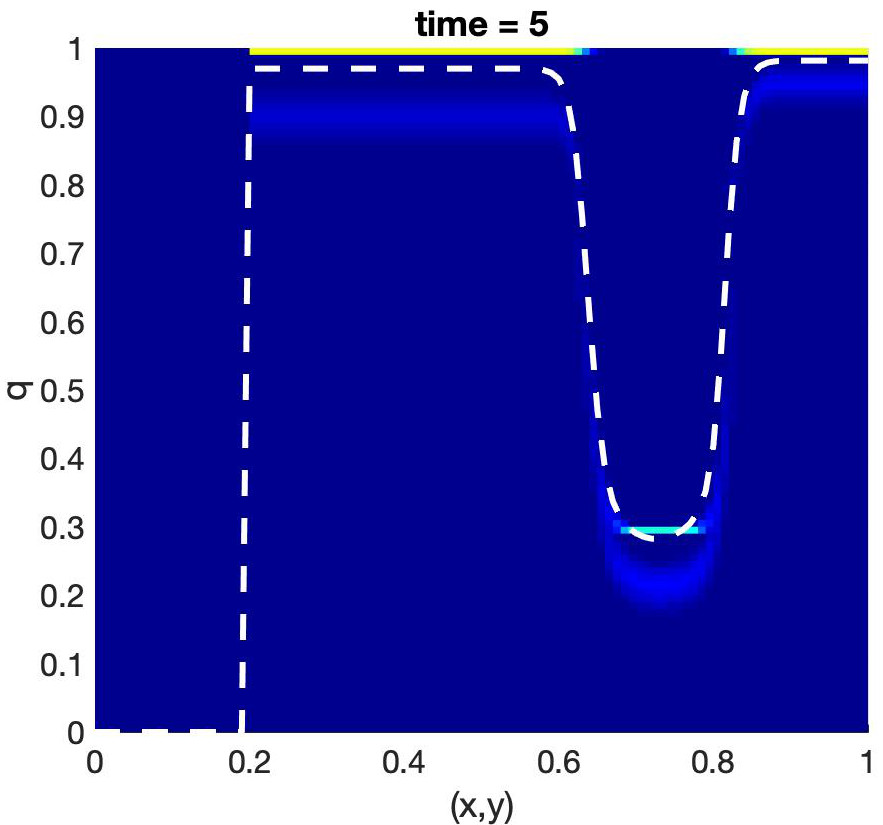}
\end{overpic}
\begin{overpic}[width=0.31\textwidth,grid=false,tics=10]{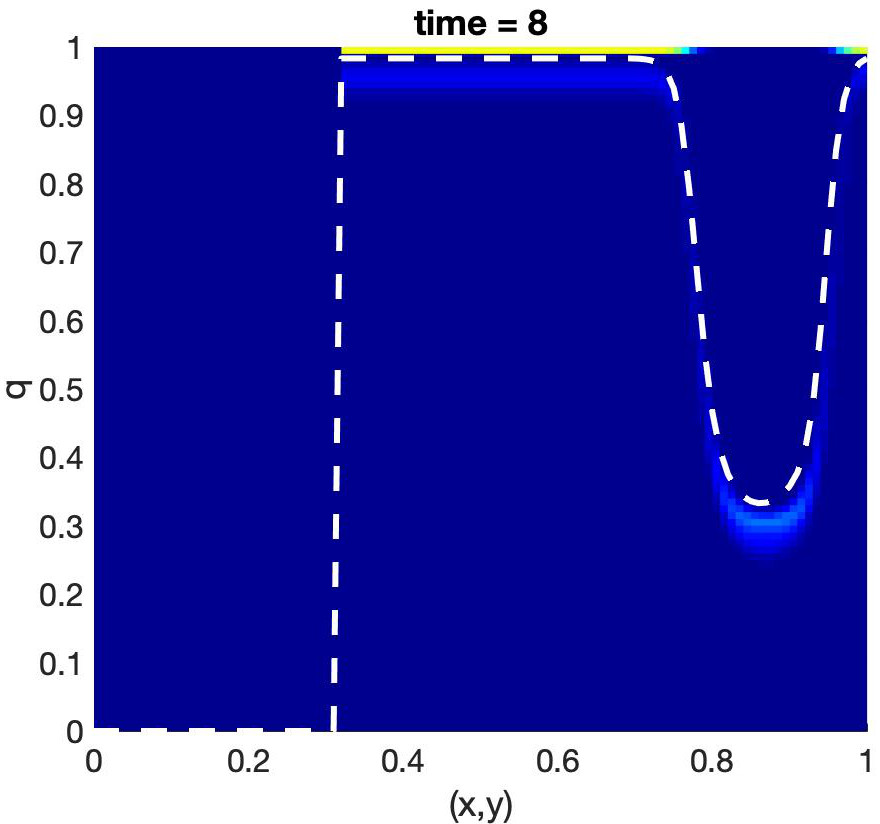}
\end{overpic}
\begin{overpic}[width=0.32\textwidth,grid=false,tics=10]{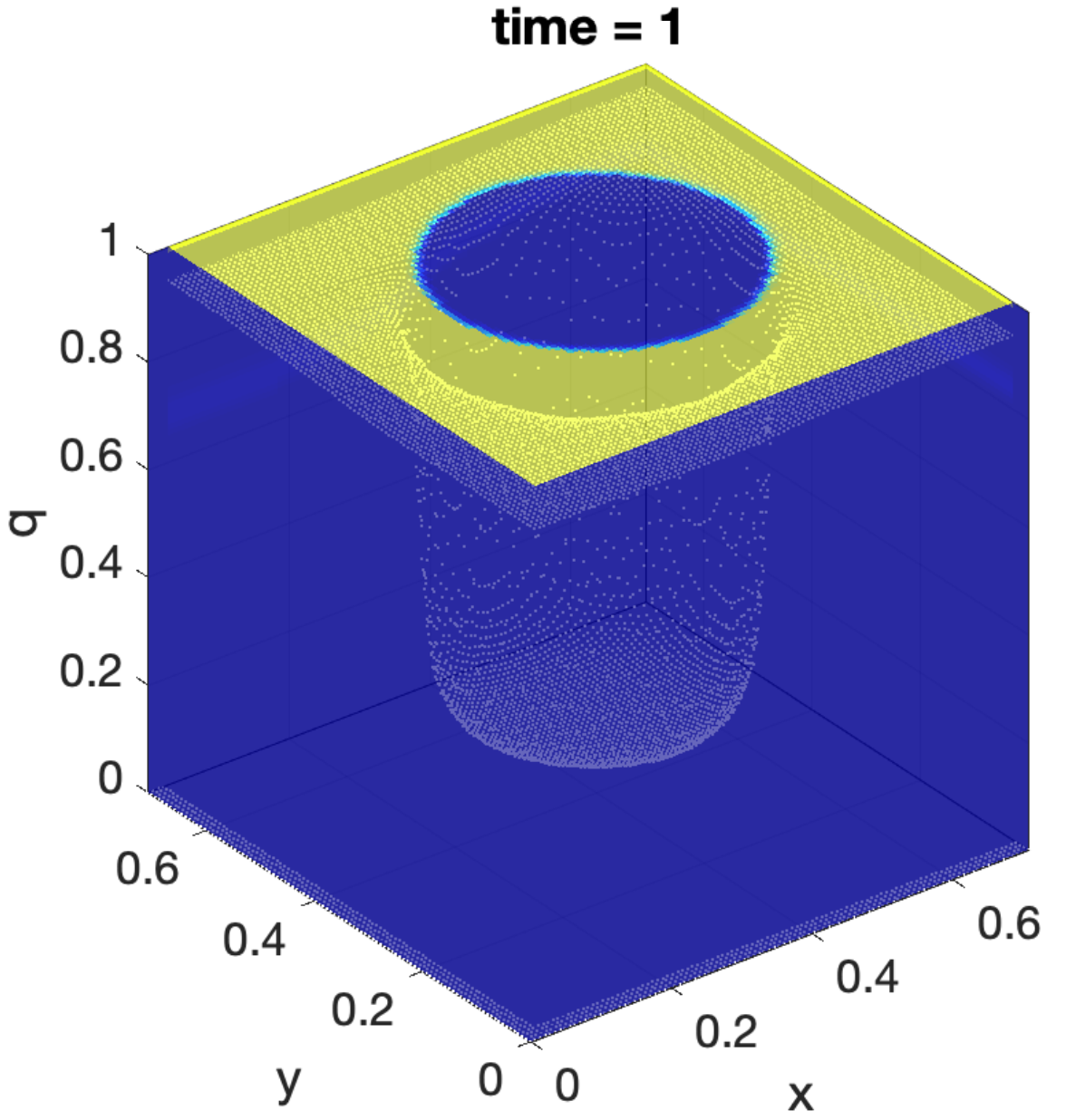}
\put(57.4,46){\color{green}\line(0,1){48.5}}
\put(57.4,46){\color{green}\line(-1,-4){9.8}}
\put(47.6,6.8){\color{green}\line(0,1){49.7}}
\put(57.1,94.2){\color{green}\line(-1,-4){9.5}}
\put(-18,45){IC2-bis}
\end{overpic} 
\begin{overpic}[width=0.32\textwidth,grid=false,tics=10]{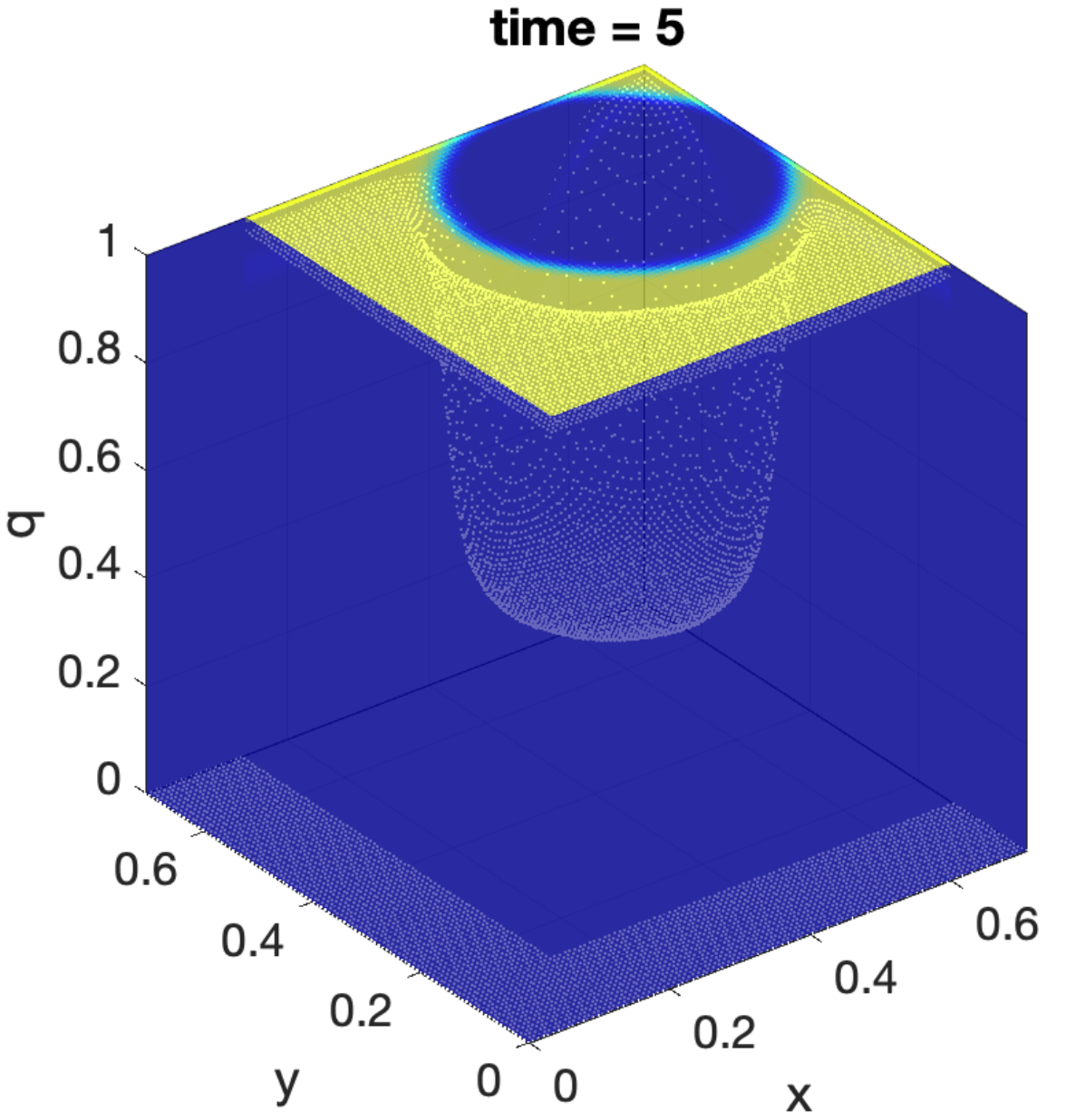}
\put(57.4,46){\color{green}\line(0,1){48.5}}
\put(57.4,46){\color{green}\line(-1,-4){9.8}}
\put(47.6,6.8){\color{green}\line(0,1){49.7}}
\put(57.1,94.2){\color{green}\line(-1,-4){9.5}}
\end{overpic}
\begin{overpic}[width=0.32\textwidth,grid=false,tics=10]{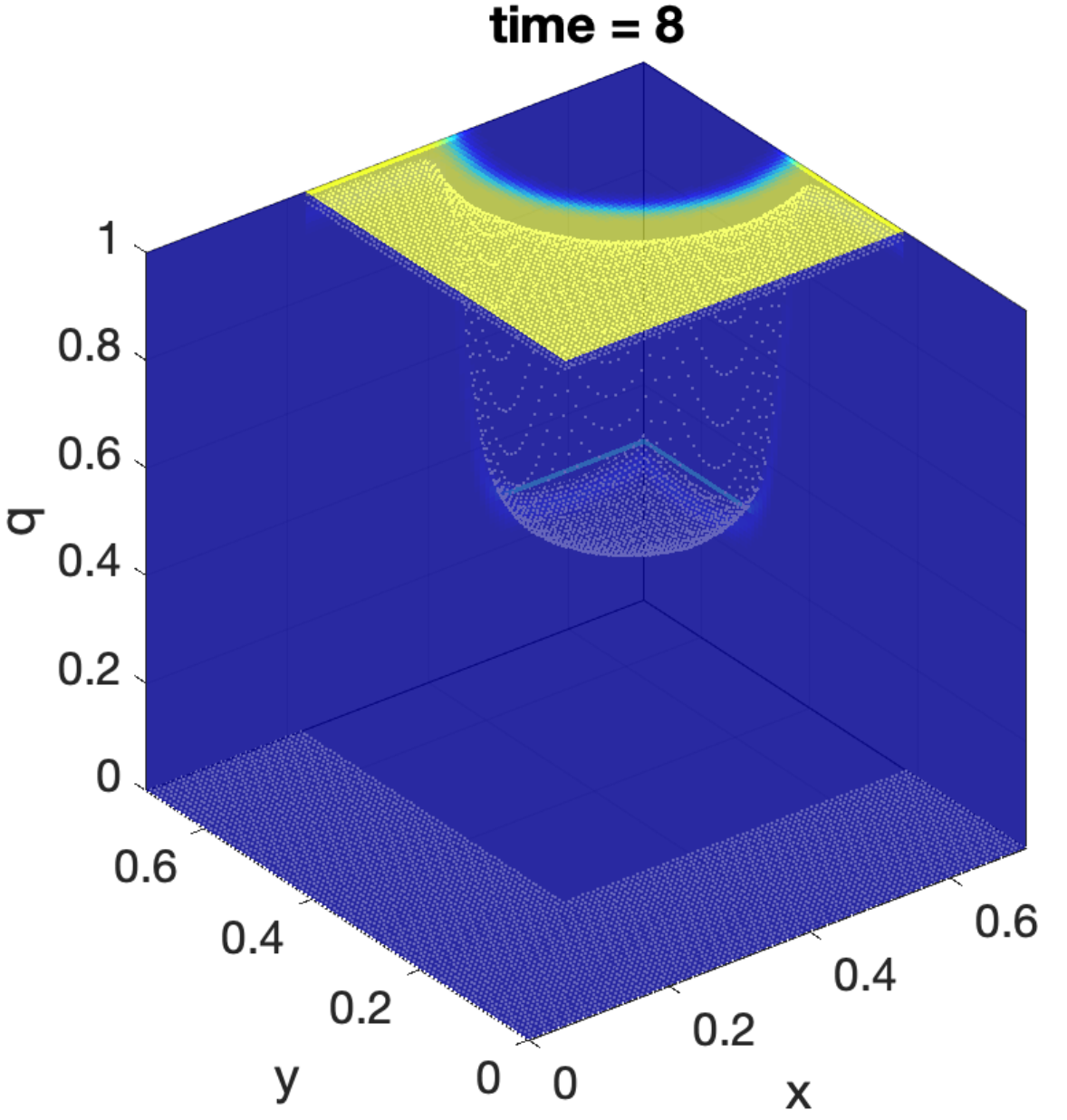}
\put(57.4,46){\color{green}\line(0,1){48.5}}
\put(57.4,46){\color{green}\line(-1,-4){9.8}}
\put(47.6,6.8){\color{green}\line(0,1){49.7}}
\put(57.1,94.2){\color{green}\line(-1,-4){9.5}}
\end{overpic}
\begin{overpic}[width=0.31\textwidth,grid=false,tics=10]{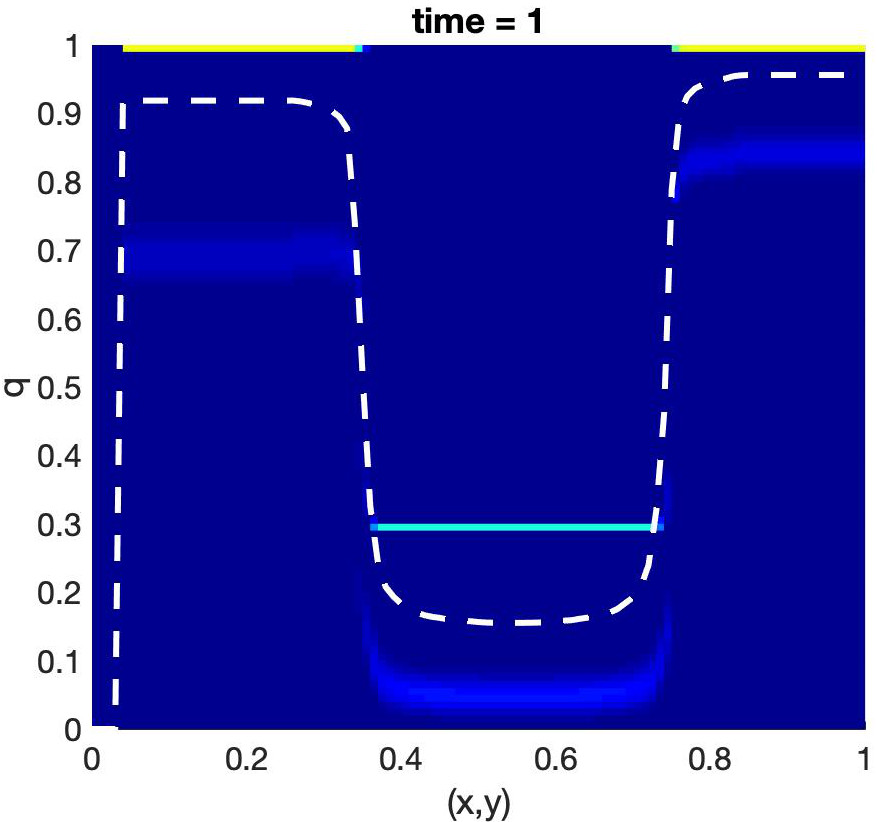}
\put(-22,43){IC2-bis}
\end{overpic} 
\begin{overpic}[width=0.31\textwidth,grid=false,tics=10]{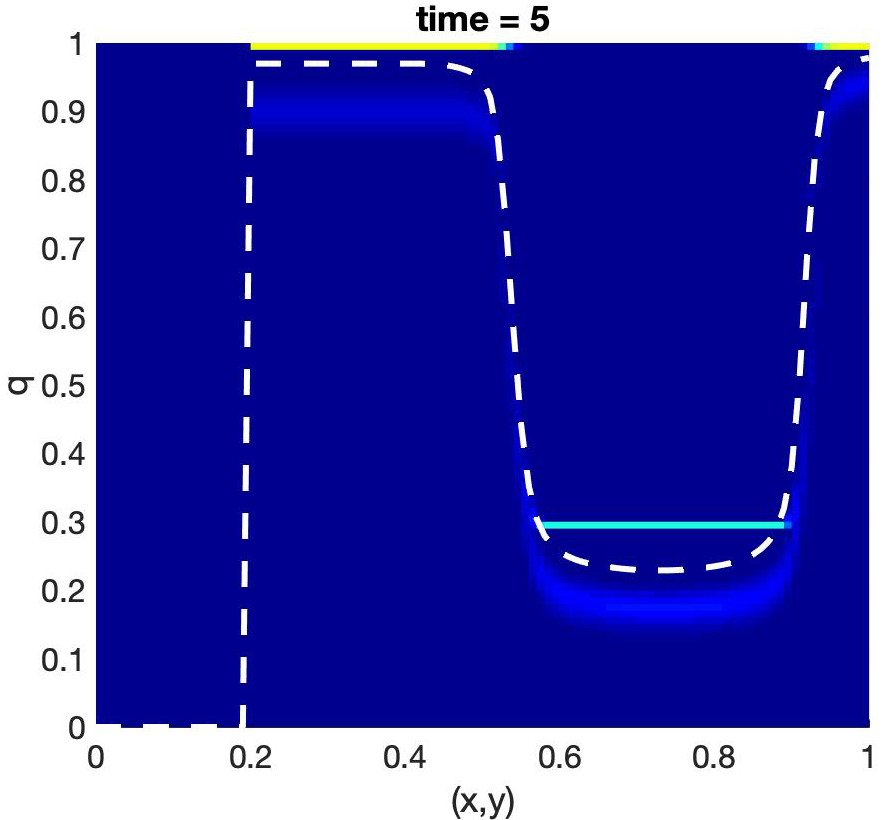}
\end{overpic}
\begin{overpic}[width=0.31\textwidth,grid=false,tics=10]{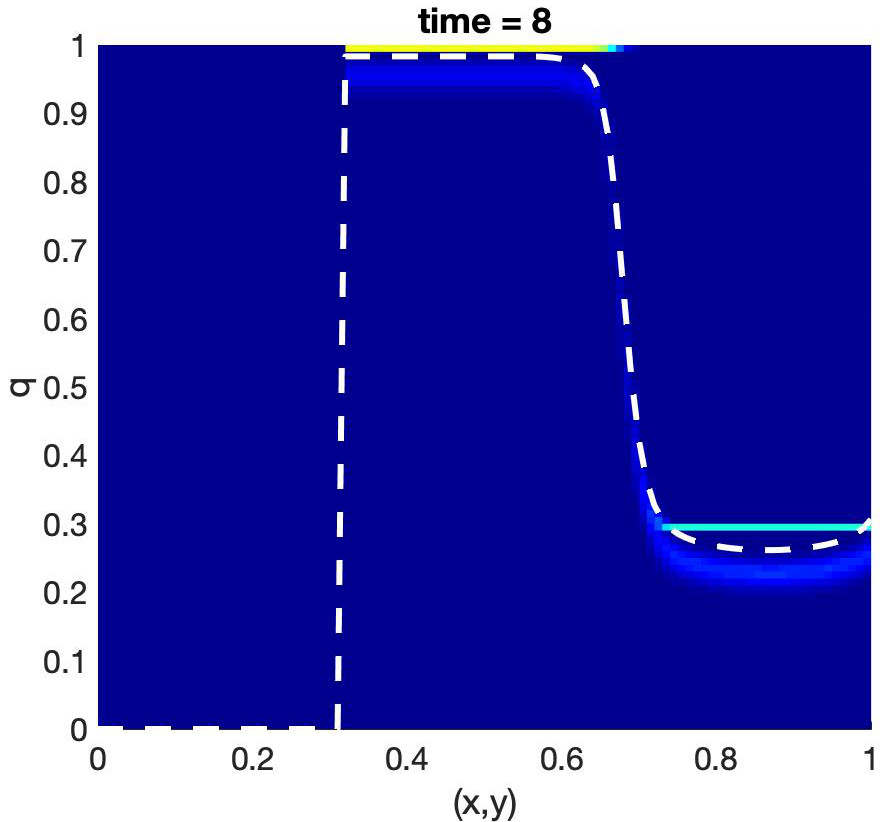}
\end{overpic}
\caption{ %\anna{Daewa, please adjust the green shade}
Tests with prescribed walking velocity: 
the first row shows the evolution of the distribution density $h$ for initial condition IC1-bis and
highlighted in green the section whose results are reported in the second row. 
The third row shows the evolution of $h$ for initial condition IC2-bis and
highlighted in green the section whose results are reported in the fourth row. 
Note that in all the subfigures time is dimensional while space is non-dimensional.
In both cases, we set $\gamma = 50$.
The white dashed line in the images on row two and four represents $q^*$.
The legend for all the images is the same as in Fig.~\ref{fig:ICbis}.} 
\label{fig:v1_ICbis}
\end{figure}

Finally, we modify the initial conditions to show that our model can handle also
more complex scenarios featuring uncertainties. IC1 and IC2 are changed to:
\begin{itemize}
\item[-] IC1-bis: people are positioned like in IC1 but the probabilities of finding people with $q = 1$
and $q = 0$ is reduced and another value of $q$ is assigned at a given $(x,y)$, i.e.~at every location 
in the walkable domain the initial distribution has two bumps in $q$. See Fig.~\ref{fig:ICbis} (left).
\item[-] IC2-bis: people are positioned like in IC2 but the probabilities of finding people with $q = 1$
and $q = 0$ is reduced and, similarly to IC1-bis, another value of $q$ is assigned at a given $(x,y)$. 
See Fig.~\ref{fig:ICbis} (right).
\end{itemize}
Fig.~\ref{fig:v1_ICbis} shows the evolution of the distribution density $h$ for initial conditions IC1-bis and
IC2-bis with $\gamma = 50$. 

\section{A more complex 2D kinetic model}\label{sec:complex}

In this section, we remove the simplifying assumption used in Sec.~\ref{sec:contagion},
i.e.~walking speed $v$ and direction $\theta$ are no longer assumed to be given. 
Instead, they result from the interaction with the surrounding people. Following \cite{Agnelli2015,kim_quaini,kim_quaini2021}, 
we account for the granular feature of the system (i.e., the fact that the distance between pedestrians can range from small to large)
by discretizing $\theta$. Instead of being continuous, variable $\theta$  can only take values in the set:
\[ I_{\theta}=\left \{ \theta_{i}= \frac{i-1}{N_d} 2\pi : i = 1, \dots, N_d \right \}, \]
where $N_d$ is the maximum number of possible directions. As for the walking speed $v$,
we assume it depends on the local level of congestion as demonstrated by experimental studies.
Thus, variable $v$ is treated as a deterministic variable. 

Let $\x=(x,y)$ denote position. For the above mentioned assumptions on $\theta$ and $v$, 
the distribution function can be written as
\begin{equation}\label{eq:f}
f(t, \x, \theta)= \sum_{i=1}^{N_d} f^i(t, \x)\delta(\theta - \theta_i),
\end{equation}
where $f^i(t, \x)$ represents the people that, at time $t$  and position $\x$, move 
with direction $\theta_i$, and $\delta$ denotes the Dirac delta function.
Like for the model in Sec.~\ref{sec:contagion},
it is convenient to switch to non-dimensional variables: position $\hat{\x}=\x/D$, walking speed $\hat{v}=v/V_M$, 
time $\hat{t}=t/T$, and distribution function $\hat{f}=f/ \rho_M$. Once again, we will  drop the hat 
for ease of notation though, with the understanding that all variables from now on are non-dimensional.
Due to the normalization of $f$ and the $f_i$ functions, the dimensionless local density is obtained 
by summing the distribution functions over the set of directions:
\begin{align}\label{eq:rho}
\rho(t, \x)=\sum_{i=1}^{N_d}f^i(t, \x).
\end{align}

The model in \cite{Agnelli2015,kim_quaini,kim_quaini2021} features two key parameters:
\begin{itemize}
\item[-] $\alpha \in [0, 1]$, which represents the quality of the walkable domain: 
if $\alpha=1$ the domain is clear, while if $\alpha=0$ an obstruction is present. Note that
$\alpha$ could be dependent on space (and possibly time) in a prescribed way. 
\item[-] $\varepsilon \in [0,1]$, which represents the overall level of stress: $\varepsilon=0$ indicates no
stress, while $\varepsilon=1$ indicates a high stress stress situation. 
\end{itemize}
Parameter $\alpha$ plays a role in determining the walking speed: where $\alpha=1$ people 
can walk at the desired speed (i.e., $V_M$), while where $\alpha=0$ people are forced to slow down or stop.
See, e.g., \cite{kim_quaini} on how to define the walking speed $v=v[\rho](t, \x, \alpha)$,
with the square brackets denoting that $v$ depends on $\rho$ in a functional way. 
Parameter $\varepsilon$ plays a role in the selection of the walking direction when interacting
with other pedestrians. Indeed, the choice of the waking direction among the possible directions in $I_{\theta}$
results from four competing factors:
\begin{enumerate}
\item The goal to reach a target, like, e.g., the gate of an airport terminal.
\item The desire to avoid collisions with the walls and other obstacles.
\item The tendency to look for less crowded areas.
\item The tendency to follow the stream or ``herd''.
\end{enumerate}
Parameter $\varepsilon$ weights between 3 (prevailing when stress is low)
and 4 (prevailing when stress is high).

In order to explain how the four factors are modeled, we need to introduce some
terminology. Interactions involve three types of active particles (or people): 
\begin{itemize}
\item[-] \textit{test particle} with state $(\x, \theta_i)$: they are representative of the whole system;
\item[-] \textit{candidate particle} with state $(\x, \theta_h)$: they can reach in probability the state of the test particle 
after individual-based interactions with the venue (walls, obstacles, or target) or with field particles; 
\item[-] \textit{field particle} with state $(\x, \theta_k)$: their interactions with candidate particles determines a 
possible change of state.
\end{itemize}
Note that if a test particle changes their state (in probability) into that of the test of the test particle as a result of the 
interactions with field people, the test particle loses their state. 

We call $\mathcal{A}_h(i)$ the probability that a candidate particle with direction $\theta_h$ 
adjusts their direction to $\theta_i$ (the direction of the test particle) due to the interaction with the venue.
The transition probability that the candidate particle changes their direction to $\theta_i$ (direction of the test particle) in the 
search for less crowded areas if their stress level is low or to $\theta_k$ (direction of the field particle)
if their stress level is high is denoted by $\mathcal{B}_{hk}(i)$.
The sets of all transition probabilities $\mathcal{A}=\{\mathcal{A}_h(i) \}_{h,i= 1, \dots, N_d}$ 
and $\mathcal{B}=\{\mathcal{B}_{hk}(i) \}_{h,k,i= 1, \dots, N_d}$
form the so-called \textit{tables of games} that model the game played by active particle interacting with the venue
and other particles, respectively. See, e.g., \cite{kim_quaini} for more details. 

In order to state the mathematical model, we need two more ingredients:
\begin{itemize}
\item[-] the \textit{interaction rate} with the venue $\mu[\rho]$ : it models the frequency of interactions 
between candidate particles and the venue and it is assume to have a functional dependence on $\rho$. In fact, 
it is easier for pedestrians to see the walls, possible obstacles, and their target if density is low. 
\item[-] the \textit{interaction rate} with other pedestrians $\eta[\rho]$: it defines the number of binary encounters per unit time.
This rate depends on $\rho$ in a functional way too since if the local density increases the interaction rate also increases.
\end{itemize}
For possible ways to set $\mu[\rho]$ and $\eta[\rho]$, see, e.g., \cite{kim_quaini}.

In order to derive the mathematical model, one has to take the balance of particles in an elementary volume
of the space of microscopic states, considering the net flow into such volume due to
transport and interactions. Let $\mathcal{J}^i[f]$ be the net balance of particles 
that move with direction $\theta_i$:
\begin{align}
&\mathcal{J}^i[f](t, \x,q)  = \mu[\rho] \left( \sum_{h = 1}^{N_d} \int_{0}^1 \mathcal{A}_h(i) f^h(t, \x,q_h) dq_h - f^i(t,\x,q) \right) \cl
&\quad +\eta[\rho] \left( \sum_{h,k = 1}^{N_d} \int_{0}^1 \int_{0}^1 \mathcal{B}_{hk}(i)  f^h(t, \x,q_h)f^k(t, \x,q_k) dq_h dq_k - f^i(t,\x,q) \rho(t, \x)\right) \label{eq:J}
\end{align}
due to interactions with the venue (first term at the left-hand side of \eqref{eq:J}) and with the surrounding particles
(second term at the left-hand side of \eqref{eq:J}).
Then, the model is given by: 
\begin{align}
\frac{\partial f^i}{\partial t} &+ \nabla \cdot \left(v (\cos \theta_i, \sin \theta_i)^T f^i(t, \x,q) \right) \cl
& = \mathcal{J}^i[f](t, \x,q) + \gamma \frac{\partial(\max \{(q^\ast-q), 0\} f^i)}{\partial q} 
\label{eq:model}
\end{align}
for $i= 1,2, \dots, N_d$.  
Problem~\eqref{eq:J}-\eqref{eq:model} is completed with equation~\eqref{eq:rho}
for the density and a suitable law $v=v[\rho](t, \x, \alpha)$ that relates the walking speed 
to the density. 

While modeling motion in 2 dimensions ($x$ and $y$), eq.~\eqref{eq:model} is a 3D problem in variables
$x, y$, and $q$. Its numerical approximation requires a carefully designed scheme to contain 
the computational cost. Here, we are only going to present ideas on how to design such a scheme. 
Numerical results will be presented in a follow-up paper. Following what we have done in \cite{kim_quaini2020b}
for emotional contagion, we will split model~\eqref{eq:model}
into subproblems that are easier to solve using operator splitting (see, e.g., \cite{glowinski2003finite}). For example, with
the Lie splitting algorithm one could break problem~\eqref{eq:model} into: (i) a pure advection problem that
features the contagion term, i.e.~the last term in eq.~\eqref{eq:model}, and (ii) a problem involving the interaction with the venue 
and other pedestrians. Then, suitable finite difference scheme will be applied for the space and
time discretization of problems (i) and (ii).

%% conclusion  or discussion %%%%%%%%%%%%%%%%%%%%%
\section{Conclusions}\label{sec:concl}
%%%%%%%%%%%%%%%%%%%%%%%%%%%%

We presented a way to design a kinetic type model to simulate the early stage of an infectious disease 
spreading in an intermediate size population occupying a confined environment 
for a short period of time. 

In order to focus on the spreading mechanism, in the first part of the paper 
we adopted a strong simplifying assumption: people's walking speed and direction are given.
The disease spreading is modeled using three main ingredients: an additional variable for the level of exposure to people 
spreading the disease, a parameter for the contagion interaction strength, 
and a kernel function that is a decreasing function of the distance between a person and a spreading individual.
For such simplified model, we propose a first-order and a second-order finite difference scheme. 
We tested our numerical approach on simple 2D problems and verified that the level of exposure evolved 
as expected, even in scenarios featuring uncertainties.

In the second part of the paper, we used ideas from the simplified model to incorporate 
disease spreading in a kinetic theory approach for crowd dynamics. The appealing 
feature of kinetic (or mesoscopic) models for crowd dynamics is the flexibility in accounting for multiple interactions
(hard to achieve in microscopic models) and heterogeneous behavior in people
(hard to achieve in macroscopic models). Modeling crowd dynamics and disease contagion
in two spatial dimensions requires the solution of a 3D problem, the additional variable being
the contagion level. The design of an efficient numerical scheme for such problem will be object
of a follow-up paper.

% 
%
%
%The obvious next step is to combine the kinetic type model for crowd dynamics in the first part
%of the paper with the disease contagion model in order to drop the simplifying assumption, i.e.~walking 
%speed and direction are provided by the model instead of being given.

\begin{acknowledgement}
This work has been partially supported by NSF through grant DMS-1620384.
\end{acknowledgement}

\bibliographystyle{plain}
\bibliography{contagion_model_update}

\end{document}